\documentclass[compress]{elsarticle}

\usepackage{textcomp}
\usepackage[english]{babel}
\usepackage[utf8]{inputenc}
\usepackage{amsmath}
\usepackage{mathrsfs,amsmath}
\usepackage{indentfirst}
\usepackage{subcaption}
\usepackage[export]{adjustbox}
\usepackage[table]{xcolor}
\usepackage{graphicx}
\usepackage{adjustbox}
\usepackage{hyperref}
\usepackage[makeroom]{cancel}
\usepackage{floatrow}
\usepackage{array}

\usepackage{appendix}
\usepackage{xcolor}
\usepackage{mathtools}
\usepackage{multirow}
\usepackage{multicol,lipsum}
\usepackage{amsfonts}
\usepackage[top=2.5cm, bottom=2cm, left=2.5cm, right=2cm]{geometry}

\usepackage{anyfontsize}
\usepackage{t1enc}

\usepackage[ruled,vlined]{algorithm2e}

\usepackage{caption}
\usepackage{subcaption}
\usepackage{verbatim}

\usepackage{ulem}

\usepackage[numbers]{natbib}
\setcitestyle{numbers,sort&compress}

\usepackage{titlesec}
\titleformat{\paragraph}
{\normalfont\normalsize\bfseries}{\theparagraph}{1em}{}
\titlespacing*{\paragraph}
{0pt}{3.25ex plus 1ex minus .2ex}{1.5ex plus .2ex}

\begin{document}
	
\begin{center}
{\textbf{\Large A State-Space Framework for trivial and Topological Metamaterial Stochastic Analysis}}

Luiz Henrique M. S. Ribeiro$^{1,2,3*}$, Abhilash Sreekumar$^3$, Vinícius F. Dal Poggetto$^4$, Vicent Romero-García$^4$, Dimitrios Chronopoulos$^3$ Carlos De Marqui Jr.$^2$, José Roberto F. Arruda$^1$
\end{center}

\noindent
 $^1$ University of Campinas, School of Mechanical Engineering, Campinas-SP, 13083-860, Brazil \\
 $^2$ University of São Paulo, São Carlos School of Engineering, Department of Aeronautical Engineering, 13566-590, São Carlos, Brazil\\
 $^3$ KU Leuven, Faculty of Engineering Technology, Department of Mechanical Engineering, LMSD Division, Ghent Campus, 9000, Belgium\\
 $^4$ Instituto Universitario de Matemática Pura y Aplicada, Departamento de Matemática Aplicada, Universitat Politècnica de València, Camino de Vera, s/n 46022 València, Spain\\
 $^*$ Corresponding author: luiz.marra@outlook.com\\

\noindent {\bf Abstract:} Topological phononic crystals and elastic metamaterials support edge states defined by global topological invariants, offering a route toward vibration-control and wave-guiding devices that remain functional in the presence of defects. However, the spatial variability inherent to manufacturing processes can perturb these invariants and compromise the very robustness that
motivates their use, so that quantifying this sensitivity is essential during the design stage. Although several stochastic approaches have been proposed to quantify the effects of manufacturing variability on periodic waveguides, efficient formulations capable of handling arbitrary spatial variations in geometry and material properties while preserving the analytical description required to evaluate topological invariants remain limited.  In this work, we first demonstrate that a previously proposed linear time-varying (LTV) formulation is mathematically equivalent to the spectral element method based on transfer matrices for the elementary rod, Saint-Venant shaft, and Euler-Bernoulli beam theories. The deterministic formulation is then extended to stochastic analyses via a stochastic linear time-varying (SLTV) framework. The proposed methodology combines Monte Carlo simulations with the stochastic Fourier series and an analytical Karhunen–Loève expansion, providing closed-form stochastic fields and their derivatives, which are required by the LTV formulation. The proposed SLTV approach enables the computation of stochastic dispersion diagrams and forced responses of one-dimensional waveguides with arbitrarily varying geometry and mechanical properties. Because the LTV-based transition matrix isolates individual wavemodes without requiring the mode-tracking step needed by conventional eigenproblem-based formulations, the framework is particularly well suited for evaluating topological invariants, specifically the Zak phase, and for assessing the robustness of the underlying topological bands under spatial variability. Numerical results obtained for various types of one-dimensional waveguides, e.g., rods, shafts, and Euler-Bernoulli beams, demonstrate the applicability of the proposed framework, showing that it provides a unified methodology for deterministic and stochastic analyses of trivial and topologically periodic waveguides under continuous spatial uncertainty.

	
\section{Introduction}

Periodic elastic metamaterials and phononic crystals can be engineered to manipulate the propagation of elastic waves through the design of their constitutive geometry and material properties \cite{arrudamanipulating}, with
applications spanning vibration isolation, seismic vibration isolation, and energy harvesting \cite{colombi2016seismic,lou2022nonlinear,chen2022ternary,zhang2023low, liu2023reprogrammable,lee2022piezoelectric}. This capability gives rise to frequency bandgaps, i.e., frequency intervals where wave propagation is prohibited, generated either by Bragg scattering, when the wavelength is
comparable to the structural periodicity \cite{liu2012wave,dal2023bioinspired}, or by locally resonant mechanisms capable of producing attenuation in the subwavelength regime \cite{liu2000locally,miranda2019flexural,dal2021flexural}. As is well established, Bloch-Floquet periodicity allows these dispersion characteristics to be obtained from the analysis of a single unit cell.

Within this broad class of periodic systems, topological phononic crystals and elastic metamaterials have attracted particular attention in recent years for their ability to support edge states protected by global topological invariants \cite{liu2020topological, zhu2023topological}, offering the prospect of vibration-control and wave-guiding devices that remain functional in the presence of defects or fabrication imperfections. These topological states arise from breaking time or spatial symmetries in periodic structures, encompassing distinct topological classes such as the Su-Schrieffer-Heeger (SSH) model \cite{carta2026flexural, zhang2025multiple}, the Quantum Anomalous, Valley, and Spin Hall effects \cite{chen2025lightweight, qi2022valley, ni2023robust, zhang2025low}, and Higher-Order Topological Insulators \cite{yang2020helical, luo2023efficient, lin2023tuning, wu2022higher}. Among these, the SSH model represents the fundamental bridge connecting conventional phononic-crystal design to topological systems, enabling the translation of topological states into practical, optimized elastic structures \cite{dong2024topological, xue2022topological}. Although practical topological metastructures are generally two- or three-dimensional, many of them are assembled from one-dimensional periodic waveguides -- such as longitudinal rods, Saint-Venant shafts, and Euler-Bernoulli or Timoshenko beams -- whose dynamic behavior provides the fundamental building blocks for more complex architectures \cite{ribeiro2022investigating}.

Realizing these designs in practice inevitably introduces spatial variability in geometry and material properties, arising from manufacturing tolerances and process-induced imperfections. Research interest in quantifying such variability has been particularly stimulated by the renewed adoption of additive manufacturing \cite{beli2019wave,souza2020bayesian,li2023algebraic, zhang2018robust}, although spatial variability of this kind is, in fact, inherent to essentially any fabrication route, whether additive or subtractive. For trivial (non-topological) phononic crystals, such variability primarily shifts bandgap edges and degrades attenuation performance, effects that can be quantified using established stochastic methods, as discussed below. For topological metamaterials, however, the consequences can be more severe, since spatial variability may directly compromise the topological invariants assumed to guarantee robustness. Indeed, new studies on the robustness of edge states in elastic topological structures are beginning to better understand the so-called topological protection \cite{ribeiro2025robustness}. In addition, in contrast to the topological protection typically invoked in the photonics literature \cite{rosiek2023observation}, spatially correlated variability in phononic crystals can further alter or suppress this energy localization \cite{ribeiro2025robustness}, underscoring the need for methods capable of quantifying how manufacturing-induced disorder affects topological invariants.

Efficiently identifying these topological invariants first requires accurate and efficient dispersion computations. The numerical computation of dispersion relations has been extensively investigated over the last few decades. Early developments relied on transfer-matrix formulations \cite{mead1970free, faulkner1985free}, followed by several numerical techniques for periodic media \cite{vazeille2024envelope}. Among them, finite element method (FEM)-based formulations combined with Bloch-Floquet theory remain the most widely adopted approach \cite{floquet1883equations,bloch1929quantenmechanik,mace2008modelling}, including the transfer matrix method (TMM) \cite{mace2005finite,zhong1995direct} and the wave and finite element method (WFEM) \cite{mace2008modelling}. Alternative formulations include the plane wave expansion method (PWEM), which is better suited for structures with piecewise homogeneous properties \cite{dal2020elastic}, and state-space formulations based on Riccati equations, which have proven effective for waveguides with complex cross-sections \cite{assis2019computing}. Within the finite-element framework, the dynamic stiffness matrix can be constructed using either conventional finite elements or spectral elements: conventional FEM requires a sufficiently refined spatial discretization to accurately capture wave propagation \cite{langer2017more}, increasing the number of eigenmodes and becoming computationally demanding \cite{rojas2024computationally}, especially at medium and high frequencies, whereas the spectral element method (SEM) employs analytical solutions in the frequency domain, providing excellent accuracy with a significantly reduced number of elements \cite{doyle1989wave,lee2001dynamic,ahmida2001spectral}. Nevertheless, spectral formulations are available only for a relatively limited class of structural models and are generally less flexible than conventional finite-element formulations when dealing with arbitrary spatial variations \cite{assis2019computing}.

Quantifying how such numerically computed dispersion relations are affected by spatial variability requires dedicated stochastic methods. Several methodologies have been proposed for this purpose, including stochastic spectral element formulations \cite{beli2019wave,machado2018spectral}, homogenization-based approaches \cite{white2022topological}, Bayesian inference methods \cite{souza2020bayesian,morris2018design,ribeiro2022bloch}, hierarchical correction strategies \cite{balla2022hierarchical}, and experimental techniques for stochastic wavenumber identification \cite{souza2020bayesian,li2022algebraic,ribeiro2022bloch,li2023wavenumber}. Despite these developments, efficient formulations capable of simultaneously handling continuously varying geometry, continuously varying material properties, and stochastic spatial variability remain relatively scarce. A further challenge arises because several wave-propagation formulations require analytical expressions not only for the spatially varying properties themselves but also for their derivatives \cite{gan2014longitudinal, claro2026numerical}; consequently, many stochastic representations based on discretized random fields cannot be directly incorporated into these formulations without additional approximation procedures. This limitation is particularly relevant for state-space approaches, such as those required to isolate individual wavemodes for topological characterization, in which the governing equations explicitly depend on the derivatives of the spatially varying properties.

Recently, the authors proposed a linear time-varying (LTV)-based formulation for the computation of dispersion diagrams and forced responses of one-dimensional waveguides with continuously varying geometry and material properties \cite{ribeiro2023computing}. The method combines the analytical state-space formulation proposed by Gan \textit{et al.} \cite{gan2014longitudinal} with numerical approximations for the transition matrix originally developed for linear time-varying systems \cite{hsu1974approximating,friedmann1977efficient}. Unlike conventional discretization techniques, the LTV formulation naturally accommodates arbitrary continuous spatial variations within a unified state-space framework while preserving the analytical description of the governing equations, directly addressing the derivative-based limitation identified above.

However, even for canonical topological models such as the SSH chain, the engineering interpretation of the Zak phase in terms of measurable displacement and force fields, together with its robustness against manufacturing-induced spatial variability, remains insufficiently investigated. Quantifying these effects relies on the precise computation of topological invariants. Nevertheless, existing numerical methods often suffer from ill-conditioning or mode-sorting ambiguities when tracking eigenmodes across the Brillouin zone---particularly within conventional FEM formulations, where paired modes create degeneracies \cite{zhu2018zak, xiao2015geometric, ribeiro2025robustness}. In this context, a state-space formulation capable of isolating individual eigenmodes within a semi-analytical framework offers a distinct computational advantage for topological characterization under uncertainty.

From a computational standpoint, the two routes available for characterizing the topological invariants of a periodic waveguide differ substantially in cost. The continuous definition of the Zak phase requires an integral of the wavemode gradient with respect to the wavenumber over the entire first Brillouin zone, combined with a further integral over the spatial coordinate of the unit cell \cite{xiao2015geometric, ribeiro2025robustness}; in its discretized form, this translates into evaluating and storing the wavemode at every wavenumber station and every spatial station considered, so that the accuracy of the estimated Zak phase improves only as both the wavenumber and the spatial discretizations are refined. For a single passband this may still be tractable, but as soon as several bands, several unit-cell configurations, or a stochastic ensemble of realizations must be evaluated, the cumulative cost of this double integration becomes considerable. The parity-based alternative avoids this cost altogether, since it only requires the parity of the Bloch eigenmode to be evaluated at the two high-symmetry points of the Brillouin zone, $k=0$ and $k=\pi/L$, rather than along the whole passband \cite{xiao2015geometric, xiao2014surface}. However, this shortcut is only as reliable as the mode identification itself: when the wavemodes are obtained from a conventional eigenproblem, as is the case for the FEM, the SEM, or the WFE, the eigensolver returns an unordered set of eigenvectors that must subsequently be tracked and matched across neighboring wavenumbers to identify which eigenvector corresponds to which physical passband  \cite{mace2008modelling, miranda2019flexural}, a task that becomes increasingly ambiguous whenever multiple modes coexist at similar frequencies, as is typically the case for flexural waveguides such as beams \cite{ribeiro2025robustness}. The LTV-based formulation circumvents this additional tracking step because the state transition matrix is built separately for each mode from the outset, so that the wavemodes are already ordered and individually addressable at the position where they are computed; consequently, the parity-inversion diagnostic can be evaluated directly at $k=0$ and $k=\pi/L$ without any auxiliary mode-sorting procedure, making the LTV-based approach considerably more efficient than FEM/SEM/WFE-based alternatives for topological characterization.

\begin{figure}[H]
    \centering
    \includegraphics[width=0.8\textwidth]{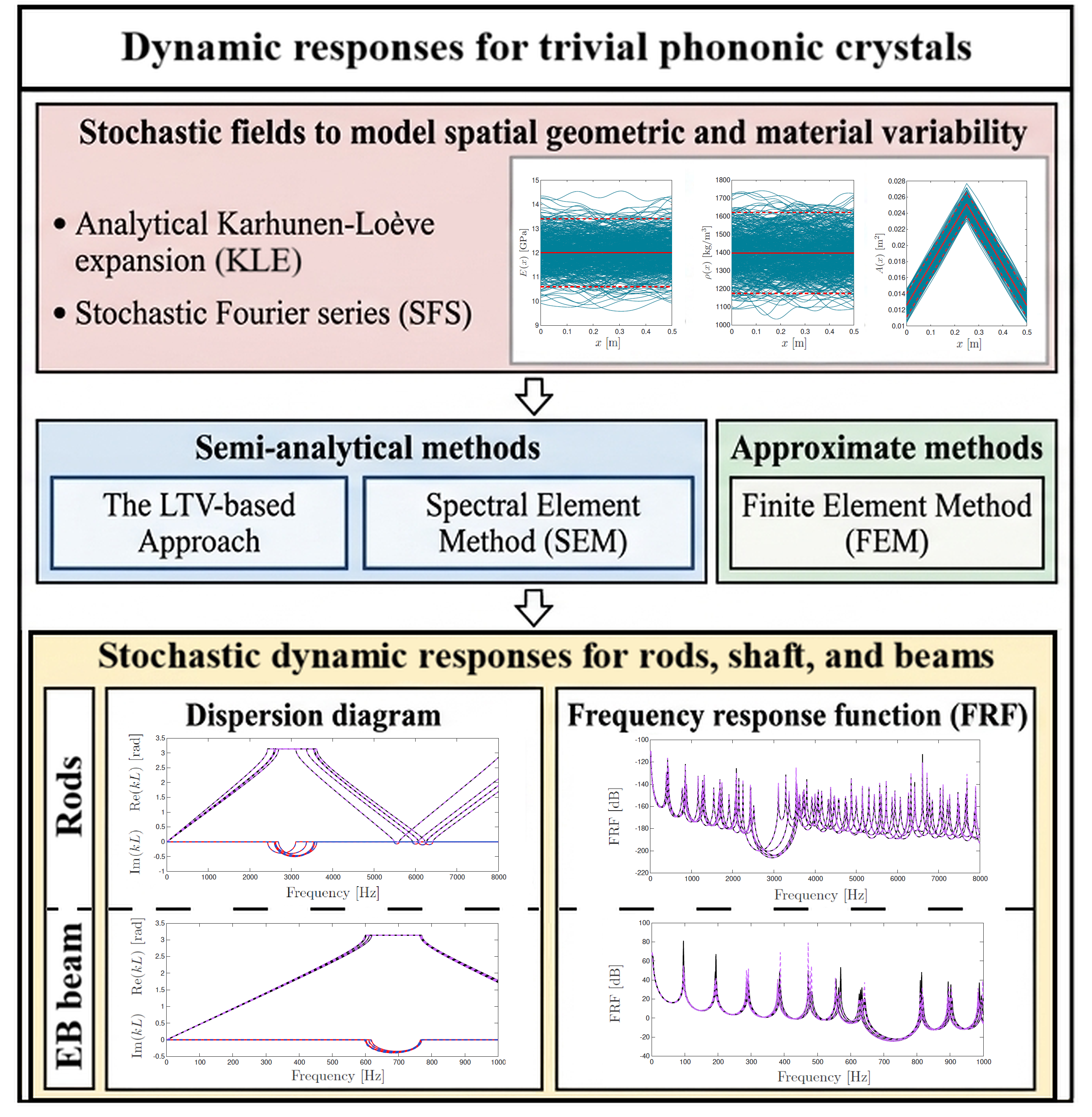}
    \caption{Schematic representation of the methodology adopted for the stochastic dynamic analysis of trivial (non-topological) phononic crystals. Spatially correlated random fields describing geometric and material variability are generated using the analytical Karhunen–Loève (KL) expansion and the stochastic Fourier series (SFS). These fields are propagated through the semi-analytical LTV-based approach and the spectral element method (SEM), and compared against the conventional finite element method (FEM), to compute the stochastic dispersion diagrams and frequency response functions (FRFs) of rods, shafts, and beams.}
    \label{fig_1}
\end{figure}

The main contributions of the present work are fourfold. First, we formally demonstrate that the proposed LTV formulation is mathematically equivalent to the transfer-matrix formulation obtained from the spectral element method for the elementary rod, Saint-Venant shaft, and Euler-Bernoulli beam theories. Second, the deterministic LTV formulation is extended to stochastic analyses, resulting in a stochastic linear time-varying (SLTV) framework. Third, two analytical stochastic field representations, namely the stochastic Fourier series (SFS) and the analytical Karhunen–Loève expansion (KLE), are incorporated into the formulation, providing closed-form stochastic fields together with their spatial derivatives, which are required by the LTV framework. Finally, the proposed methodology is employed to compute stochastic dispersion diagrams and forced responses of one-dimensional periodic waveguides subjected to simultaneous variability in geometry and material properties, as illustrated in Fig. \ref{fig_1}, explicitly evaluating the robustness of topological invariants under continuous spatial variability and disorder.

The remainder of this paper is organized as follows. Section~\ref{Sec2} presents the deterministic LTV modeling and demonstrates its equivalence with the spectral element method. Section~\ref{Sec3} introduces the Zak phase computation using the LTV-based approach. Section~\ref{Sec4} provides deterministic results for conventional and topological phononic crystals. Section~\ref{Sec5} describes the statistical analyses through analytical stochastic field representations based on the SFS and KLE. Section~\ref{Sec6} presents the stochastic results for different one-dimensional structural elements and evaluates the robustness of the topological structures under spatial variability. Finally, the concluding remarks are given in Section~\ref{Sec7}.

\section{Deterministic modeling} \label{Sec2}

Here, we first briefly review the approach to compute dispersion diagrams and
forced responses for one-dimensional unit cells with arbitrarily varying
geometry and mechanical properties presented by the authors in
\cite{ribeiro2023computing}. Then, algebraic manipulations of the equivalence
between the results obtained using the LTV-based method and the SEM using
transfer matrices is presented for the elementary rod, Saint-Venant shaft, and
Euler-Bernoulli beam theories.

\subsection{The LTV-based model}\label{Sec2.1}

In this section, the LTV system approach is applied to a one-dimensional
structure with spatially varying geometric and mechanical properties. It is
worth clarifying the terminology at the outset: although the formulation is
termed linear time-varying (LTV), consistent with its origin in systems and
control theory, this designation does not refer to material or geometric
nonlinearity, nor does it involve an explicit dependence on time. Instead, the
spatial coordinate $x$ plays the role analogous to the independent
("time-like") variable of classical LTV systems, so that "time-varying" here
denotes spatially varying coefficients; the formulation operates entirely in
the frequency (reciprocal) domain.

\begin{figure}[H]
    \centering 

    \begin{subfigure}{0.325\textwidth} 
        \captionsetup{justification=raggedright, singlelinecheck=false}
        \caption{} \label{Fig_3a}
        \includegraphics[width=\textwidth]{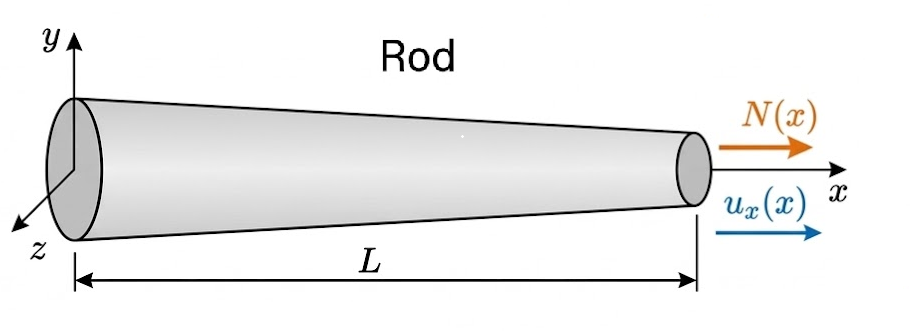} 
    \end{subfigure}
    \begin{subfigure}{0.325\textwidth}
        \captionsetup{justification=raggedright, singlelinecheck=false}
        \caption{} \label{Fig_3b}
        \includegraphics[width=\textwidth]{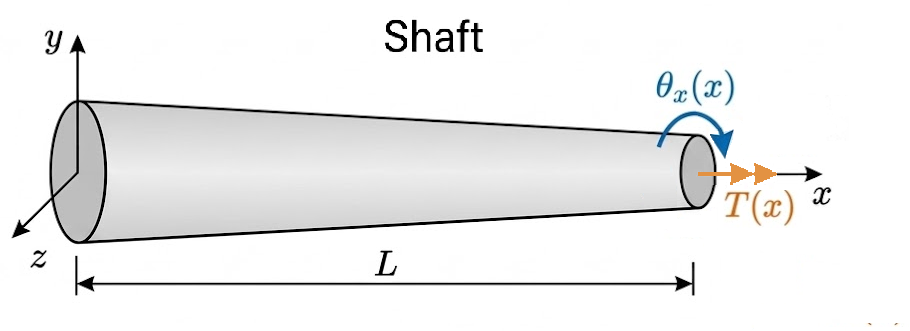}
    \end{subfigure}
    \begin{subfigure}{0.325\textwidth}
        \captionsetup{justification=raggedright, singlelinecheck=false}
        \caption{} \label{Fig_3c}
        \includegraphics[width=\textwidth]{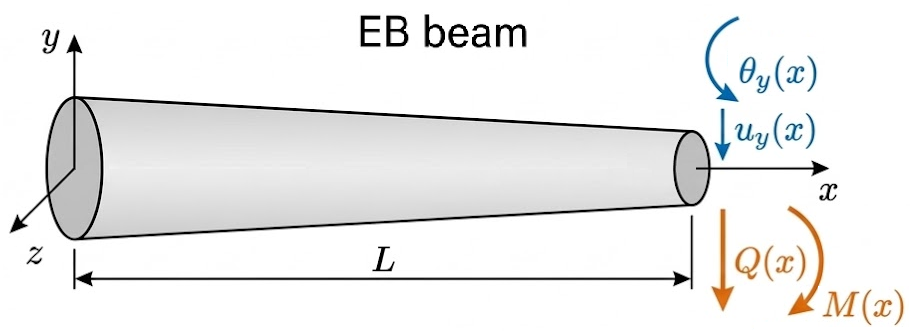}
    \end{subfigure}

    \caption{Schematic comparison of fundamental internal loads (orange) and displacement (blue) for three tapered bar models of length $L$: Rod (a), Shaft (b), and EB beam (c). The $(x, y, z)$ coordinate system is consistent across all representations.}
    \label{Fig_3}
\end{figure}

The spatially varying geometric characteristics are described by a
cross-sectional area $A(x)$, shaft shear coefficient $K_S(x)$, polar moment of
inertia of the cross-section $J(x)$, and second moment of inertia with respect
to the neutral axis $I(x)$.
The material properties are given by the complex Young's modulus $E(x)=E_u(x)(1+i\eta_E)$, where $\eta_E$ is Young's modulus loss factor, $E_u(x)$ is the undamped Young's modulus, and $i$ is the imaginary number; $\rho(x)$ is the mass density, and $G(x)=G_u(x)(1+i\eta_G)$ is the complex shear modulus, where $\eta_G$ is the shear modulus loss factor and $G_u(x)$ the undamped shear modulus.
As shown in Fig.~\ref{Fig_3}, internal forces are denoted as the longitudinal force $N(x)$, the torsional moment $T(x)$, the flexural moment $M(x)$, and the transverse shear force $Q(x)$. Displacements of the unit cells of length $L$ are defined as longitudinal $u_x(x)$, torsional angle $\theta_x(x)$, transverse displacement $u_y(x)$, and rotation angle $\theta_y(x)$. All spatially varying properties are assumed periodic over the unit cell of length $L$, i.e., $E(x+L)=E(x)$, and analogously for $A(x)$, $K_S(x)$, $J(x)$, $I(x)$, $\rho(x)$, and $G(x)$, consistent with the Bloch-Floquet framework adopted throughout this work; in particular, the boundary values of these properties coincide, e.g., $E(0)=E(L)$ and $A(0)=A(L)$.

Consider a general one-dimensional structural element with field displacements $\mathbf{u}(x)$. A first-order state-space representation is obtained by collecting $\mathbf{u}(x)$ and its spatial derivatives up to order $q$ into the state vector $\mathbf{z}(x) = \begin{Bmatrix}\mathbf{u}(x) & \normalsize{\frac{\partial^q \mathbf{u}(x)}{\partial x^q}}\end{Bmatrix}^t$, $q \in \mathbb{N}$ \cite{gan2014longitudinal}, stated as 
\begin{equation}\label{Eq.1}
    \normalsize{\frac{\partial \mathbf{z}(x)}{\partial x}}=  \mathbf{H}(x)\mathbf{z}(x) \, ,
\end{equation}	
where $\mathbf{H}(x)$ encodes the equilibrium (governing) and constitutive relations of the structural element, and is a periodic matrix with the same periodicity as $\mathbf{z}(x)$.

Because $\mathbf{z}(x)$ collects displacement derivatives that are not, by themselves, physically measurable or directly usable for imposing boundary or continuity conditions, it is further related to a vector containing the
 isplacements and the associated internal forces, $\mathbf{y}(x) = \begin{Bmatrix}
    \mathbf{u}(x) & \mathbf{F}(x) \end{Bmatrix}^t$, via
\begin{equation}\label{Eq.2}
    \mathbf{y}(x) = \mathbf{G}(x)\mathbf{z}(x) \, ,
\end{equation}
where $\mathbf{G}(x)$, similarly to $\mathbf{H}(x)$, contains information on the material and geometrical properties of the unit cell. This relation is essential because the physical continuity and equilibrium conditions coupling neighboring unit cells -- and, ultimately, the Bloch periodic boundary condition introduced below -- must be enforced on $\mathbf{y}(x)$, i.e., on displacements and internal forces, rather than on the abstract state vector $\mathbf{z}(x)$. The specific forms of $\mathbf{u}(x)$, $\mathbf{G}(x)$, and
$\mathbf{H}(x)$ for the elementary rod, Saint-Venant shaft, and Euler-Bernoulli beam theories are given in Section~\ref{Sec2.2}.

A state $\mathbf{z}(x)$ evaluated at a reference coordinate $x_0$ can be related to the state at another, general coordinate $x$ through a transition matrix \cite{fivel2022analysis}
	\begin{equation}\label{Eq.3}
		\mathbf{z}(x) = \boldsymbol{\phi}(x,x_0) \mathbf{z}(x_0),
	\end{equation}
where the reference coordinate $x_0$ is arbitrary and need not coincide with the origin of the unit cell; in the remainder of this work it is conveniently taken as $x_0=0$, the unit-cell edge.

The solution of this system is approximated by discretizing the interval $[x_0,x]$ into $N$ equally spaced sub-intervals of length $\Delta_n = (x-x_0)/N$ ($n=1,\dots,N$), such that the $n$-th sub-interval spans $[x_{n-1},x_n]$, with $x_N \equiv x$. Approximating $\mathbf{H}(x)$ by its average value $\mathbf{C}_n$ over each sub-interval (Eq.~\eqref{Eq.5}) reduces Eq.~\eqref{Eq.1} to a piecewise-constant-coefficient linear system, whose exact solution over each sub-interval is a matrix exponential; the transition matrix is then obtained as the ordered product of these exponentials \cite{hsu1974approximating,friedmann1977efficient}
\begin{align}\label{Eq.4}
    \boldsymbol{\phi}(x,x_0)\approx& \ \text{expm}(\Delta_1\mathbf{C}_1)\ \dots\ \text{expm}(\Delta_{N-1}\mathbf{C}_{N-1})\  \text{expm}(\Delta_N\mathbf{C}_N)\  = \prod_{n=1}^{N}\text{expm}(\Delta_n\mathbf{C}_n),
\end{align}
where the product is ordered such that the sub-interval closest to the reference coordinate $x_0$ ($n=1$) is applied first to $\mathbf{z}(x_0)$ and therefore appears as the rightmost factor, consistent with the composition rule in Eq.~\eqref{Eq.3}; the sub-interval closest to $x$ ($n=N$) is applied last and appears as the leftmost factor. \footnote{This ordering should be verified against the specific matrix/vector convention used in the numerical implementation.} In the subsections that follow, this general discretization is specialized to
the unit-cell edges, $x_0=0$ and $x=L$, so that the sub-interval length reduces
to $\Delta_n = L/N$ for all $n=1,\dots,N$. Setting $x=L$ in Eq.~\eqref{Eq.4} recovers the transition matrix between the two edges of the unit cell used in the remainder of this work, where expm$(\cdot)$ denotes a matrix exponential, and for the $n$-th interval ($n \in [1,N]$) the constant matrix $\mathbf{C}_n$ can be obtained via \cite{friedmann1977efficient}
\begin{equation}\label{Eq.5}
    \mathbf{C}_n = \normalsize{\frac{1}{\Delta_n}}\int_{x_{n-1}}^{x_n} \mathbf{H}(s) \, ds.
\end{equation}
	
The matrix exponential of $\Delta_n\mathbf{C}_n$ is obtained through the eigenvalue decomposition $\boldsymbol{\psi}_C\boldsymbol{\Lambda}_C\boldsymbol{\psi}_C^{-1}$, where $\boldsymbol{\psi}_C$ is the matrix whose columns are the eigenvectors of $\Delta_n\mathbf{C}_n$, and $\boldsymbol{\Lambda}_C$ is the corresponding diagonal matrix containing the exponential of its eigenvalues \cite{golub2013matrix}. Further details of the solution are given in \cite{ribeiro2023computing}.

Once the transition matrix is obtained, it can be used in the form $\boldsymbol{\phi}(L,0)$ (see Eq.~\eqref{Eq.3}) to relate $\mathbf{z}(L)$ and $\mathbf{z}(0)$. Hence, the transfer matrix $\mathbf{T}$, relating displacements and internal forces at the two edges of a one-dimensional element, can be computed for a given frequency $\omega$ using
\begin{align}\label{Eq.6}
			\mathbf{y}(L)= \mathbf{G}(L)\mathbf{z}(L)
			=\mathbf{G}(L)\boldsymbol{\phi}(L,x_0)  \mathbf{z}(0)
			=\mathbf{T} \mathbf{y}(0),
\end{align}
where $\mathbf{T} = \mathbf{G}(L) \boldsymbol{\phi}(L,x_0) \mathbf{G}^{-1}(0)$. The term $\mathbf{G}(0)^{-1}$ arises directly from Eq.~\eqref{Eq.2} evaluated at $x=0$, which is inverted to express the state vector in terms of the physical quantities at the unit-cell origin, $\mathbf{z}(0)=\mathbf{G}(0)^{-1}\mathbf{y}(0)$, before propagating it through $\boldsymbol{\phi}(L,x_0)$; the resulting state at $x=L$ is then converted back to displacements and internal forces via $\mathbf{G}(L)$.

By combining the previous relation with the Bloch(-Floquet) periodic boundary condition relating the displacement and force state vectors at the two edges of the unit cell, $\mathbf{y}(L)=e^{ikL}\mathbf{y}(0)$, where $k$ is the (generally complex) Bloch wavenumber, one obtains an eigenproblem whose eigenvalues contain the wavenumbers for a given frequency $\omega$
\cite{beli2016uncertainty}:
	\begin{equation}\label{Eq.7}
		\mathbf{T} \mathbf{y}(0) = e^{ikL} \mathbf{y}(0),
	\end{equation}
with the eigenvectors of this eigenproblem, $\mathbf{y}(0)$, corresponding to the wavemode shapes, hereafter referred to as wavemodes. In the following sections, we demonstrate the equivalence between solutions obtained using this approach and the SEM.

\subsection{Equivalence between the LTV formulation and the SEM}\label{Sec2.2}

The LTV-based formulation presented in the previous section provides the transfer matrix relating the displacement and internal-force state vectors at the two ends of a one-dimensional structural element. Although the numerical results reported in our previous work \cite{ribeiro2023computing} showed excellent agreement with the SEM, a demonstration of the equivalence between both formulations has not yet been established. Such a demonstration is particularly relevant because the SEM is one of the most accurate and widely adopted methodologies for wave propagation analyses in one-dimensional waveguides. Establishing that both formulations produce the same transfer matrix therefore provides a validation of the proposed LTV framework.

It is worth noting that this equivalence is a validation exercise, not the primary motivation for adopting the LTV-based approach: as shown in Section~\ref{Sec3}, once the dynamic-stiffness eigenproblem underlying the SEM is assembled for a spatially varying unit cell, its eigenvectors must still be numerically sorted and tracked across the Brillouin zone to identify individual wavemodes, whereas the LTV-based state-space formulation isolates these wavemodes directly, without requiring this additional tracking step. The computational advantage of the LTV-based approach over the SEM therefore lies specifically in this feature, which becomes particularly relevant for the topological characterization addressed in Section~\ref{Sec3}, rather than in the accuracy or generality of the deterministic dispersion computation established here.

The objective of this section is to demonstrate that the transfer matrix obtained from the LTV formulation is mathematically identical to that derived from the SEM for the elementary rod, Saint-Venant shaft, and Euler--Bernoulli beam theories. The algebraic manipulations are carried out independently for each structural element by comparing the analytical expressions of their transfer matrices. It will be shown that the state transition matrix computed from the LTV formulation using the exponential-product approximation of Eq.~\eqref{Eq.4} exactly reproduces the transfer matrix obtained from the corresponding spectral element formulation.

It should be emphasized that the equivalence established herein relies on the computation of the transition matrix through the exponential-product approximation presented in Eq.~\eqref{Eq.4}, which preserves the analytical structure of the governing equations within each discretized interval. Consequently, these algebraic manipulations does not directly apply to alternative numerical integration procedures, such as Runge--Kutta, Adams--Bashforth \cite{atkinson2009numerical}, or Poincaré--Lindstedt methods \cite{verhulst2007periodic}, which provide different approximations of the state transition operator. The manipulations begin with the elementary rod theory, followed by the Saint-Venant shaft and the Euler-Bernoulli beam theories.

\subsubsection{Elementary rod theory ($r$)}
	
Starting from the equation of motion of a rod in the frequency domain, one obtains \cite{lee2009spectral,qiao2023inverse, nunes2022exact}
	\begin{equation}\label{Eq.8}
		\normalsize{\frac{\partial^2 U_x(x)}{\partial x^2}} + \left( \normalsize{\frac{1}{\frac{A(x)}{\frac{\partial A(x)}{\partial x}}}} + \normalsize{\frac{1}{\frac{E(x)}{\frac{\partial E(x)}{\partial x}}}}\right) \normalsize{\frac{\partial U_x(x)}{\partial x}} + \normalsize{\omega^2\frac{\rho(x)}{E(x)}}U_x(x) = 0.
	\end{equation}

In this case, $\mathbf{z}_r(x)=\begin{Bmatrix} U_x(x) & \normalsize{\frac{\partial U_x(x)}{\partial x}}\end{Bmatrix}^t$, $\mathbf{y}_r(x) = \begin{Bmatrix}U_x(x) &  N(x)\end{Bmatrix}^t$, the internal axial force is expressed as $N(x)=E(x)A(x)\normalsize{\frac{\partial U_x(x)}{\partial x}}$, and matrices $\mathbf{H}_r(x)$ and $\mathbf{G}_r(x)$ can be written as \cite{ribeiro2023computing}
\begin{equation}\label{Eq.9}
	\mathbf{H}_r(x) = \begin{bmatrix}
		0 & 1\\
		-\normalsize{\omega^2\frac{\rho(x)}{E(x)}} & -\left( \normalsize{\frac{1}{\frac{A(x)}{\frac{\partial A(x)}{\partial x}}}} + \normalsize{\frac{1}{\frac{E(x)}{\frac{\partial E(x)}{\partial x}}}}\right)
	\end{bmatrix} \hspace{0.5cm} \text{and} \hspace{0.5cm} \mathbf{G}_r(x)=
	\begin{bmatrix}
		1 & 0\\
		0 & E(x)A(x)
	\end{bmatrix}.
\end{equation}
	
In order to show that the LTV-based method yields equivalent physical wave
propagation results to the SEM, recall that both methods approximate the
continuously varying waveguide by $N$ homogeneous segments with constant local
material and geometric parameters (Section~\ref{Sec2.1}). The wavenumber for
the $n$-th rod homogeneous segment is $k_{r,n} = \omega\sqrt{\rho_n/E_n}$
\cite{arruda2007investigating}.
		
For the $n$-th homogeneous segment, Eq.~(\ref{Eq.5}) simplifies to
\begin{equation}\label{Eq.10}
	\mathbf{C}_{r,n} = \begin{bmatrix}
		0 & 1\\
		-k_{r,n}^2 & 0
	\end{bmatrix},
\end{equation}
and its matrix exponential is given by
\begin{equation}\label{Eq.11}
	\text{expm}(\Delta_n\mathbf{C}_{r,n}) = \boldsymbol{\psi}_{C_r,n}\boldsymbol{\Lambda}_{C_r,n}\boldsymbol{\psi}_{C_r,n}^{-1},
\end{equation}
where $\boldsymbol{\Lambda}_{C_r,n}$ contains the exponential of the eigenvalues $\pm i k_{r,n}\Delta_n$, and $\boldsymbol{\psi}_{C_r,n}$ denotes the matrix of eigenvectors:
\begin{equation}\label{Eq.12}
	\boldsymbol{\Lambda}_{C_r,n} = 
	\begin{bmatrix} e^{-ik_{r,n}\Delta_n} & 0 \\
		0 & e^{ik_{r,n}\Delta_n}
	\end{bmatrix}, \hspace{0.5cm} 
	\boldsymbol{\psi}_{C_r,n} = \begin{bmatrix} i/k_{r,n} & -i/k_{r,n} \\
	1 & 1  
	\end{bmatrix}, \hspace{0.5cm} \boldsymbol{\psi}_{C_r,n}^{-1} = \begin{bmatrix} -i k_{r,n}/2 & 1/2 \\
	i k_{r,n}/2 & 1/2  
	\end{bmatrix}.
\end{equation}
	
Evaluating the exponential via Euler's identity yields the local state transition operator $\mathbf{M}_{r,n}$:
\begin{equation}\label{Eq.13}
    \mathbf{M}_{r,n} = \text{expm}(\Delta_n\mathbf{C}_{r,n}) = \begin{bmatrix}
		\cos(k_{r,n}\Delta_n) & \frac{1}{k_{r,n}} \sin(k_{r,n}\Delta_n)\\
		-k_{r,n}\sin(k_{r,n}\Delta_n) &  \cos(k_{r,n}\Delta_n)
	\end{bmatrix}.
\end{equation}

Thus, for an $N$-segment discretization, the global transfer matrix derived from the LTV formulation is given by
\begin{equation}\label{Eq.14}
	\mathbf{T}_{\text{LTV}_r} \approx \mathbf{G}_{r,1} \left( \prod_{n=1}^{N} \mathbf{M}_{r,n} \right) \mathbf{G}_{r,N}^{-1} = 
	\begin{bmatrix}
		1 & 0\\
		0 & E_1 A_1
	\end{bmatrix}
	\left( \prod_{n=1}^{N} \mathbf{M}_{r,n} \right)
	\begin{bmatrix}
		1 & 0\\
		0 & \frac{1}{E_N A_N}
	\end{bmatrix}.
\end{equation}

On the other hand, the transfer matrix for the $n$-th homogeneous segment computed via SEM is defined as \cite{arruda2007investigating}
\begin{equation}\label{Eq.15}
	\mathbf{T}_{\text{SEM}_r,n} = \begin{bmatrix}
			\cos(k_{r,n}\Delta_n) & \frac{1}{E_n A_n k_{r,n}} \sin(k_{r,n}\Delta_n)\\
		-(E_n A_n k_{r,n})\sin(k_{r,n}\Delta_n) &  \cos(k_{r,n}\Delta_n)
	\end{bmatrix}.
\end{equation}
By factoring out the constitutive boundary operator $\mathbf{G}_{r,n} = \text{diag}(1, E_n A_n)$, Eq.~(\ref{Eq.15}) can be rewritten precisely as the state transition matrix wrapped by the mechanical properties of the element:
\begin{equation}\label{Eq.16}
	\mathbf{T}_{\text{SEM}_r,n} = \mathbf{G}_{r,n} \mathbf{M}_{r,n} \mathbf{G}_{r,n}^{-1}.
\end{equation}

Consequently, the discretized global SEM transfer matrix for the unit cell takes the form
\begin{equation}\label{Eq.17}
	\mathbf{T}_{\text{SEM}_r} \approx \prod_{n=1}^{N} \left( \mathbf{G}_{r,n} \mathbf{M}_{r,n} \mathbf{G}_{r,n}^{-1} \right).
\end{equation}

In the continuum limit ($\Delta_n \to 0$), the variation between adjacent properties vanishes, $\mathbf{G}_{r,n}^{-1} \mathbf{G}_{r,n+1} \to \mathbf{I}$, collapsing the internal product of Eq.~(\ref{Eq.17}) into $\mathbf{G}_{r,1} \left( \prod_{n=1}^{N} \mathbf{M}_{r,n} \right) \mathbf{G}_{r,N}^{-1}$. Furthermore, since the unit cell is periodic (Section~\ref{Sec2.1}), the mechanical properties at the boundaries are equal, $E_1 = E_N$ and $A_1 = A_N$, which implies $\mathbf{G}_{r,1} = \mathbf{G}_{r,N}$. Under this condition, $\mathbf{T}_{\text{LTV}_r}$ and $\mathbf{T}_{\text{SEM}_r}$ are related to the core state transition product $\mathbf{M}_{\text{tot}} = \prod_{n=1}^{N} \mathbf{M}_{r,n}$ via a similarity transformation:
\begin{equation}\label{Eq.18}
	\mathbf{T}_{\text{LTV}_r} = \mathbf{G}_{r,1} \, \mathbf{M}_{\text{tot}} \, \mathbf{G}_{r,1}^{-1}.
\end{equation}
Since similarity transformations preserve matrix traces, determinants, and
spectral properties, $\mathbf{T}_{\text{LTV}_r}$ and $\mathbf{T}_{\text{SEM}_r}$
share exactly the same eigenvalues; consequently, the characteristic equations
$\det(\mathbf{T} - e^{ikL}\mathbf{I}_2) = 0$ associated with Eq.~(\ref{Eq.7})
are identical for both formulations, and their roots -- the Bloch wavenumbers
$k$ that define the dispersion relation -- are therefore also identical.\footnote{This
equivalence is established purely from algebraic properties (the similarity
transformation) shared by both discretized formulations, and does not rely on
any curve-fitting or data-driven approximation step; consequently, no
additional assumption regarding causality or passivity of the resulting
transfer matrix is introduced beyond what is already implicit in the
underlying local homogeneous solutions common to both the LTV-based approach
and the SEM.} This proves that the LTV-based approach and the SEM yield
identical dispersion relations and wave propagation characteristics across the
waveguide, completing the demonstration.

\subsubsection{Saint-Venant shaft theory ($s$)}

In the case of Saint-Venant shafts, the equation of motion in the frequency domain yields \cite{rao2019vibration, teimoori2016saint}
\begin{equation}\label{Eq.18b}
		\normalsize{\frac{\partial^2 \Theta_x(x)}{\partial x^2}} + \left( \normalsize{\frac{1}{\frac{G(x)}{\frac{\partial G(x)}{\partial x}}}} + \normalsize{\frac{1}{\frac{K_S(x)}{\frac{\partial K_S(x)}{\partial x}}}}\right) \normalsize{\frac{\partial \Theta_x(x)}{\partial x}} + \normalsize{\omega^2\frac{\rho(x)J(x)}{G(x)K_S(x)}}\Theta_x(x) = 0.
\end{equation}
	
The vectors $\mathbf{z}_s(x)=\begin{Bmatrix} \Theta_x(x) & \normalsize{\frac{\partial \Theta_x(x)}{\partial x}}\end{Bmatrix}^t$ and $\mathbf{y}_s(x) = \begin{Bmatrix}\Theta_x(x) &  M_x(x)\end{Bmatrix}^t$ are used to obtain \cite{ribeiro2023computing}
\begin{equation}\label{Eq.19}
	\mathbf{H}_s(x) = \begin{bmatrix}
			0 & 1\\
			-\normalsize{\omega^2\frac{\rho(x)J(x) }{G(x)K_S(x)}} & -\left( \normalsize{\frac{1}{\frac{G(x)}{\frac{\partial G(x)}{\partial x}}}} + \normalsize{\frac{1}{\frac{K_S(x)}{\frac{\partial K_S(x)}{\partial x}}}}\right)
	\end{bmatrix} \hspace{0.5cm} \text{and} \hspace{0.5cm} \mathbf{G}_s(x)=\begin{bmatrix}
		1 & 0\\
		0 & G(x)K_S(x)
	\end{bmatrix}.
\end{equation}	

Analogously to the derivation done for rods, and again discretizing the unit
cell into $N$ homogeneous segments (Section~\ref{Sec2.1}), the wavenumber for
the $n$-th homogeneous Saint-Venant shaft segment with mechanical properties
$\rho_n$ and $G_n$, and geometric parameters $J_n$ and $K_{S,n}$ is given by
$k_{s,n} = \omega\sqrt{\frac{\rho_nJ_n}{G_nK_{S,n}}}$ \cite{arruda2007investigating}. Hence, for the $n$-th homogeneous segment, the matrix $\mathbf{C}_{s,n}$ is
\begin{equation}\label{Eq.20}
	\mathbf{C}_{s,n} = \begin{bmatrix}
		0 & 1\\
		-k_{s,n}^2 & 0
	\end{bmatrix}.
\end{equation}
Evaluating the matrix exponential for this constant segment yields the local state transition operator $\mathbf{M}_{s,n}$:
\begin{equation}\label{Eq.21}
	\mathbf{M}_{s,n} = \text{expm}(\Delta_n\mathbf{C}_{s,n}) = \begin{bmatrix}
		\cos(k_{s,n}\Delta_n) & \frac{1}{k_{s,n}} \sin(k_{s,n}\Delta_n)\\
		-k_{s,n}\sin(k_{s,n}\Delta_n) &  \cos(k_{s,n}\Delta_n)
	\end{bmatrix}.
\end{equation}

For an $N$-segment discretization, the global transfer matrix obtained via the LTV-based method is given by wrapping the global state transition matrix with the boundary constitutive operators:
\begin{equation}\label{Eq.22}
	\mathbf{T}_{\text{LTV}_s} \approx \mathbf{G}_{s,1} \left( \prod_{n=1}^{N} \mathbf{M}_{s,n} \right) \mathbf{G}_{s,N}^{-1} = \begin{bmatrix}
		1 & 0\\
		0 & G_1 K_{S,1}
	\end{bmatrix}
	\left( \prod_{n=1}^{N} \mathbf{M}_{s,n} \right)
	\begin{bmatrix}
		1 & 0\\
		0 & \frac{1}{G_N K_{S,N}}
	\end{bmatrix}.
\end{equation}
	
On the other hand, the transfer matrix calculated via SEM for the $n$-th homogeneous Saint-Venant shaft segment is \cite{ribeiro2023computing}
\begin{equation}\label{Eq.23}
	\mathbf{T}_{\text{SEM}_s,n} = \begin{bmatrix}
		\cos(k_{s,n}\Delta_n) & \frac{1}{G_n K_{S,n} k_{s,n}} \sin(k_{s,n}\Delta_n)\\
		-(G_n K_{S,n} k_{s,n})\sin(k_{s,n}\Delta_n) &  \cos(k_{s,n}\Delta_n)
	\end{bmatrix}.
\end{equation}
Just as demonstrated for the elementary rod theory, Eq. (\ref{Eq.23}) can be analytically factored into the state transition matrix transformed by the elemental geometric and material properties $\mathbf{G}_{s,n} = \text{diag}(1, G_n K_{S,n})$:
\begin{equation}\label{Eq.24}
	\mathbf{T}_{\text{SEM}_s,n} = \mathbf{G}_{s,n} \mathbf{M}_{s,n} \mathbf{G}_{s,n}^{-1}.
\end{equation}

Consequently, the global transfer matrix of the spatially varying unit cell computed by SEM in a discretized manner is
\begin{equation}\label{Eq.25}
	\mathbf{T}_{\text{SEM}_s} \approx \prod_{n=1}^{N} \left( \mathbf{G}_{s,n} \mathbf{M}_{s,n} \mathbf{G}_{s,n}^{-1} \right).
\end{equation}

In the continuum limit ($\Delta_n \to 0$), the adjacent internal property matrices cancel out as $\mathbf{G}_{s,n}^{-1} \mathbf{G}_{s,n+1} \to \mathbf{I}$. Since the unit cell is periodic ($G_1 = G_N$ and $K_{S,1} = K_{S,N}$, Section~\ref{Sec2.1}), the LTV and SEM transfer matrices become mathematically linked through a similarity transformation: $\mathbf{T}_{\text{LTV}_s} = \mathbf{G}_{s,1} ( \prod \mathbf{M}_{s,n} ) \mathbf{G}_{s,1}^{-1}$. As in the rod case, since similar matrices share identical eigenvalues and characteristic polynomials, both approaches yield precisely the same dispersion relations.

\subsubsection{Euler-Bernoulli beam theory ($Eb$)}

The dynamic equation of the Euler-Bernoulli beam can be used to obtain \cite{machado2018spectral, banerjee2020influence}
\begin{equation}\label{Eq_EB}
		\normalsize{\frac{\partial^4}{\partial x^4}} U_y(x) = -2\left( \normalsize{\frac{1}{\frac{E(x)}{\frac{\partial E(x)}{\partial x}}}} + \normalsize{\frac{1}{\frac{I(x)}{\frac{\partial I(x)}{\partial x}}}}\right) \normalsize{\frac{\partial^3 U_y(x)}{\partial x^3}} -\left( \normalsize{\frac{1}{\frac{E(x)}{\frac{\partial^2 E(x)}{\partial x^2}}}} + \normalsize{\frac{1}{\frac{I(x)}{\frac{\partial^2 I(x)}{\partial x^2}}}} +2 \normalsize{\frac{1}{\frac{E(x)}{\frac{\partial E(x)}{\partial x}}}} \normalsize{\frac{1}{\frac{I(x)}{\frac{\partial I(x)}{\partial x}}}}  \right) \normalsize{\frac{\partial^2 U_y(x)}{\partial x^2}} + \omega^2 \normalsize{\frac{\rho(x)A(x)}{E(x)I(x)}}U_y(x).
\end{equation}

Using the corresponding vectors for the LTV-based formulation $\mathbf{z}_{Eb}(x)=\begin{Bmatrix} U_y(x) & \normalsize{\frac{\partial U_y(x)}{\partial x}}  & \normalsize{\frac{\partial^2 U_y(x)}{\partial x^2}}  & \normalsize{\frac{\partial^3 U_y(x)}{\partial x^3}} \end{Bmatrix}^t$ and $\mathbf{y}_{Eb}(x) = \begin{Bmatrix}U_y(x) & \normalsize{\frac{\partial U_y(x)}{\partial x}} & Q(x) & M(x) \end{Bmatrix}^t$,
flexural moment and transverse force are written, respectively, as
$M(x)=-E(x)I(x)\normalsize{\frac{\partial^2 U_y(x)}{\partial x^2}}$ and $Q(x)=-\normalsize{\frac{\partial M(x)}{\partial x}}$ \cite{bittencourt2014computational, lustosa2021euler}.
One obtains the necessary $\mathbf{H}(x)$ and $\mathbf{G}(x)$ matrices for the Euler-Bernoulli beam as \cite{ribeiro2023computing} 
\begin{equation}\label{Eq.27}
	\mathbf{H}_{Eb}(x) = \begin{bmatrix}
		0 & 1 & 0 & 0 \\
		0 & 0 & 1 & 0 \\
		0 & 0 & 0 & 1 \\
		\omega^2 \normalsize{\frac{\rho(x)A(x)}{E(x)I(x)}} & 0 & -\left( \normalsize{\frac{1}{\frac{E(x)}{\frac{\partial^2 E(x)}{\partial x^2}}}} + \normalsize{\frac{1}{\frac{I(x)}{\frac{\partial^2 I(x)}{\partial x^2}}}} +2 \normalsize{\frac{1}{\frac{E(x)}{\frac{\partial E(x)}{\partial x}}}} \normalsize{\frac{1}{\frac{I(x)}{\frac{\partial I(x)}{\partial x}}}}  \right)  & -2\left( \normalsize{\frac{1}{\frac{E(x)}{\frac{\partial E(x)}{\partial x}}}} + \normalsize{\frac{1}{\frac{I(x)}{\frac{\partial I(x)}{\partial x}}}}\right)
	\end{bmatrix}
\end{equation}
and
\begin{equation}\label{Eq.28}
	\mathbf{G}_{Eb}(x)= \begin{bmatrix}
		1 & 0 & 0 & 0\\
		0 & 1 & 0 & 0\\
		0 & 0 & \left( \normalsize{\frac{\partial E(x)}{\partial x}} I(x) + \normalsize{\frac{\partial I(x)}{\partial x}} E(x)\right) & E(x)I(x) \\
		0 & 0 & -E(x)I(x) & 0
	\end{bmatrix}.
\end{equation}

Analogously to the procedure applied to rods and shafts, the $n$-th homogeneous Euler-Bernoulli beam segment, whose wavenumber is given by $k_{Eb,n}=\left(\omega^2 \frac{\rho_nA_n}{E_nI_n}\right)^{1/4}$, yields
\begin{equation}\label{Eq.29}
	\mathbf{C}_{Eb,n} = 
	\begin{bmatrix}
		0 & 1 & 0 & 0 \\
		0 & 0 & 1 & 0 \\
		0 & 0 & 0 & 1 \\
		k_{Eb,n}^4 & 0 & 0  & 0
	\end{bmatrix}, \hspace{0.5cm}
	\mathbf{G}_{Eb,n}= \begin{bmatrix}
		1 & 0 & 0 & 0\\
		0 & 1 & 0 & 0\\
		0 & 0 & 0 & E_nI_n \\
		0 & 0 & -E_nI_n & 0
	\end{bmatrix}.
\end{equation}
The analytical matrix exponential, leading to the local state transition operator $\mathbf{M}_{Eb,n}$, is formally evaluated as:
\begin{equation}\label{Eq.30}
\mathbf{M}_{Eb,n} = \text{expm}(\Delta_n\mathbf{C}_{Eb,n}) =  \boldsymbol{\psi}_{C_{Eb},n}\boldsymbol{\Lambda}_{C_{Eb},n}\boldsymbol{\psi}_{C_{Eb},n}^{-1},
\end{equation}
with expanded results presented in \ref{AppendixA}.
 
The spectral element method for Euler-Bernoulli beams \cite{oh2004dynamics, lee2009spectral} assumes the following wave solution for the frequency components of transverse displacements:
\begin{equation}\label{Eq.31}
U_y(x) = A(\omega)e^{-ik_{Eb,n}x} + B(\omega)e^{-k_{Eb,n}x} + C(\omega)e^{-ik_{Eb,n}(L-x)} + D(\omega)e^{-k_{Eb,n}(L-x)},
\end{equation}
where $A$, $B$, $C$, and $D$ are complex amplitudes. Thus, the slope can be written as 
\begin{equation}\label{Eq.32}
\theta_y(x) = -ik_{Eb,n}A(\omega)e^{-ik_{Eb,n}x} -k_{Eb,n}B(\omega)e^{-k_{Eb,n}x} +  ik_{Eb,n}C(\omega)e^{-ik_{Eb,n}(L-x)} + 
k_{Eb,n}D(\omega)e^{-k_{Eb,n}(L-x)}.
\end{equation}
	
For this structural element, the bending moment is expressed as
\begin{equation}\label{Eq.33}
M_y(x) = E_nI_nk_{Eb,n}^2\left( -A(\omega)e^{-ik_{Eb,n}x} + B(\omega)e^{-k_{Eb,n}x}  -C(\omega)e^{-ik_{Eb,n}(L-x)} +  D(\omega)e^{-k_{Eb,n}(L-x)}\right)
\end{equation}
and the transverse force as
\begin{equation}\label{Eq.34}
V_y(x) = E_nI_nk_{Eb,n}^3\left( iA(\omega)e^{-ik_{Eb,n}x}  -B(\omega)e^{-k_{Eb,n}x}  -iC(\omega)e^{-ik_{Eb,n}(L-x)} +  D(\omega)e^{-k_{Eb,n}(L-x)}\right).
\end{equation}

The displacements and slopes at the edges of the spectral element are related to the complex amplitudes using
\begin{equation}\label{Eq.35}
\begin{Bmatrix}
	U_y(0)\\
	\theta_y(0)\\
	U_y(L)\\
	\theta_y(L)
	\end{Bmatrix} =
	\mathbf{A}
	\begin{Bmatrix}
		A(\omega)\\
		B(\omega)\\
		C(\omega)\\
		D(\omega)
\end{Bmatrix},
\end{equation}
where
\begin{equation}\label{Eq.36}
	\mathbf{A} =
	\begin{bmatrix}
		1 & 1 & e^{-ik_{Eb,n}L} & e^{-k_{Eb,n}L}\\
		-ik_{Eb,n} & -k_{Eb,n} & ik_{Eb,n}e^{-ik_{Eb,n}L} & k_{Eb,n}e^{-k_{Eb,n}L}\\
		e^{-ik_{Eb,n}L} & e^{-k_{Eb,n}L} & 1 & 1 \\
		-ik_{Eb,n}e^{-ik_{Eb,n}L} & -k_{Eb,n}e^{-k_{Eb,n}L} & ik_{Eb,n} & k_{Eb,n}
	\end{bmatrix}.
\end{equation}

The bending moments and transverse forces at the edges of the spectral element can be expressed as
\begin{equation}\label{Eq.37}
	\begin{Bmatrix}
			Q(0)\\
			M(0)\\
			Q(L)\\
			M(L)
	\end{Bmatrix} =
			\mathbf{F}
	\begin{Bmatrix}
			A(\omega)\\
			B(\omega)\\
			C(\omega)\\
			D(\omega)
	\end{Bmatrix},
\end{equation}
where
\begin{equation}\label{Eq.38}
		\mathbf{F}
        = E_nI_nk_{Eb,n}^2
		\begin{bmatrix}
			ik_{Eb,n} & -k_{Eb,n} & -ik_{Eb,n}e^{-ik_{Eb,n}L} & k_{Eb,n}e^{-k_{Eb,n}L}\\
			1 & -1 & e^{-ik_{Eb,n}L} & -e^{-k_{Eb,n}L}\\
			-ik_{Eb,n}e^{-ik_{Eb,n}L} & k_{Eb,n}e^{-k_{Eb,n}L} & ik_{Eb,n} & -k_{Eb,n} \\
			-e^{ik_{Eb,n}L} & e^{-k_{Eb,n}L} & -1 & 1
	\end{bmatrix},
\end{equation}
with the expanded results for the dynamic stiffness matrix, obtained when Eq.~\eqref{Eq.35} is substituted into Eq.~\eqref{Eq.37}, presented in \ref{AppendixA}. 

By algebraically manipulating the dynamic stiffness relations mapping the state vector at $x=0$ to the state vector at $x=L$, one extracts the precise analytical SEM transfer matrix $\mathbf{T}_{\text{SEM}_{Eb},n}$ for the $n$-th homogeneous segment. Because the structural equations of motion fundamentally couple the kinematics to the internal forces via the flexural rigidity $E_n I_n$, this transfer matrix exact solution is mathematically equivalent to the state transition matrix $\mathbf{M}_{Eb,n}$ enclosed by the boundary geometric operator $\mathbf{G}_{Eb,n}$:
\begin{equation}\label{Eq.39}
	\mathbf{T}_{\text{SEM}_{Eb},n} = \mathbf{G}_{Eb,n} \mathbf{M}_{Eb,n} \mathbf{G}_{Eb,n}^{-1}.
\end{equation}

Accordingly, the complete discrete global SEM transfer matrix across the spatially varying beam is formulated as
\begin{equation}\label{Eq.40}
	\mathbf{T}_{\text{SEM}_{Eb}} \approx \prod_{n=1}^{N} \left( \mathbf{G}_{Eb,n} \mathbf{M}_{Eb,n} \mathbf{G}_{Eb,n}^{-1} \right).
\end{equation}
Once again, taking the limit as $\Delta_n \to 0$ collapses the intermediate constitutive mismatch, leaving $\mathbf{T}_{\text{SEM}_{Eb}}$ and $\mathbf{T}_{\text{LTV}_{Eb}}$ tied by the boundary matrices $\mathbf{G}_{Eb,1}$ and $\mathbf{G}_{Eb,N}$. For a periodic unit cell ($E_1 I_1 = E_N I_N$), this relationship simplifies strictly to a similarity transformation, ensuring identical equation characteristics. Consequently, both approaches theoretically lead to exactly the same dispersion relations, wavemodes, and forced responses.

This result is significant for two reasons. First, it provides a rigorous theoretical validation of the proposed LTV formulation by establishing its structural equivalence with the SEM, which is widely recognized as one of the most accurate frameworks for wave propagation analysis in one-dimensional waveguides. Second, unlike the conventional SEM, whose exact analytical formulations are restricted to piecewise homogeneous segments and struggle with continuous arbitrary profiles, the proposed LTV framework natively accommodates spatially varying fields through its state-space representation. This generalized characteristic establishes a robust foundation for the stochastic extension explored in the following sections, where spatial uncertainty is embedded directly into the structural operators without fundamentally altering the computational architecture.

\section{Zak phase computation using the LTV-based approach}\label{Sec3}

\begin{figure}[H]
    \centering
    \includegraphics[width=0.8\textwidth]{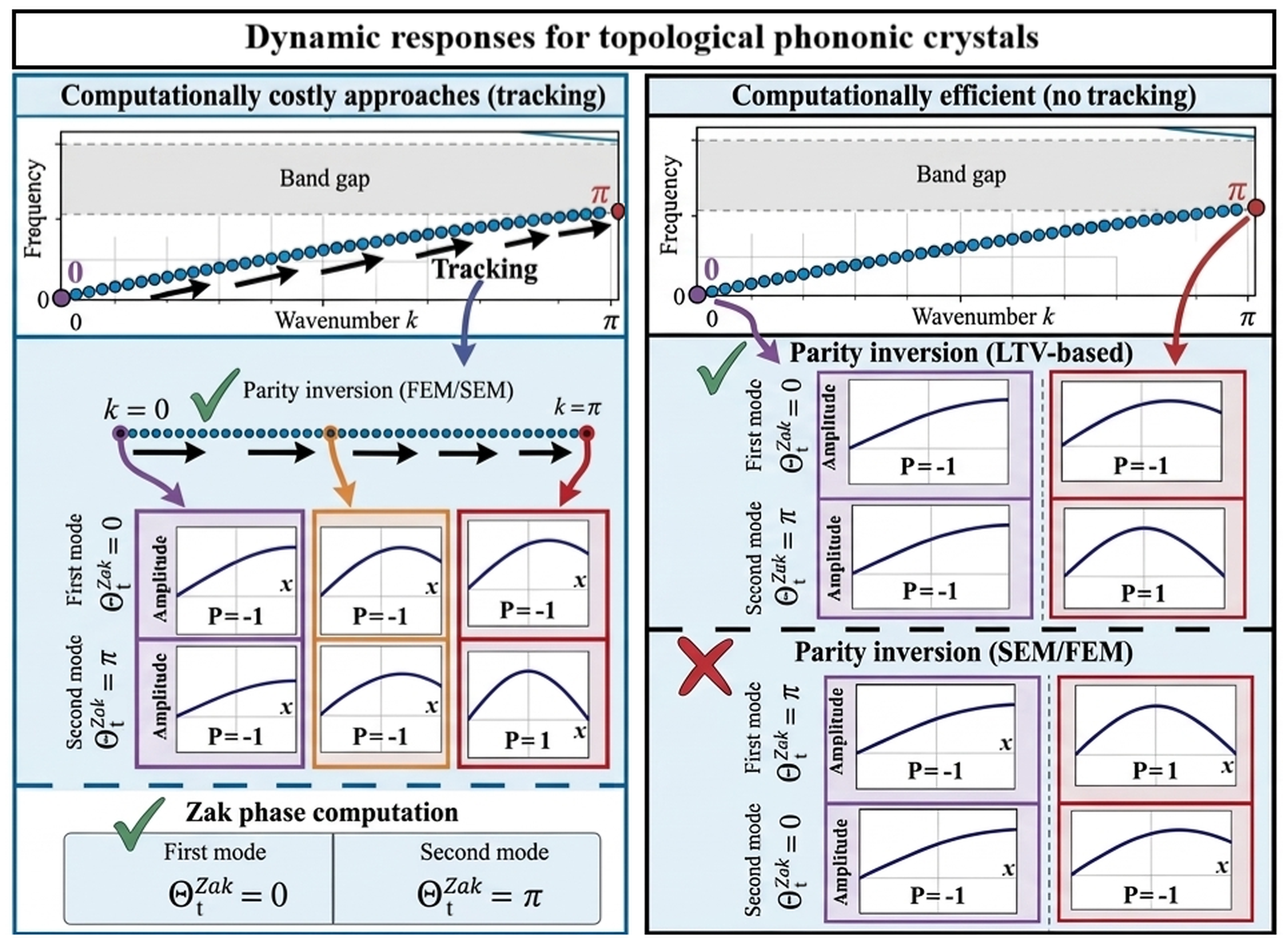}
    \caption{Comparison between the computational strategies available for evaluating the topological invariants of periodic waveguides via the parity-inversion criterion. Left: to obtain the correct parity assignment, the SEM/FEM formulation requires tracking the wavemodes across the wavenumber discretization from $k=0$ to $k=\pi/L$, which increases the associated computational cost. Bottom: resulting Zak phase computed for the first and second passbands, $\Theta_t^{\text{Zak}}=0$ and $\Theta_t^{\text{Zak}}=\pi$, respectively. Right: when applied directly at the high-symmetry points of the Brillouin zone ($k=0$ and $k=\pi/L$), the LTV-based approach correctly identifies the parity of each wavemode without requiring mode tracking (check mark), whereas the conventional SEM/FEM formulation misidentifies the wavemodes when tracking is not performed, leading to an incorrect parity assignment (cross mark). }
    \label{Fig_02}
\end{figure}

Before presenting the equations used to compute the Zak phase, it is worth
emphasizing the computational trade-off between the two available strategies.
Evaluating $\Theta_t^{\text{Zak}}$ directly from its definition demands
discretizing both the spatial domain of the unit cell and the wavenumber domain
along the entire $t^{\text{th}}$ passband \cite{xiao2015geometric}, so the
computational cost grows with the refinement of both discretizations. The
parity-inversion criterion, in contrast, only requires the wavemode parity to be
assessed at the boundaries of the Brillouin zone ($k=0$ and $k=\pi/L$)
\cite{ma2019topological, huang2021recent}; its accuracy is therefore not linked
to how finely the band is discretized, but rather to whether the wavemode used at
these two points is unambiguously identified, a guarantee that conventional
eigenproblem-based formulations (FEM, SEM, WFE) cannot offer without an
additional mode-tracking step \cite{mace2008modelling, zhu2018zak,
ribeiro2025robustness}. The benefits of the proposed approach and comparison with
the existing methods are illustrated in Fig.~\ref{Fig_02}.

In this section, a computation method for the Zak phase using the LTV-based approach is presented. We first review the physical meaning of the Zak phase and the two strategies available for its evaluation; the wavemode computation required by the LTV-based approach to carry out either strategy is then presented in Section~\ref{Sec3.2}. Beyond validating the LTV formulation against the SEM, the algebraic manipulations of Section~\ref{Sec2.2} also yield the explicit local operators $\mathbf{H}_{T,r}(x)$, $\mathbf{H}_{T,s}(x)$, and $\mathbf{H}_{T,Eb}(x)$ that are used directly below to compute these wavemodes.

\subsection{Zak Phase Computation}\label{Sec3.1}

For both the SEM and LTV-based approach, it is necessary to compute the inversion diagram to initially identify where each passband begins and ends.
During this computation, the wavemodes computed from Eq.~\eqref{Eq.7} are used
as the initial state at $x_0$, $u_t(x_0,k)$, which enforces the Bloch periodic
boundary condition already implicit in the eigenproblem of Eq.~\eqref{Eq.7}.
This initial wavemode is then propagated along the unit cell, from $0$ to $L$,
using the LTV-based transition-matrix approach detailed in
Section~\ref{Sec3.2} (Eq.~\eqref{Eq_44}).

The topological invariant analysis was proven to be equivalent to the inversion
diagram for deterministic cases, and the main goal is to find the topological
transition points in the reciprocal space \cite{xiao2015geometric,
ma2019topological}. Obtaining the inversion diagram only requires a parametric
sweep of a single geometric or material modulation parameter, tracking how the
bandgaps close and reopen; computing the Zak phase directly from its definition
is comparatively more demanding, since it additionally requires evaluating the
double integral in Eq.~\eqref{Eq_42} for every passband and at every value of
the modulation parameter. The continuous formulation for the Zak phase (also
referred to in part of the literature as the geometric phase, owing to its
origin as a special case of the Berry phase) can be defined as
\cite{xiao2015geometric}
	\begin{equation}\label{Eq_42}
		\Theta_t^{\text{Zak}} = \int_{-\pi/L}^{\pi/L} \left[ i \int_{0}^{L} \left(u_{t}^*(x,k)  \frac{\partial u_{t}(x,k)}{\partial k} dx \right) \right] dk ,
	\end{equation}
	where $u_{t}(x,k)$ is the continuous wavemode as function of $x$ and $k$ at
    the $t^\text{th}$ passband, and $^{*}$ denotes the complex conjugate
    \cite{ms2023robustness}. Equation~\eqref{Eq_42} represents an integral along
    the unit cell (from $x=0$ to $x=L_x$) and along the FBZ (from
    $k_x=-\pi/L_x$ to $k_x=\pi/L_x$). The resulting Zak phase can assume only
    the discrete values $0$ or $\pi$.
	
	However, in practice, the Zak phase can be numerically computed using the
    wavemodes in a discrete way in both frequency and space, as shown in
    \ref{AppendixB} \cite{xiao2015geometric}. This is done by using the
    wavemode $u_{t,k_s,x_n}$ associated with the $t^\text{th}$ passband,
    $s^\text{th}$ wavenumber and at spatial position $x_n$, which gives
    \cite{zhu2018zak}, 
	\begin{equation}\label{Eq1}
		\Theta_t^{\text{Zak}} = \left\lbrace - \text{Im}  \left[ 2\sum_{s=1}^{M} \text{ln} \left( \sum_{t=1}^{N} u_{t,k_s,x_n}^*  u_{t,k_{s+1},x_n} \Delta x \right) \right] \right\rbrace \text{ mod }  2\pi,
	\end{equation}
	where $M$ is the discretization size in the irreducible Brillouin zone
    (IBZ) in the wavenumber domain, and $N$ is the discretization size in the
    unit cell (spatial domain). The leading factor of 2 accounts for the
    contribution of the remaining half of the full Brillouin zone, which is
    recovered from the IBZ by symmetry. As $M\to\infty$, Eq.~\eqref{Eq1}
    converges to the continuous definition in Eq.~\eqref{Eq_42}. It is also
    important to note that, as the wavenumber has an associated frequency,
    once the frequency along the passband is identified, for example, by
    measuring the magnitude of the imaginary part of the wavenumber, the
    integral can be done using the wavemodes for a given frequency range
    \cite{ribeiro2025robustness}.

    Alternatively, for unit cells that possess spatial inversion symmetry, the
    Zak phase $\Theta_t^{\text{Zak}}$ can be evaluated in a highly efficient
    and robust deterministic manner, bypassing the need for numerical
    integration over the entire Brillouin zone. Under inversion symmetry, the
    topological invariant of the $t$-th passband is strictly determined by the
    parity (spatial symmetry) of the Bloch displacement field. This parity
    only needs to be evaluated at the two high-symmetry points of the
    reciprocal space, namely the zone center ($k=0$) and the zone boundary
    ($k=\pi/L$) \cite{xiao2015geometric, chaunsali2017demonstrating}.
    
    Let $P_t(k)$ define the eigenvalue parity of the $t$-th Bloch mode at a
    specific wavenumber $k$, where $P_t(k) = 1$ denotes a symmetric mode (even
    parity) and $P_t(k) = -1$ represents an antisymmetric mode (odd parity).
    The Zak phase of the $t$-th passband is then mathematically determined by
    \cite{huang2021recent, ma2019topological}:
    \begin{equation}\label{Eq_ParityZak}
    \Theta_t^{\text{Zak}} = \begin{cases}0, & \text{if } P_t(0) = P_t(\pi/L) \\ \pi, & \text{if } P_t(0) \neq P_t(\pi/L)\end{cases}
    \end{equation}
    Equation~\eqref{Eq_ParityZak} implies that if a wavemode preserves its
    geometric parity from the beginning to the end of the passband, its Zak
    phase is trivially $0$. 

    Conversely, if parity inversion occurs across the Brillouin zone, the band
    accumulates a topological phase of $\pi$. Furthermore, the topological
    classification of a specific bandgap is not governed solely by its
    adjacent band, but rather by the cumulative topological history of all
    underlying bulk bands. The topological invariant of the $t$-th bandgap,
    denoted as $\Omega_t^{\text{gap}}$, is defined by the cumulative sum of the
    Zak phases of all passbands located below that respective gap
    \cite{xiao2015geometric}:
    \begin{equation}\label{Eq_GapInvariant}
    \Omega_t^{\text{gap}} = \left( \sum_{i=1}^{t} \Theta_i^{\text{Zak}} \right) \pmod{2\pi}.
    \end{equation}
    If $\Omega_t^{\text{gap}} = 0$, the $t^{\text{th}}$ bandgap is classified as
    topologically trivial. If $\Omega_t^{\text{gap}} = \pi$, the bandgap is non-trivial, ensuring the emergence of a protected, localized interface
    state when the metastructure is truncated or paired with a topologically
    distinct counterpart.

\subsection{Wavemode computation using the LTV-based approach}\label{Sec3.2}

The wavemode computation required above is carried out using a state-space
representation distinct from the one introduced in
Sections~\ref{Sec2.1}--\ref{Sec2.2}. There, the state vector $\mathbf{z}(x)$
collected the generalized displacements and their spatial derivatives, and the matrix
$\mathbf{G}(x)$ was subsequently used to recover the physical field vector $\mathbf{y}(x)$ containing displacements and internal forces via Eq.~\eqref{Eq.2}. For the wavemode computation, it is instead convenient to formulate the governing equations directly in terms of $\mathbf{y}(x)$ from the outset, by directly combining the local momentum equilibrium and constitutive relations of each structural element (rods, shafts, or beams). This yields a new, independently defined state-space operator $\mathbf{H}_T(x)$, such that
\begin{equation}
    \frac{\partial \mathbf{y}(x)}{\partial x} = \mathbf{H}_{T}(x)\mathbf{y}(x),
\end{equation}
which governs physical displacements and internal forces directly.

A key advantage of formulating the system directly in terms of $\mathbf{y}(x)$ is that local balance laws and constitutive relations depend strictly on local field values rather than their spatial rates of change. Consequently, all spatial derivatives of material and geometric properties—such as $\partial A/\partial x$ and $\partial E/\partial x$ for rods, $\partial K_S/\partial x$ and $\partial G/\partial x$ for shafts, or $\partial I/\partial x$ and $\partial E/\partial x$ for beams—disappear from $\mathbf{H}_T(x)$. Mathematically, differentiating $\mathbf{y}(x) = \mathbf{G}(x)\mathbf{z}(x)$ with respect to $x$ requires the product rule:
\begin{equation}
    \frac{\partial \mathbf{y}}{\partial x} = \frac{\partial \mathbf{G}}{\partial x}\mathbf{z} + \mathbf{G}\frac{\partial \mathbf{z}}{\partial x} = \underbrace{\left( \frac{\partial \mathbf{G}}{\partial x}\mathbf{G}^{-1} + \mathbf{G}\mathbf{H}\mathbf{G}^{-1} \right)}_{\mathbf{H}_T(x)} \mathbf{y}(x).
\end{equation}
Here, the derivative term of the transformation matrix, $\frac{\partial \mathbf{G}}{\partial x}\mathbf{G}^{-1}$, generates spatial logarithmic derivatives of the local structural rigidities—for instance, $+\left( \frac{A'}{A} + \frac{E'}{E} \right)$ for rods, $+\left( \frac{K_S'}{K_S} + \frac{G'}{G} \right)$ for shafts, or $+\left( \frac{I'}{I} + \frac{E'}{E} \right)$ for beams. These terms exactly cancel the corresponding negative property-derivative terms appearing in $\mathbf{G}\mathbf{H}\mathbf{G}^{-1}$, resulting in an operator $\mathbf{H}_T(x)$ completely free of spatial derivative terms for all three structural theories.

Furthermore, because $\mathbf{H}_T(x)$ is built directly on $\mathbf{y}(x)$, no matrix inversion of $\mathbf{G}(x)$ is required at any stage: the wavemode transition matrix is obtained by locally averaging and exponentiating $\mathbf{H}_T(x)$, following exactly the same procedure already used for $\mathbf{H}(x)$ in Section~\ref{Sec2.1} (Eqs.~\eqref{Eq.4}--\eqref{Eq.5}).

Hence, for a given initial wavemode at a specified wavenumber $k$ along the first Brillouin zone (FBZ)—the fundamental domain of the reciprocal space over which the dispersion relation is uniquely defined, $k \in [-\pi/L,\pi/L]$—the generalized wavemode profile is computed from $x_0 = 0$ to $x$, when this spatial domain is divided into $N$ sub-intervals of length $\Delta_1,\dots,\Delta_N$ via the same discretization scheme of Eq.~\eqref{Eq.4}, applied here to $\mathbf{H}_T(x)$ and $\mathbf{y}(x)$:
\begin{align}\label{Eq_44}
    \boldsymbol{\phi}(x_0,x)\approx& \ \text{expm}(\Delta_N\mathbf{C}_N)\ \dots\ \ \text{expm}(\Delta_{2}\mathbf{C}_{2}) \ \text{expm}(\Delta_1\mathbf{C}_1) = \prod_{n=N}^{1}\text{expm}(\Delta_n\mathbf{C}_n),
\end{align}
where the ordering follows the same convention established in Eq.~\eqref{Eq.4}: the sub-interval closest to the reference coordinate $x_0$ ($n=1$) is applied first and appears as the rightmost factor, while the sub-interval closest to $x$ ($n=N$) is applied last and appears as the leftmost factor.

We note that Eq.~\eqref{Eq_44} is computed for all the discretized positions from $x=0$ to $x=L$ to obtain the spatial distribution of the wavemode for a given frequency $\omega$.

Depending on the dynamic problem under consideration, the governing state-space matrices $\mathbf{H}_{T}(x)$ for the elementary rod (longitudinal), Saint-Venant shaft (torsional), and Euler-Bernoulli beam (flexural) theories are respectively given by
\begin{subequations}
\begin{equation}
\mathbf{H}_{T,r}(x) = 
\begin{bmatrix} 
0 & \frac{1}{E(x)A(x)} \\ 
-\rho(x) A(x) \omega^2 & 0 
\end{bmatrix},
\end{equation}
\begin{equation}
\hspace{0.5 cm} 
\mathbf{H}_{T,s}(x) =  
\begin{bmatrix} 
0 & \frac{1}{G(x)K_S(x)} \\ 
-\rho(x) J(x) \omega^2 & 0 
\end{bmatrix},
\end{equation}
\begin{equation}
\mathbf{H}_{T,Eb}(x) = 
\begin{bmatrix}
0 & 1 & 0 & 0 \\
0 & 0 & 0 & -\frac{1}{E(x)I(x)} \\
\omega^2 \rho(x) A(x) & 0 & 0 & 0 \\
0 & 0 & -1 & 0
\end{bmatrix},
\end{equation}
\end{subequations}
where the wavemodes are computed for the physical state vectors $\mathbf{y}_r(x) = \begin{Bmatrix}U_x(x) & N(x)\end{Bmatrix}^t$, $\mathbf{y}_s(x) = \begin{Bmatrix}\Theta_x(x) & M_x(x)\end{Bmatrix}^t$, and $\mathbf{y}_{Eb}(x) = \begin{Bmatrix}U_y(x) & \frac{\partial U_y(x)}{\partial x} & Q(x) & M(x) \end{Bmatrix}^t$ for rods, Saint-Venant shafts, and Euler-Bernoulli beams, respectively. 

The wavemodes computed via this state-space formulation are naturally tracked and separated according to the structure of $\mathbf{H}_{T}(x)$. This offers a clear advantage compared to wavemode computation via SEM (see \ref{AppendixB}), where an additional mode-tracking step is required to identify the specific band modes needed for the Zak phase calculation.

\section{Deterministic results}\label{Sec4}

\subsection{Conventional Phononic Crystals}

This section presents the results obtained concerning dispersion diagrams and forced responses for one-dimensional deterministic spatially varying geometry and mechanical properties.
The assumed spatial variation of geometric and mechanical properties are presented in Table~\ref{Table_1}. These functions were selected to be smooth and continuous, with an amplitude of variation large enough to produce a clearly visible shift in the resulting bandgaps, so as to demonstrate that the LTV-based approach naturally accommodates arbitrary continuous spatial variations without requiring additional discretization refinement; they are not intended to represent a specific manufacturing process. [Authors to confirm/expand: the specific numerical values were chosen to place the resulting bandgaps within a representative, easily visualized frequency range for each structural element.] Although Table~\ref{Table_1} indicates which stochastic field representation (SFS or KLE) will later be used to model random variability about each of these functions in Section~\ref{Sec6}, the functions themselves are entirely deterministic in the present section and are treated here as prescribed, known spatial profiles.
The dispersion diagrams are computed using the transfer matrix using the LTV-based approach and SEM, as described in the previous section.

\begin{table}[H]
\setlength{\extrarowheight}{-7pt}
    \begin{tabular}{llll}
	\hline
	Structural element                  & Property & $f(x)$                                      & Unit  \\ \hline
	\multirow{4}{*}{Circular elementary rod}     & $E(x)$        & 12                                            & GPa   \\
	       	& $\rho(x)$      & 1400                                          & kg/m$^3$ \\
			& $A(x)$        & $\pi/125 - (2\pi \mid(x - 1/4)\mid)/125$              & m$^2$    \\ 
			& $L$        & 0.5  & m    \\ 
			\hline
			\multirow{4}{*}{Square Saint-Venant Shaft} & $G(x)$        & $4\text{cos}(2x)+10$                                & GPa   \\
			& $\rho(x)$      & $1000+300(x-\pi/2)$                            & kg/m$^3$ \\
			& $b(x)$        & 0.0050                                        & m     \\ 
			& $L$        & $\pi$  & m    \\ 
			\hline
			\multirow{4}{*}{Circular Euler-Bernoulli beam}            & $E(x)$        & $2\text{sin}(2x) + 8$                                & GPa   \\
			& $\rho(x)$      & $1200 - 300(x - \pi/4)^2$      & kg/m$^3$ \\
			& $r(x)$        & $(3(x - \pi/4)^2)/200 + 1/200$ & m     \\ 
			& $L$        & $\pi/2$  & m    \\ 
			\hline
\end{tabular}
\caption{Proposed functions for the geometric and mechanical properties. The elementary rod and the Euler-Bernoulli beam are modeled via SFS; the Saint-Venant shaft is modeled using KLE.}\label{Table_1}
\end{table}

\begin{figure}[H]
    \centering

    \begin{minipage}{.48\textwidth}
        \begin{subfigure}{\textwidth}
            \captionsetup{justification=raggedright, singlelinecheck=false}
            \caption{} \label{F_rod_2}
            \includegraphics[width=\textwidth]{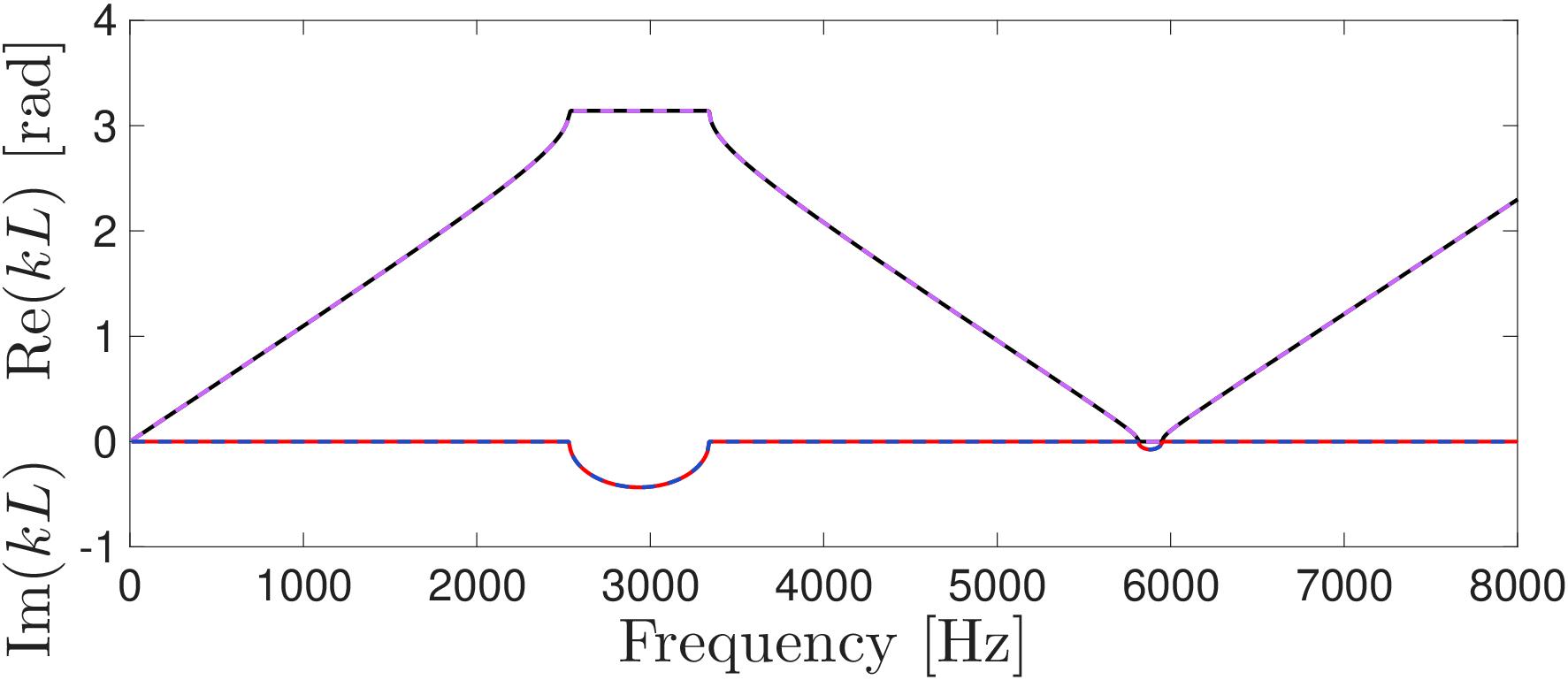}
        \end{subfigure}

        \vspace{0.3cm}

        \begin{subfigure}{\textwidth}
            \captionsetup{justification=raggedright, singlelinecheck=false}
            \caption{} \label{F_shaft_2}
            \includegraphics[width=\textwidth]{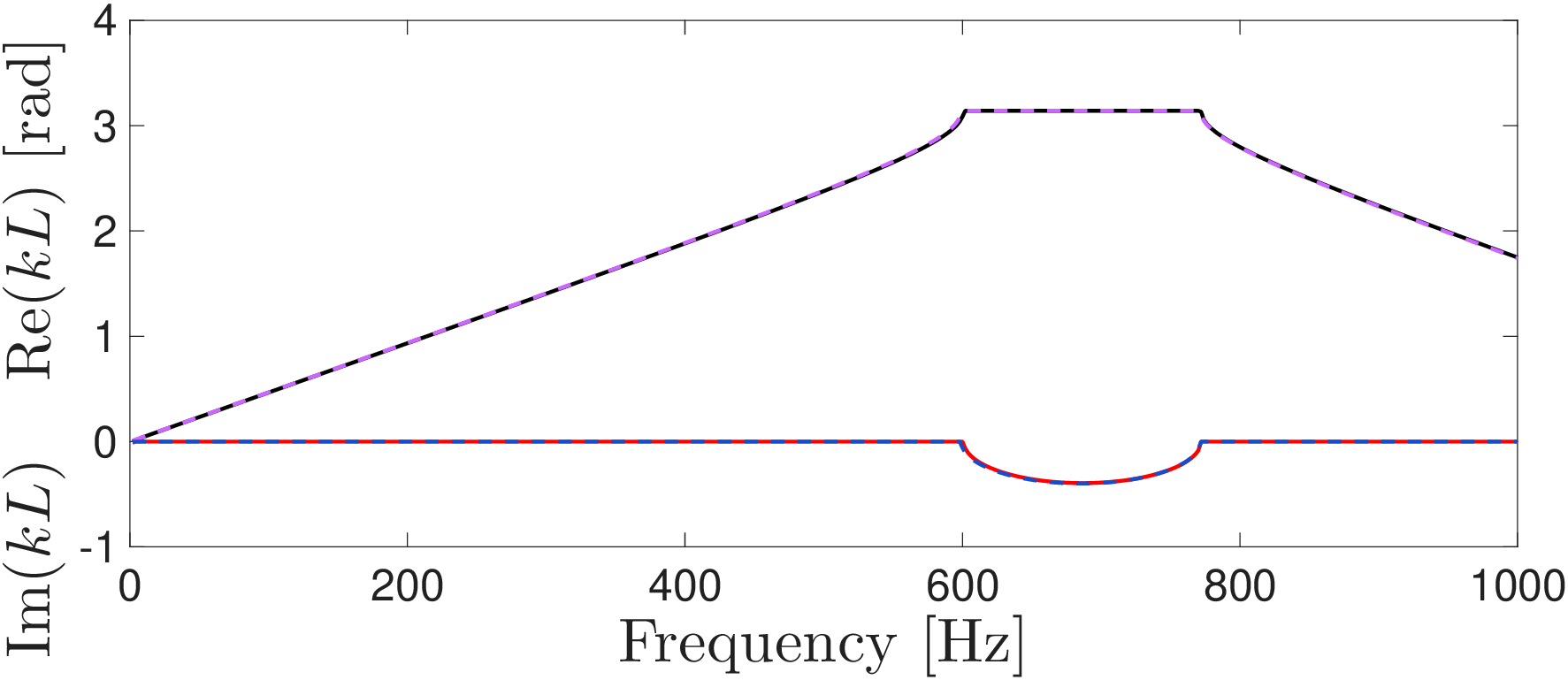}
        \end{subfigure}
       
        \vspace{0.3cm}

        \begin{subfigure}{\textwidth}
            \captionsetup{justification=raggedright, singlelinecheck=false}
            \caption{} \label{F_EB_2}
            \includegraphics[width=\textwidth]{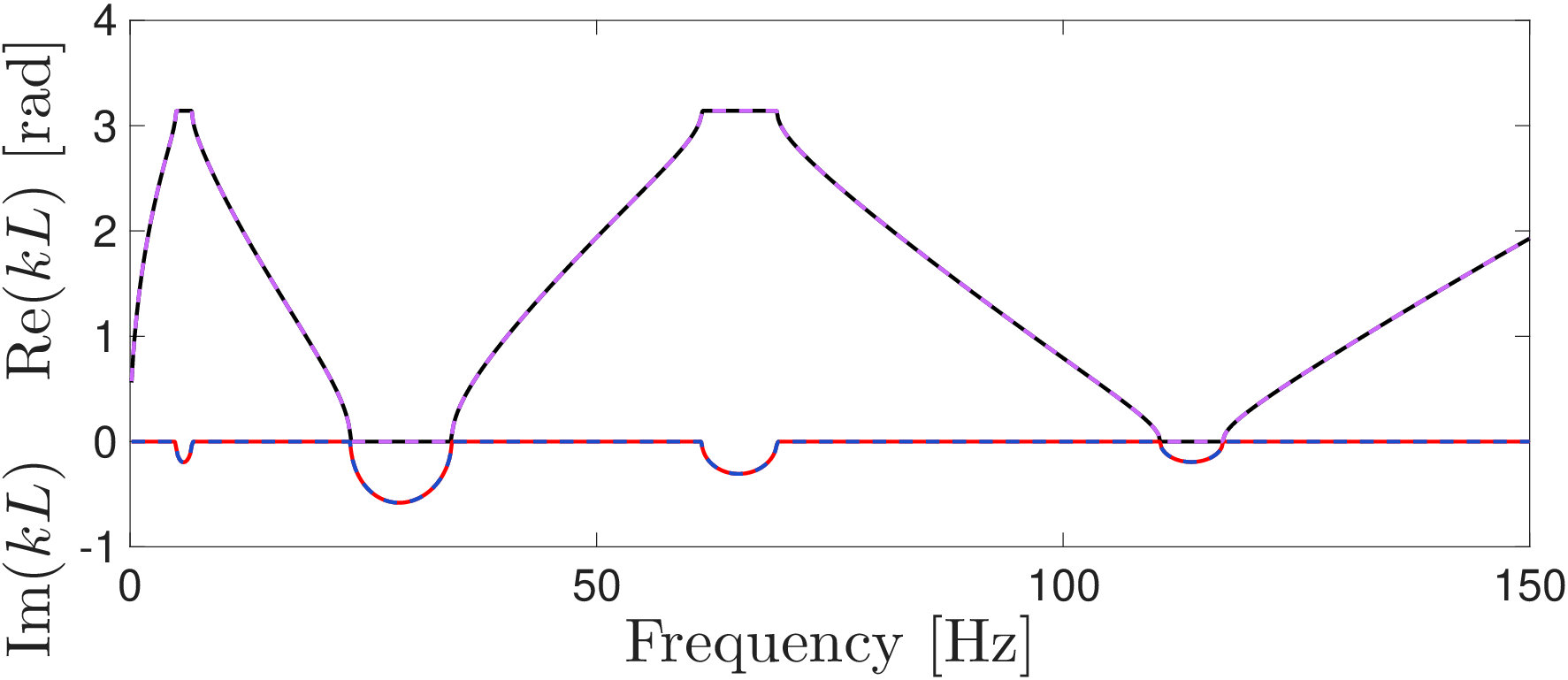}
        \end{subfigure}
    \end{minipage}\hfill
    \begin{minipage}{.48\textwidth}
        \begin{subfigure}{\textwidth}
            \captionsetup{justification=raggedright, singlelinecheck=false}
            \caption{} \label{F_rod_23}
            \includegraphics[width=\textwidth]{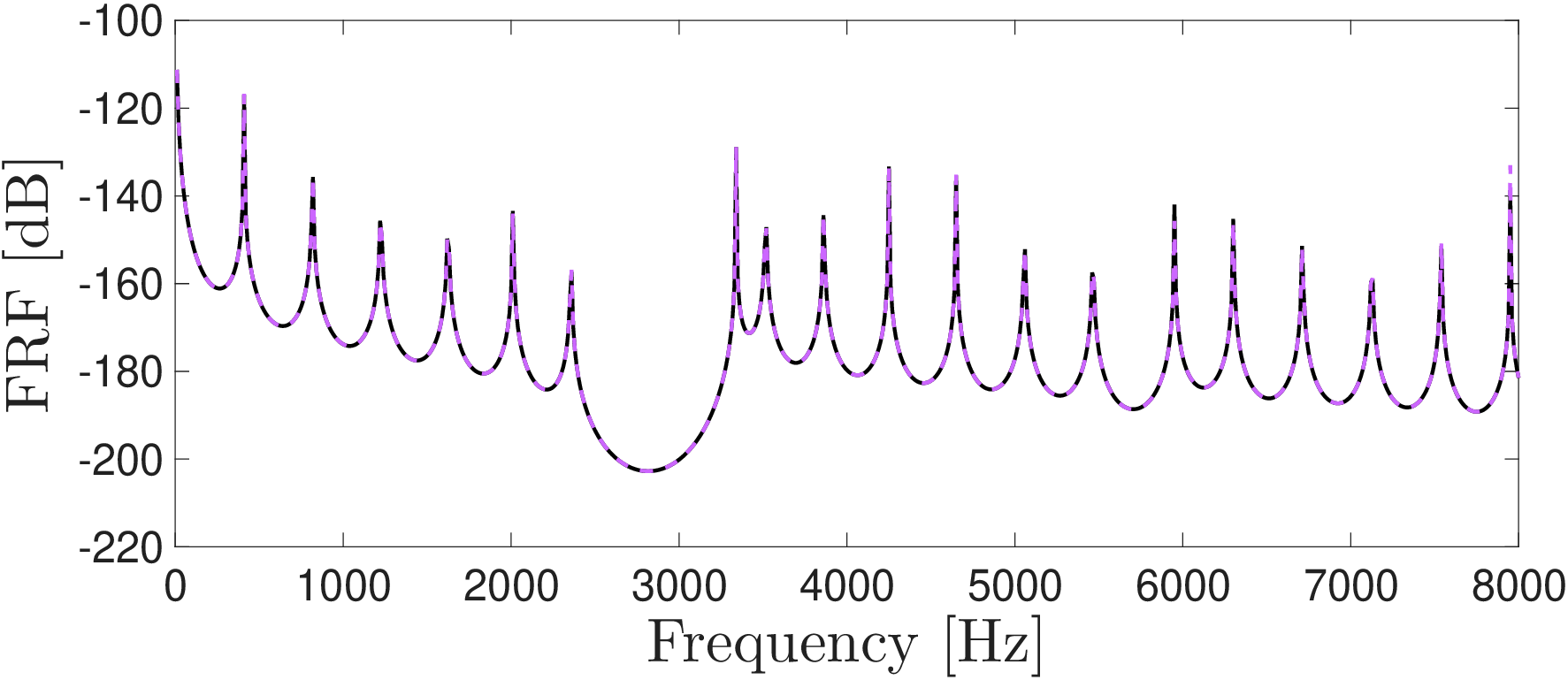}
        \end{subfigure}

        \vspace{0.3cm}

        \begin{subfigure}{\textwidth}
            \captionsetup{justification=raggedright, singlelinecheck=false}
            \caption{} \label{F_shaft_23}
            \includegraphics[width=\textwidth]{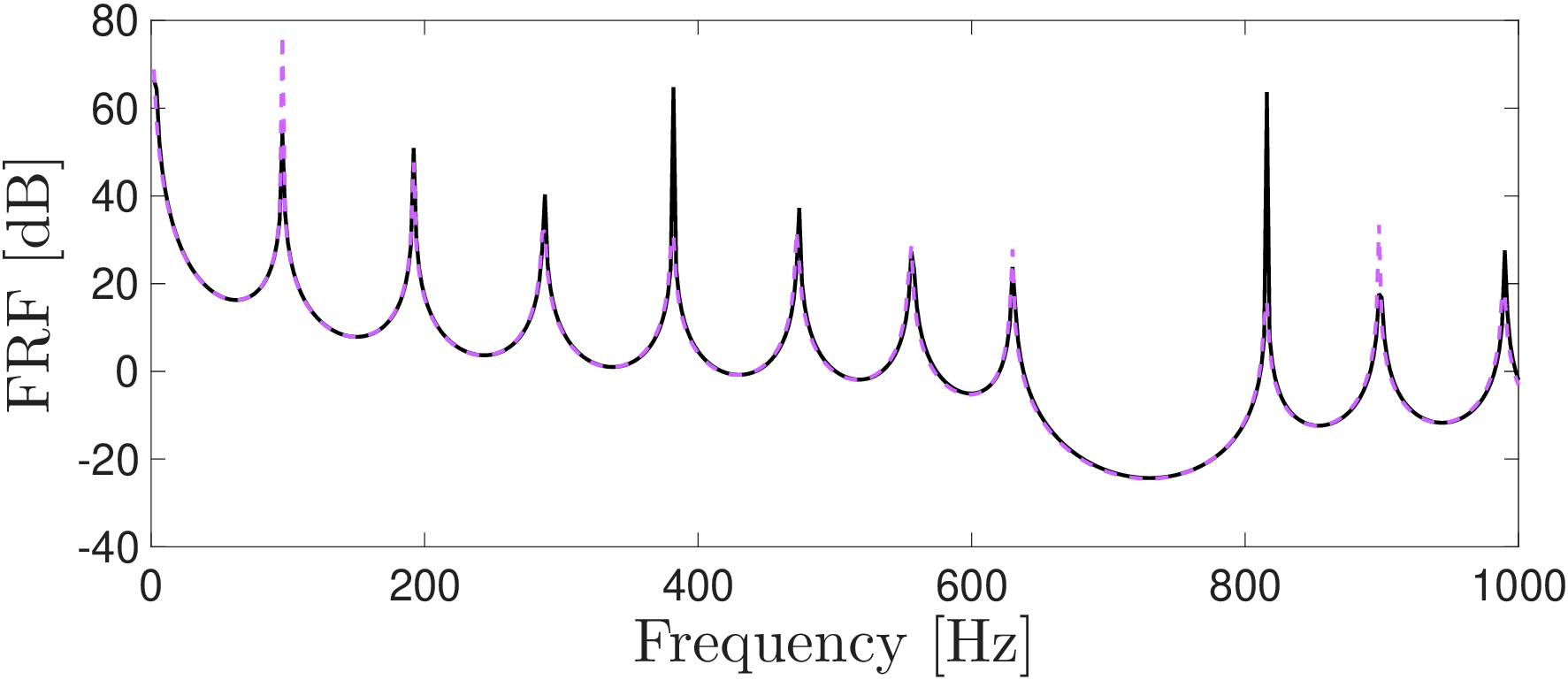}
        \end{subfigure}
       
        \vspace{0.3cm}

        \begin{subfigure}{\textwidth}
            \captionsetup{justification=raggedright, singlelinecheck=false}
            \caption{} \label{F_EB_23}
            \includegraphics[width=\textwidth]{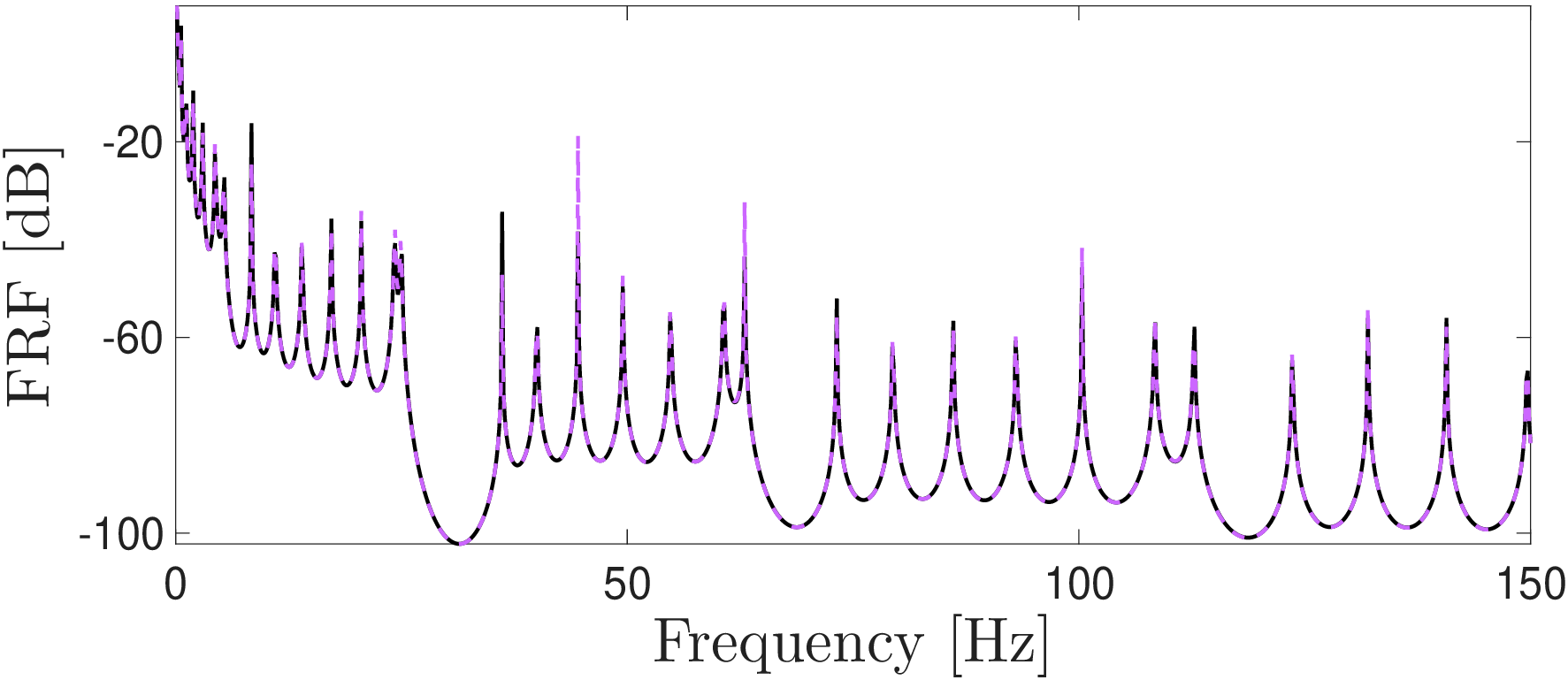}
        \end{subfigure}
    \end{minipage}
 
    \caption{Left panels: comparison between dispersion diagrams computed via the SEM (black and red continuous lines) and the LTV-based method (blue and purple dashed lines) for unit cells made of (a) rod, (b) shaft, and (c) Euler-Bernoulli beam. Right panels: comparison between FRFs computed using the SEM (black continuous line) and the LTV-based method (purple dashed line) for a finite structure made of 7 unit cells.}
    \label{Fig_2}
\end{figure}

The dispersion diagrams computed for the elementary rod, Saint-Venant shaft, and Euler-Bernoulli beam are presented in the left panels of Figs.~\ref{Fig_2}. The computed dispersion diagrams indicate bandgaps where the real part of the wavenumber $k$ is equal to $0$ or $\pi$ (Re$(kL) = 0 \pm \pi$), i.e., with an imaginary part of the wavenumber different from zero, $\text{Im}(kL) \neq 0$. The bandgap indicates frequency ranges where no vibration modes are present and only evanescent parts of the wave solutions exist, thus resulting in significant decreases in the frequency response functions (FRFs).

The forced responses were computed under free-free boundary conditions, with the structure excited at one end and the response measured at the other end (transfer receptance, in m/N), for a finite structure made of seven unit cells. For the SEM, the direct assembly was used; to compute the equivalent response via recursive condensation combined with the LTV-based approach, see \cite{ribeiro2023computing}. The resulting FRFs for the elementary rod, Saint-Venant shaft, and Euler-Bernoulli beam are presented in the bottom panels of Figs.~\ref{F_rod_23}--\ref{F_EB_23}, respectively. In all cases, the agreement between results is exact, thus providing numerical evidence of the equivalence between both methods.

After validating the deterministic LTV-based approach by comparing it with the SEM using deterministic functions, stochastic results are now presented.

\subsection{SSH Topological Phononic Crystals}

An important graph that can be used for topological phononic crystal design is the inversion diagram. In this graph, indications of the bandgap, such as the natural frequencies of a metastructure made of several unit cells or the imaginary (or real) part of the wavenumber computed from a unit cell, are plotted against a varying parameter, which is $\Delta_L$ for the length-modulated waveguides and $\Delta_A$ for the area-modulated rod. For a given modulation parameter, the frequencies marking the beginning and the end of each passband are computed from the dispersion diagram, corresponding to the frequencies at which $\text{Im}(kL)$ transitions between zero (propagating band) and nonzero (evanescent bandgap). The Zak phase is then computed for each passband at a discrete, finite set of sampled values of the corresponding modulation parameter, obtained by sweeping the modulation parameter over its admissible range to construct the inversion diagram. There is usually an agreement between the inversion diagram and the Zak phase inversion, i.e., when the bandgap closes and reopens, the Zak phase changes its value. The computed Zak phase gives information about the topological characteristics of the immediate next bandgap.

\begin{figure}[H]
		\renewcommand\figurename{Fig.}
		
    \includegraphics[width=0.6\textwidth]{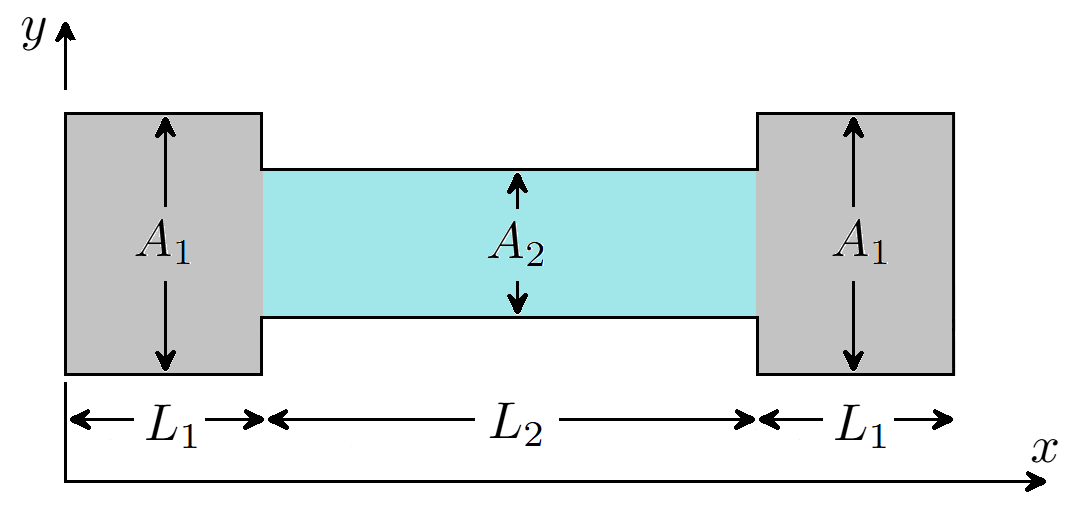}
    \caption{Sketch of the proposed unit cell used in the band inversion diagram and Zak phase computation. The unit cell is composed of two segments with different cross-sectional areas and lengths.}
		\label{Fig_cell}
	\end{figure}

Consider now a three-segment SSH-type unit cell of total length $L=L_1+L_2$, composed of two edge segments of length $L_1$ and cross-sectional area $A_1$, and a middle segment of length $L_2$ and cross-sectional area $A_2$, as illustrated in Fig.~\ref{Fig_cell}. The edge and middle segments share the same material properties, differing only in cross-sectional area ($A_1 \neq A_2$), so that the dimerization of the unit cell -- analogous to the alternating stiffness of the canonical SSH chain -- is introduced purely through geometry. For the length-modulated waveguides, the modulation parameter $\Delta_L=\frac{L_1-L_2}{2}$ is defined, with $-L/2 \leq\Delta_L\leq L/2$, while keeping $A_1$ and $A_2$ fixed; sweeping $\Delta_L$ redistributes length between the edge and middle segments without changing the total unit-cell length $L$, and $\Delta_L=0$ corresponds to the undimerized configuration in which $L_1=L_2$. For the rod, the segment lengths are kept fixed at $L_1=L_2=L/2$, while the cross-sectional areas are modulated through $\Delta_A=A_2-A_1$, such that
$A_1=\frac{A_{\Sigma}-\Delta_A}{2}$ and $A_2=\frac{A_{\Sigma}+\Delta_A}{2}$, where $A_{\Sigma}=A_1+A_2$ is kept constant. Thus, sweeping $\Delta_A$ redistributes the cross-sectional area between the edge and middle segments without changing the total unit-cell length, and $\Delta_A=0$ corresponds to the configuration with $A_1=A_2$. 
Moreover, all inversion diagrams and Zak phase computations were conducted assuming isotropic nylon elements ($E = 4\text{ GPa}$, $G = 1.5\text{ GPa}$, $\nu = 0.333$, $\rho = 1200\text{ kg/m}^3$) with a total length of $L = 1\text{ m}$ for the rod and  $L = 0.1\text{ m}$ for the shaft and beam. 

\begin{figure}[H]
    \centering

    \begin{subfigure}{0.8\textwidth}
        \captionsetup{justification=raggedright, singlelinecheck=false}
        \caption{} \label{F_Top_rod_2}
        \includegraphics[width=\textwidth]{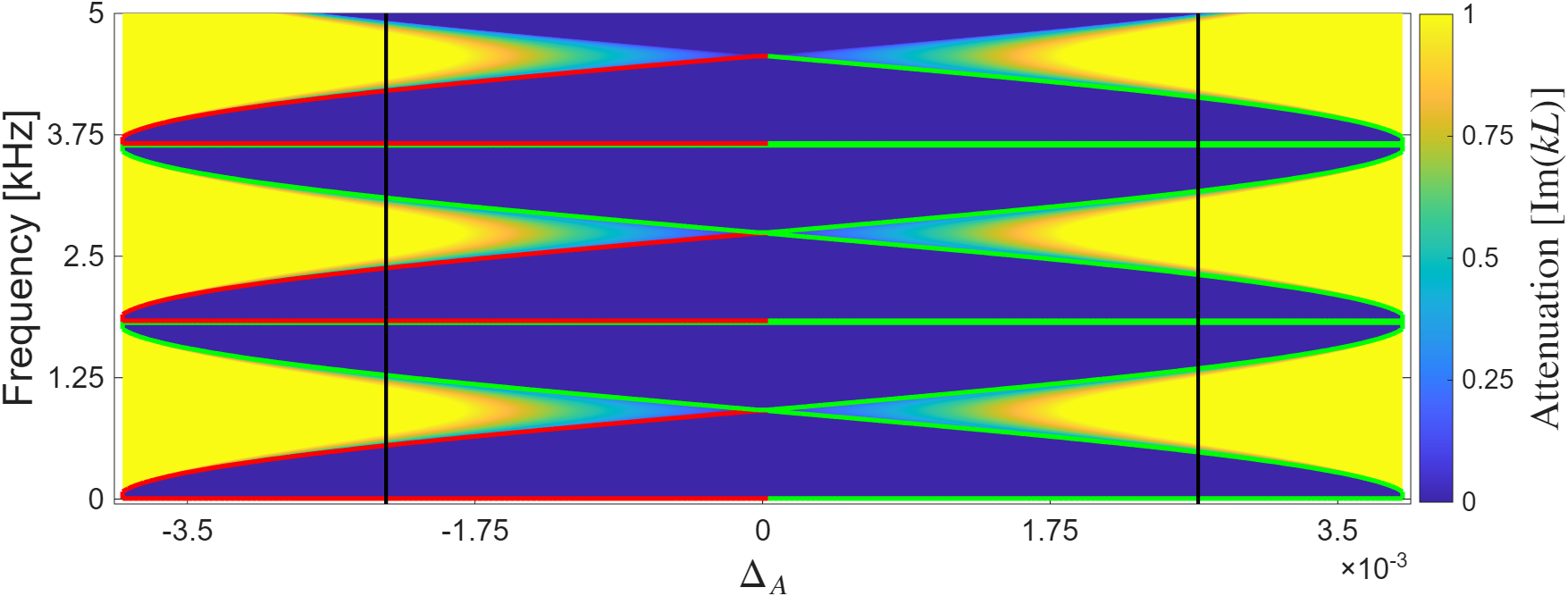}
    \end{subfigure}

    \vspace{0.4cm}

    \begin{subfigure}{0.80\textwidth}
        \captionsetup{justification=raggedright, singlelinecheck=false}
        \caption{} \label{F_Top_shaft_2}
        \includegraphics[width=\textwidth]{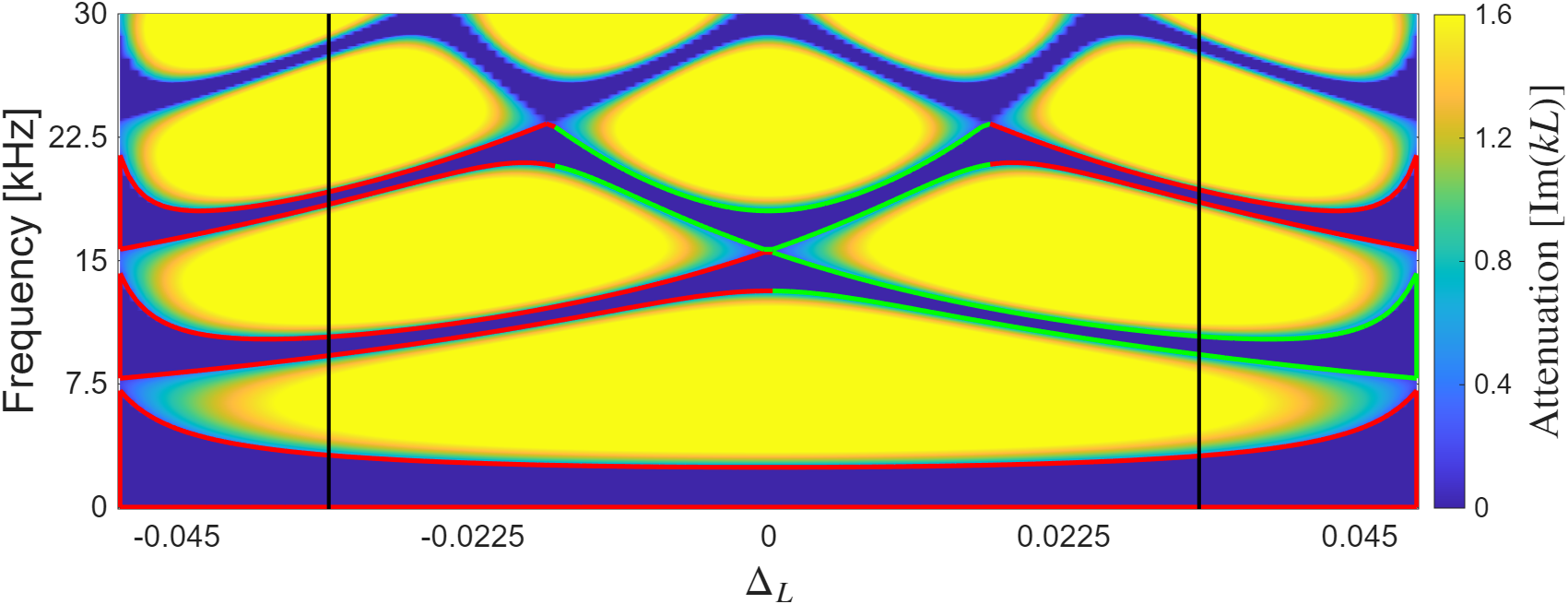}
    \end{subfigure}
       
    \vspace{0.4cm}

    \begin{subfigure}{0.80\textwidth}
        \captionsetup{justification=raggedright, singlelinecheck=false}
        \caption{} \label{F_Top_EB_2}
        \includegraphics[width=\textwidth]{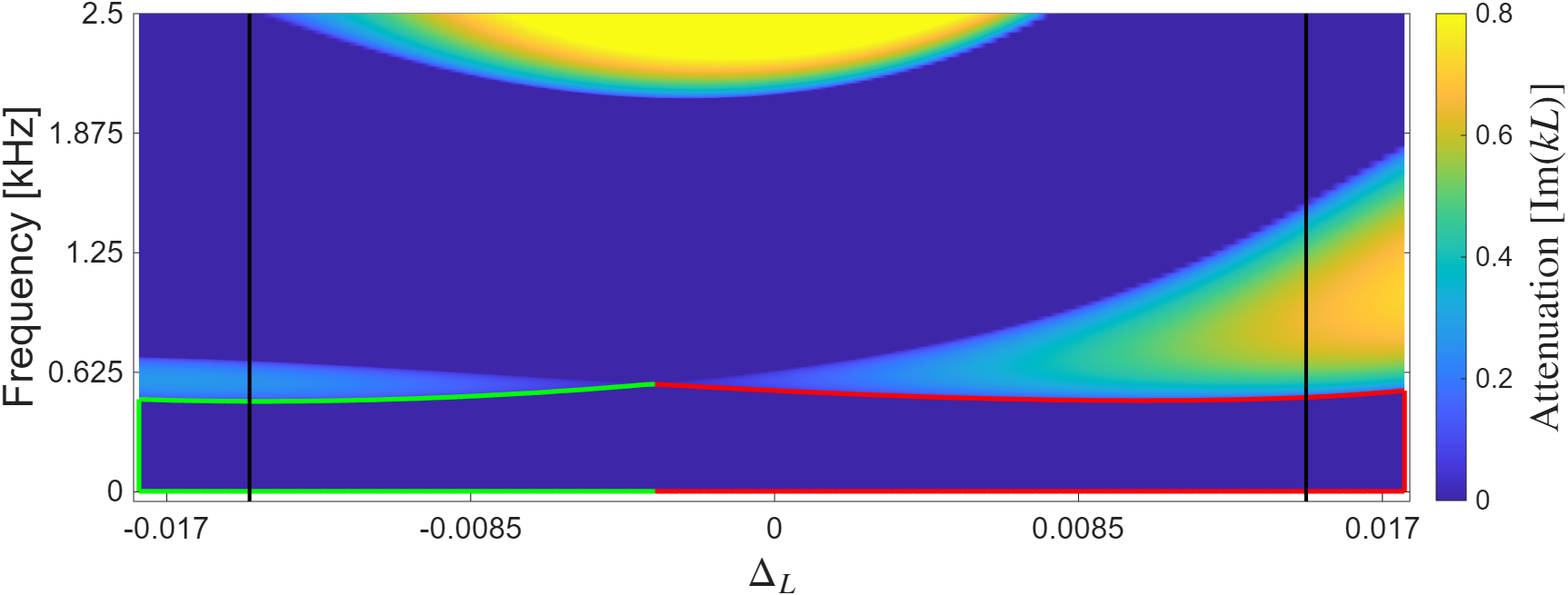}
    \end{subfigure}

    \caption{Inversion diagrams and Zak phase evaluation computed via the LTV-based approach for different structural elements: (a) rod, (b) shaft, and (c) Euler-Bernoulli beam. Each panel shows the attenuation, defined as the imaginary part of the wavenumber $\text{Im}(kL)$, as a two-dimensional map over the modulation parameter $\Delta_L$ (horizontal axis) and frequency (vertical axis): dark-blue regions correspond to propagating passbands ($\text{Im}(kL)\approx0$), while yellow/warm regions correspond to bandgaps ($\text{Im}(kL)>0$). The continuous curves trace the passband-bandgap boundary as $\Delta_L$ is swept, colored red where the passband immediately below that boundary has a non-trivial Zak phase ($\Theta_t^{\text{Zak}}=\pi$), and green where it is trivial ($\Theta_t^{\text{Zak}}=0$). The vertical black lines indicate the specific values of $\Delta_L$ selected as representative unit cells for the stochastic analyses of Section~\ref{Sec6}.}
    \label{Fig_Top_red}
\end{figure}

    The results illustrate the influence of the modulation parameter on the topological phase transitions (bandgap closing and reopening), highlighting the corresponding Zak phase inversions along the passband boundaries, as shown in Fig.~\ref{Fig_Top_red}. Fig.~\ref{Fig_Top_red} should be read as follows: for a fixed $\Delta_L$ (a vertical slice through each map), the sequence of dark-blue (passband) and yellow (bandgap) regions along the frequency axis reproduces the corresponding dispersion diagram, while the color of the curve bounding each bandgap indicates the topological character of the passband immediately below it. Up to 5~kHz, the rod element exhibits three topological phase transition points (at approximately 0.9, 2.7, and 4.5 kHz, shown in Fig.~\ref{F_rod_2}) symmetric along the $x$-axis, alongside two trivial gap-closing points (around 1.8 and 3.6~kHz) where band touching occurs. For positive $\Delta_L$ values, all bands display trivial topological properties ($\Theta_t^{\text{Zak}} = 0$). Conversely, for negative $\Delta_L$ values, the bands located below each topological transition point exhibit non-trivial properties ($\Theta_t^{\text{Zak}} = \pi$).

    For the shaft element (Fig.~\ref{F_Top_shaft_2}), while the inversion diagram remains symmetric with respect to the modulation origin ($\Delta_L = 0$), a more complex topological profile is observed due to the emergence of an additional transition point within each passband up to 30~kHz. Unlike the rod, the topological phase transitions (bandgap closing and reopening) do not occur exclusively at $\Delta_L = 0$, but shift to distinct non-zero modulation values across different frequency bands. Consequently, the band inversion trajectory becomes dependent on the specific magnitude of $\Delta_L$, resulting in non-trivial gap regions that vary dynamically with the modulation level.

    Regarding the Euler-Bernoulli beam (Fig.~\ref{F_Top_EB_2}), within the low-frequency spectrum (up to 2.5~kHz), only a single topological phase transition point is present. In contrast to both the rod and shaft configurations, the beam's inversion diagram loses symmetry with respect to the modulation parameter ($\Delta_L = 0$), demonstrating a non-symmetric bandgap closing and reopening trajectory relative to the transition point. 

    Despite these distinct symmetry profiles and transition trajectories across the three waveguides, a unified topological behavior is observed. In all cases, evaluating the Zak phase ($\Theta_t^{\text{Zak}}$) along the passband contour located immediately below a given bandgap consistently dictates the topological nature of the adjacent gap. A non-trivial Zak phase ($\Theta_t^{\text{Zak}} = \pi$) strictly coincides with the bandgap inversion regions identified in the diagrams, confirming the suitability of the LTV-based framework for topological characterization.

    For a better understanding of specific unit cells on different sizes of the transition points, the topological properties of the band of the unit cells are defined by the continuous vertical lines, which will be explained. 

\section{Statistical analyses}\label{Sec5}

The LTV-based approach requires an analytical dynamic equation and first derivatives for the functions that describe the geometry and mechanical properties of the unit cell.
As a consequence, analytical equations are also required for the samples of the stochastic processes that will represent these spatially varying and, now assumed, statistical properties.
Therefore, in the current section, the SFS and the analytical KLE are presented to simulate these stochastic fields.
	
\subsection{Stochastic fields using Fourier series}

Let $\{Y_x; 0 \leq x \leq L\}$ be a Gaussian random field defined over the unit
cell. The field is decomposed as
\begin{equation}
	Y_x = f_x + Z_x,
\end{equation}
where $f_x$ is the deterministic field mean and $Z_x$ is the zero-mean stochastic
fluctuation. The field is assumed to be \emph{homoscedastic}, i.e., the variance of
$Z_x$ is assumed constant and equal to $\sigma^2$ for every position $x$ within the
unit cell, and weakly stationary, i.e., the correlation between the field values at
any two positions $x_1$ and $x_2$ depends only on their separation, so that the
correlation function $\varphi(x_1,x_2)$ can be written as a function of a single
variable, $\varphi(x)$, with $x = x_1 - x_2$. For simplicity, and in the absence of
any other characteristic length scale in the problem, the correlation length is
assumed here to be equal to the unit-cell length $L$.

Under these assumptions, the following relation holds \cite{jha2013simulating}:
\begin{equation}\label{Eq.50}
	\text{Var}\left( Z_x \, \text{exp}\left( \frac{i2j\pi x}{L}\right)\right) = \sigma^2 \varphi(x).
\end{equation}

The zero-mean field $Z_x$ can be approximated by a zero-mean Fourier series with
terms $a_j$ \cite{jha2013simulating}
\begin{equation}
	Z_x\,\text{exp}\left( \frac{i2j\pi x}{L}\right) = \sum_{j=1}^{\infty} a_j \, \text{exp}\left( \frac{i2j\pi x}{L}\right),
\end{equation}
where $a_j$ are mutually independent Gaussian variables. It is also necessary to
compute the variance $\sigma^2_j$ of the terms $a_j$ that makes the local variance,
i.e., for a given $x$ value, equal to $\sigma^2$.

For a given $j$, one can define $b_j=\text{exp}(i2j\pi x / L)$ and use the
statistical property that the variance of the sum of scalars times random variables
is the sum of all covariances of the two-by-two terms \cite{welch1956linear,
bonamente2017statistics}. Hence, one may write
\begin{equation}
	\text{Var}\left( \sum_{j=1}^\infty a_j b_j \right) = \sum_{j=1}^\infty \sum_{i=1}^\infty \text{Cov}(a_j b_j, a_i b_i).
\end{equation}
Since $a_j$ is zero-mean, $\text{Var}(a_j)=\text{Cov}(a_j,a_j)=\sigma_j^2$. Also,
since the terms $a_j$ are mutually independent, $\text{Cov}(a_jb_j,a_ib_i)=0$ for
$j\neq i$. Because the field is homoscedastic, and approximating the solution by a
truncated version using $J$ terms, it is possible to write
\begin{equation}\label{Eq.52}
	\text{Var}\left( Z_x\text{exp}\left( \frac{i2j\pi x}{L}\right)\right)  = \sigma^2 \text{exp}\left( \frac{i2j\pi x}{L}\right) \approx \sum_{j=1}^{J}\sigma_j^2 \text{exp}\left( \frac{i2j\pi x}{L}\right).
\end{equation}
Here, $J$ denotes the number of retained terms in the truncated Fourier series and
is independent of the number of spatial intervals $N$ used to discretize the unit
cell in the LTV-based transition-matrix computation.

Thus, the local variance is given by Eq.~\eqref{Eq.52}. However, the inverse problem
is of practical interest here: given a prescribed field variance $\sigma^2$ and
correlation function $\varphi$, the coefficient variances $\sigma_j^2$ must be
determined. This can be solved by combining Eqs.~\eqref{Eq.50} and \eqref{Eq.52},
relating the variances $\sigma$ and $\sigma_j$ through
\begin{equation}\label{Eq.53}
	\sigma^2 \varphi(x) = \sum_{j=1}^{J}\sigma_j^2 \text{exp}\left( \frac{i2j\pi x}{L}\right),
\end{equation}
where two interpretations arise. The first one is that this field is a zero-mean
Gaussian field with local variance $\sigma$. Alternatively, it can also be
interpreted as a Fourier series representing $\sigma^2 \varphi(x)$, with terms
$\sigma_j^2$, which can be computed using
\begin{equation}
	\sigma_j^2=\int_{-L/2}^{L/2}\sigma^2\varphi(\tau)\text{exp}\left( \frac{i2j\pi \tau}{L} \right) d\tau.
\end{equation}

The preceding integral yields the relation for the variances $\sigma_j^2$ of the
terms $a_j$ that correspond to the local variance $\sigma^2$, which is the same for
any value of $x$ due to the assumed homoscedasticity of the field.

Finally, it is important to recall that the LTV-based approach requires the
derivatives of the functions that represent the spatially varying geometries or
mechanical properties of the unit cell. Thus, since it is assumed that
$f(x)+\sum_{j=1}^{\infty} a_j \, \text{exp}\left( i2j\pi x / L\right)$ represents the
spatially varying field for the property, its derivative can be calculated as
\begin{equation}
	Y_x' = f'(x)+\sum_{j=1}^{\infty} \left( \frac{i2j\pi}{L}\right) a_j \text{exp}\left( \frac{i2j\pi x}{L}\right),
\end{equation}
which makes the current method suitable for obtaining stochastic results from the
LTV-based approach when compared to the combination of a statistical function with a
kernel smoother, either using the discretized expansion optimal linear estimator
(EOLE) or the discrete KLE. This is because both expansions do not have closed-form
expressions for their samples; for details, please see \cite{daquantifying}.

\subsection{Analytical Karhunen–Loeve expansion}

A different approach to model a stochastic process with samples given by analytical
functions is the KLE \cite{de2015uncertainty}. As introduced in Section~5.1, the
stochastic field is again decomposed as $Y_x = f_x + Z_x$, where $f_x$ is the field
mean and $Z_x$ is the zero-mean stochastic field. In the KLE, the stochastic term is
obtained as the infinite expansion
\begin{equation}\label{Eq.57}
	Z_x = \sum_{s=1}^{\infty} \sqrt{\lambda_s}\,\chi_s(x)\, W_s,
\end{equation}
where $W_s$ is a random variable, and $\chi_s(x)$ and $\lambda_s$ represent,
respectively, the $s$-th eigenfunction and eigenvalue of the eigenproblem
\begin{equation}\label{Eq.KLE_eig}
	\int_{-L}^{L}\varphi(x_1, x_2)\,\chi_s(x_2)\,dx_2 = \lambda_s\,\chi_s (x_1), \hspace{0.5cm} -L \leq x_1 \leq L.
\end{equation}
In Eq.~\eqref{Eq.KLE_eig}, the correlation function $\varphi(x_1,x_2)$ plays the
role of the continuous covariance kernel, or operator, for which the eigenproblem is
solved: this is a Fredholm integral eigenvalue problem, the infinite-dimensional
counterpart of the eigenvalue decomposition of a discrete covariance matrix, whose
solutions are the eigenfunctions $\chi_s(x)$ and eigenvalues $\lambda_s$. The local
variance of the zero-mean field, assumed to be homoscedastic, is the same variance
$\sigma^2$ as the one chosen for $W_s$, which is here assumed to follow a normal
distribution.

In practice, a finite number of terms is used in Eq.~\eqref{Eq.57} to approximate
the solution for the field as
\begin{equation}
	Y_x \approx f_x + \sum_{s=1}^{S} \sqrt{\lambda_s}\,\chi_s(x)\, W_s,
\end{equation}
where $S$ is the number of retained KLE terms, again independent of the number of
spatial intervals $N$ used elsewhere in the LTV-based formulation.

The solutions adopted here follow \cite{xiu2010numerical}. The domain is defined
for $-L \leq x \leq L$ for mathematical convenience, but only $0 \leq x \leq L$ is
used in the simulations. The correlation length is assumed to be equal to the
unit-cell length $L$, and the correlation between any two positions $x_1$ and $x_2$
is assumed to decay exponentially with their separation, following
\begin{equation}
	\varphi(x_1,x_2)=\exp\left(-\frac{\mid x_1-x_2 \mid}{L}\right).
\end{equation}

Under this assumption, the analytical eigenvalues are given by
\begin{equation}
	\lambda_s=
	\begin{dcases}
		\frac{2L}{1+L^2a_s^2}, & \text{for even } s,\\
		\frac{2L}{1+L^2b_s^2}, & \text{for odd } s,
	\end{dcases}
\end{equation}
with the corresponding eigenfunctions
\begin{equation}
	\chi_s(x)=
	\begin{dcases}
		\frac{\text{sin}(a_s x)}{\sqrt{L-\frac{\text{sin}(2 a_sL)}{2a_s}}}, & \text{for even } s,\\
		\frac{\text{sin}(b_s x)}{\sqrt{L-\frac{\text{sin}(2 b_sL)}{2b_s}}}, & \text{for odd } s,
	\end{dcases}
\end{equation}
where the terms $a_s$ and $b_s$ are obtained from the solution of the transcendental
equations \cite{xiu2010numerical}
\begin{equation}
	\begin{dcases}
		La_s + \text{tan}(La_s) = 0, & \text{for even } s,\\
		1 - Lb_s \text{tan}(Lb_s) = 0, & \text{for odd } s.
	\end{dcases}
\end{equation}

These statistical results are used in the next section to illustrate the SLTV
methodology obtained when combining the LTV-based method with these two approaches
for computing stochastic fields. This method can compute stochastic dispersion
diagrams and forced responses for one-dimensional structures with spatially varying
geometry and mechanical properties. Furthermore, it is as efficient as the SEM for
dealing with mid- to high-frequencies in trivial phononic crystals, but it is more
efficient than FEM and SEM when dealing with SSH topological crystals.

\section{Stochastic results}\label{Sec6}

The stochastic analysis results obtained in this section assume the variability levels presented in Table \ref{Table_2}. The SFS method is used to simulate the stochastic processes representing the geometric and mechanical properties of the proposed unit cells for rod and Euler-Bernoulli beam models. Analogously, the analytical KLE approach is used for Saint-Venant shaft and Timoshenko beam models. It is worth noting that both stochastic field representations are, in principle, equally applicable to any of the structural elements considered in this work; the assignment of the SFS to the rod and beam models and of the KLE to the shaft and Timoshenko beam models is arbitrary and adopted here solely to demonstrate, across the four structural elements, that the proposed SLTV-based framework accommodates either stochastic field representation interchangeably. In both cases, these methods are applied in combination with Monte Carlo simulations \cite{shonkwiler2009explorations, rubinstein2016simulation}.
The statistical results concerning the structure attenuation are obtained using the minimum value of the imaginary part of the computed wavenumber values for a given sample. A statistical inference procedure was then applied to the obtained statistical results to compute the minimum attenuation that occurs for at least 95\% of the samples for each frequency.
The following subsections present the results obtained for the various types of elements. 

\begin{table}[H]
\setlength{\extrarowheight}{-5pt}
    \begin{tabular}{lllll}
		\hline
		Method & Structural element & Property & Local standard deviation  & Local variance  \\ \hline
			\multirow{3}{*}{SFS} &
			\multirow{3}{*}{Circular elementary rod}     & $E(x)$ [Pa]        &  $7.07\times 10 ^{8}$ & $5\times 10 ^{17}$   \\
			& & $\rho(x)$ [kg/m$^3$] &  109.54 & $1.2 \times 10^4$ \\
			& & $A(x)$ [m$^2$] & $7.07 \times 10^{-4}$  & $5\times 10^{-7}$   \\ \hline
			\multirow{3}{*}{KLE} &
			\multirow{3}{*}{Square Saint-Venant Shaft} & $G(x)$ [Pa] & $5\times 10^{8}$ & $2.5\times 10^{17}$  \\
			& & $\rho(x)$  [kg/m$^3$]  & 31.63  & $1\times 10^3$  \\
			& & $b(x)$  [m]  &  $1.05 \times 10^{-4}$ & $1.11\times 10^{-8}$ \\ \hline
			\multirow{3}{*}{SFS} &
			\multirow{3}{*}{Circular Euler-Bernoulli beam}            & $E(x)$ [Pa]  &  $7.07\times 10 ^{8}$ & $5\times 10^{17}$ \\
			& & $\rho(x)$  [kg/m$^3$]    & 82.16  & $6.75\times 10^3$ \\
			& & $r(x)$ [m]  & $2.79 \times 10^{-4}$  & $7.81\times 10 ^{-8}$ \\ 
		\hline
    \end{tabular}
\caption{Proposed local variance and standard deviation to be included in the stochastic models for the geometric and mechanical properties.}\label{Table_2}
\end{table}
	
\subsection{Elementary rod theory}

\begin{figure}[H]
    \centering 
		
    \begin{minipage}{.28\textwidth}
        \begin{subfigure}{\textwidth}
            \captionsetup{justification=raggedright, singlelinecheck=false}
            \caption{} \label{F_field_1a} 
            \includegraphics[width=\textwidth]{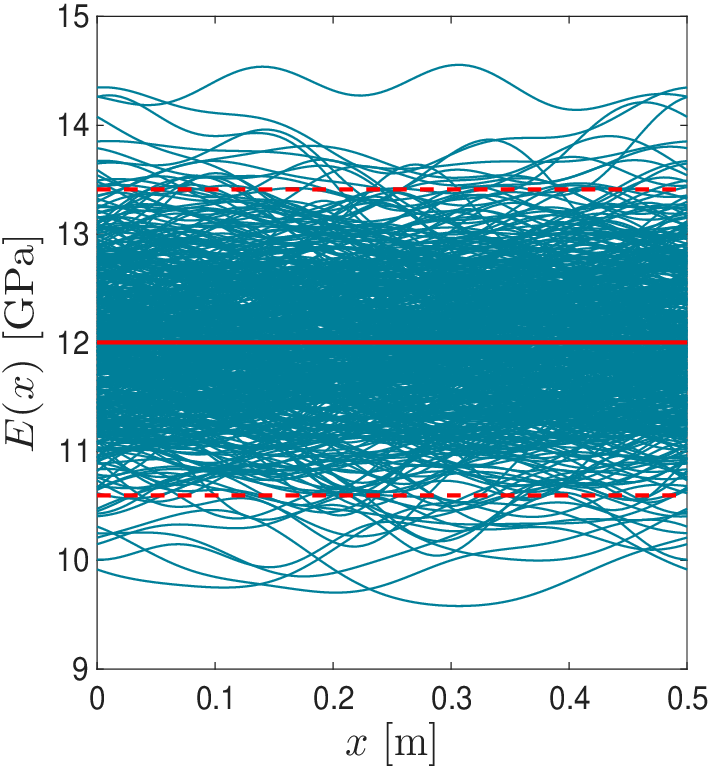}
        \end{subfigure}
        
        \vspace{0.3cm} 

        \begin{subfigure}{\textwidth}
            \captionsetup{justification=raggedright, singlelinecheck=false}
            \caption{} \label{F_field_1b}
            \includegraphics[width=\textwidth]{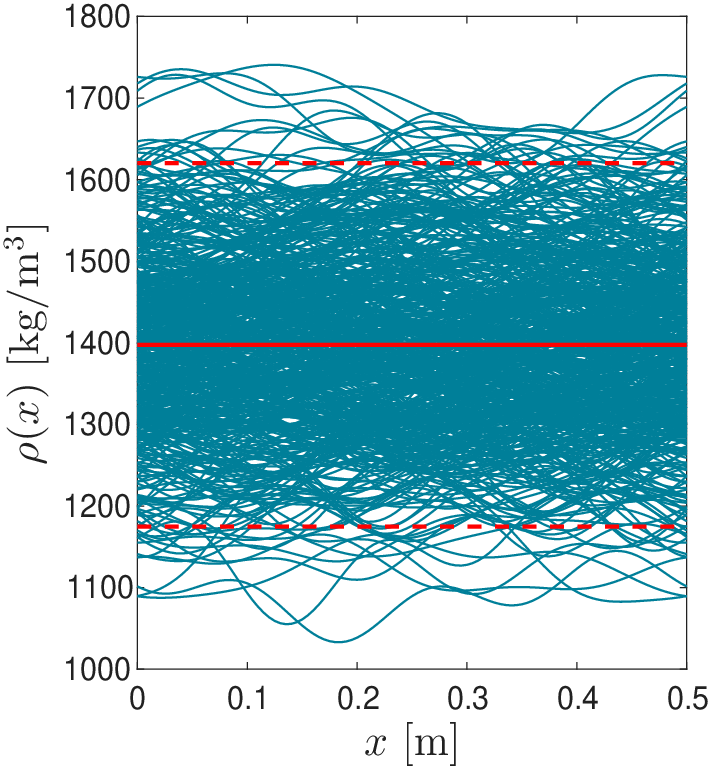}
        \end{subfigure}

        \vspace{0.3cm}

        \begin{subfigure}{\textwidth}
            \captionsetup{justification=raggedright, singlelinecheck=false}
            \caption{} \label{F_field_1c}
            \includegraphics[width=\textwidth]{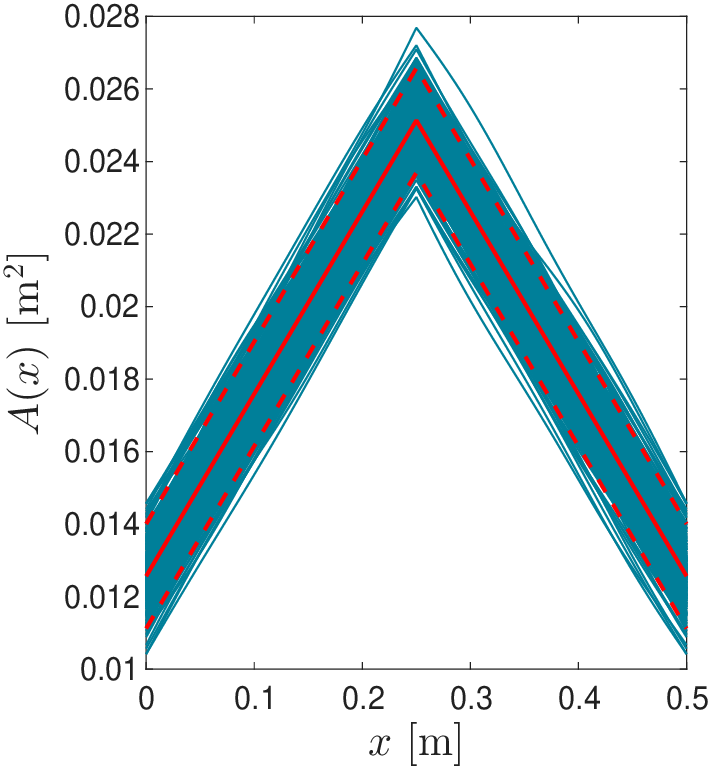}
        \end{subfigure}
    \end{minipage}\hfill 
    \begin{minipage}{.69\textwidth}
        \begin{subfigure}{\textwidth}
            \captionsetup{justification=raggedright, singlelinecheck=false}
            \caption{} \label{F_rod_4}
            \includegraphics[width=\textwidth]{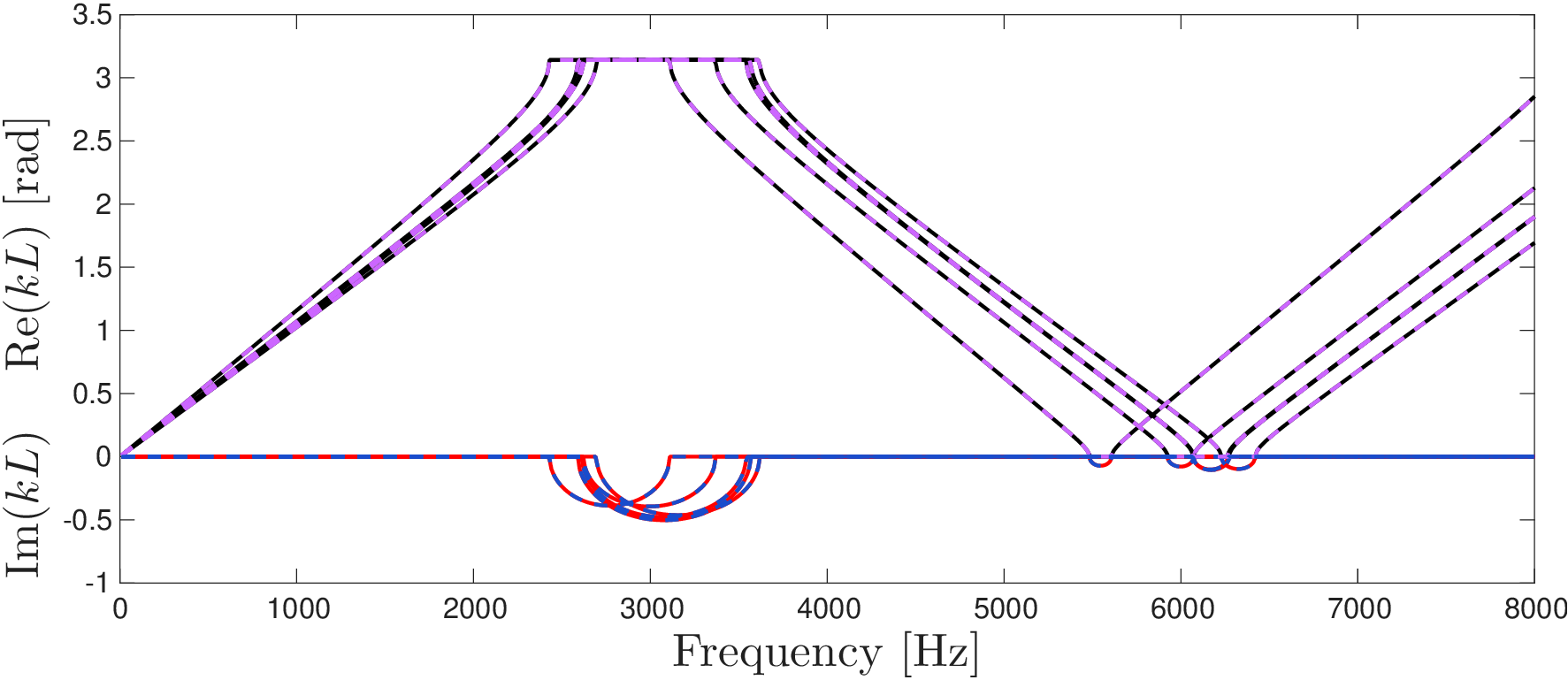}
        \end{subfigure}
        
        \vspace{0.3cm} 

        \begin{subfigure}{\textwidth}
            \captionsetup{justification=raggedright, singlelinecheck=false}
            \caption{} \label{F_rod_5}
            \includegraphics[width=\textwidth]{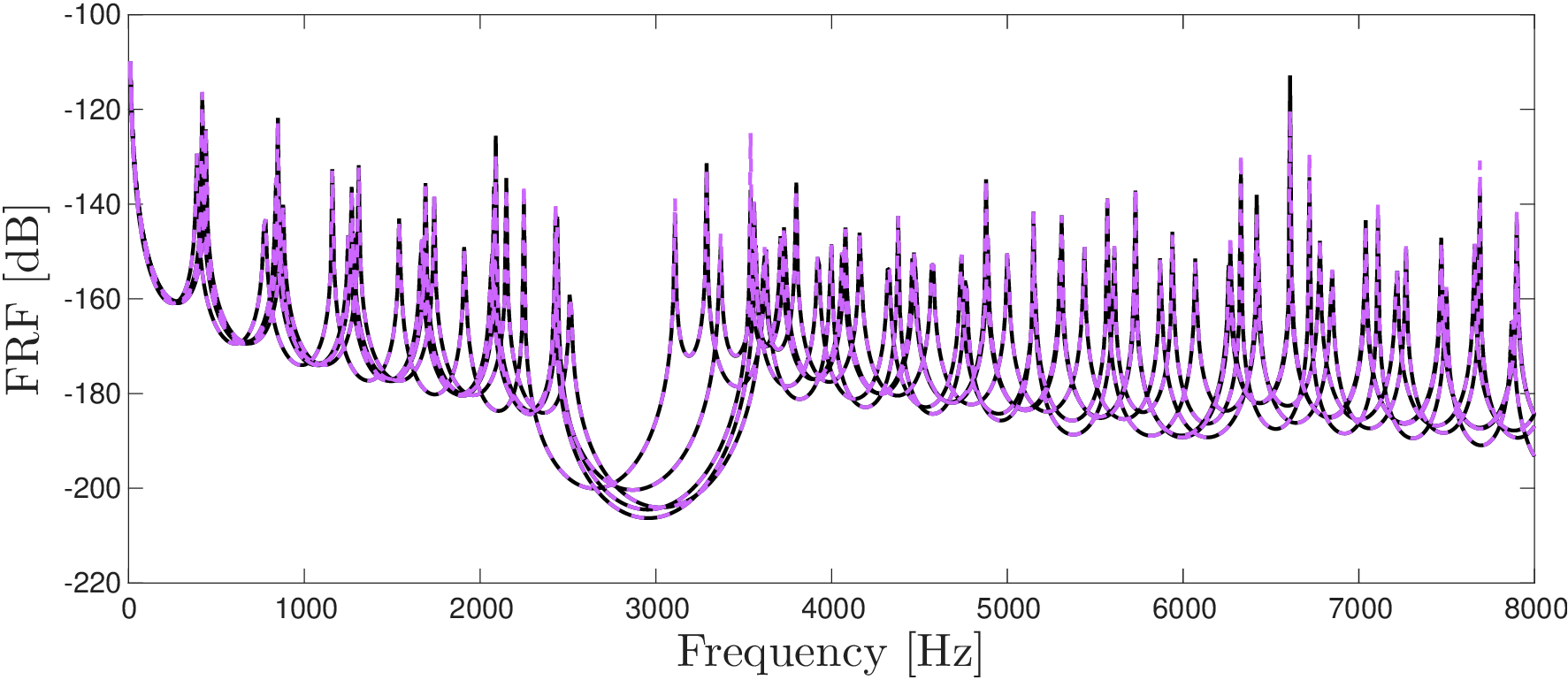}
        \end{subfigure}

        \vspace{0.3cm}

        \begin{subfigure}{\textwidth}
            \captionsetup{justification=raggedright, singlelinecheck=false}
            \caption{} \label{F_rod_6}
            \includegraphics[width=\textwidth]{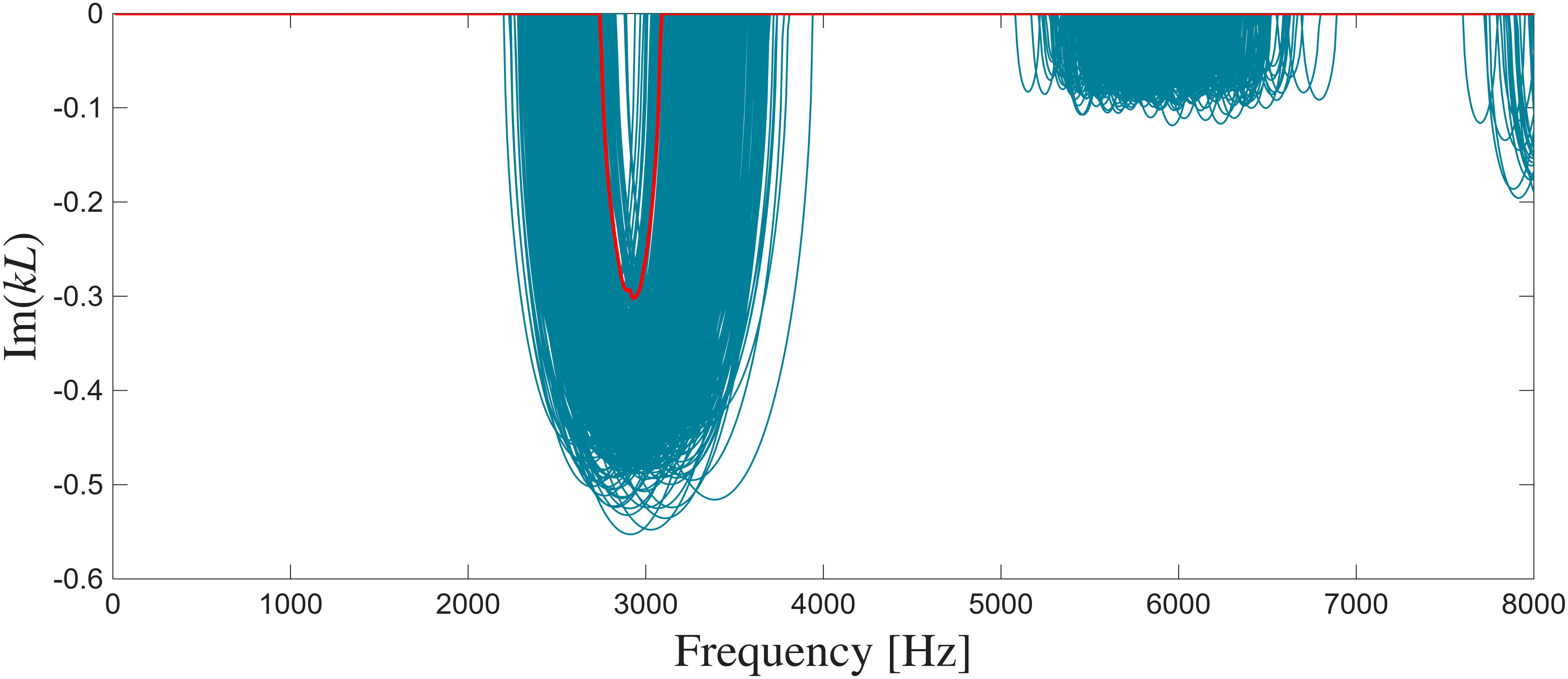}
        \end{subfigure}
    \end{minipage}
    
    \caption{Simulated stochastic fields (samples in purple, mean as a blue continuous line and 95\% confidence interval as blue dashed lines) for the spatially varying (a) Young's modulus, (b) mass density, (c) and cross-sectional area.
    (d) Illustration of five samples of the stochastic dispersion diagram computed via the SEM (black and red continuous lines) compared with the one computed from the SLTV-based method (purple and blue dashed lines) for the elementary rod unit cell and (e) five samples of the stochastic FRF computed via the SEM (black continuous lines) compared with the one computed from the SLTV-based method (purple dashed lines).
    (f) Illustration of five hundred samples of the stochastic dispersion diagram computed via the SLTV-based method (green lines) and the 95\% confidence interval robust attenuation band (red line)  for elementary rod unit cells using the spatially varying properties. The blue line represents the inference performed on the stochastic results, indicating the attenuation that will occur in 95\% of the cases.}
    \label{F_disp_1}
\end{figure}

The spatially varying geometry and mechanical properties following the stochastic fields assumed for the rod element are illustrated in Figs. \ref{F_field_1a}--\ref{F_field_1c}. Using the SFS method, five samples of the stochastic dispersion diagram for the rod element obtained using the SLTV-based approach are compared to the stochastic dispersion diagrams obtained using the SEM and the TMM in Fig. \ref{F_rod_4}.
Analogously, five samples of the stochastic FRFs are shown in Fig. \ref{F_rod_5} to illustrate their equivalence with the results obtained using the SEM and the conventional assembling for the stochastic modeling for an elementary rod. 

The stochastic dispersion relation for an elementary rod and the inference performed on this stochastic result to present the robust bandgap is presented in Fig. \ref{F_rod_6}. Comparing the presented stochastic minimum attenuation with its deterministic version (Fig. \ref{F_rod_23}), it is possible to notice that a robust first bandgap is present. However, the second bandgap is not robust against the assumed variability in the geometry and mechanical properties.

\subsection{Saint-Venant shaft theory}

\begin{figure}[H]
    \centering 
		
    \begin{minipage}{.28\textwidth}
        \begin{subfigure}{\textwidth}
            \captionsetup{justification=raggedright, singlelinecheck=false}
            \caption{} \label{F_field_2a} 
            \includegraphics[width=\textwidth]{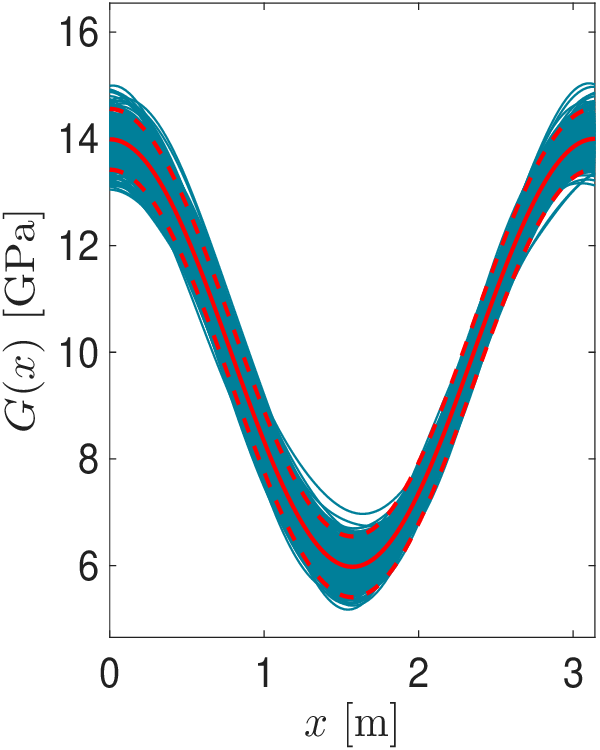}
        \end{subfigure}
        
        \vspace{0.3cm} 

        \begin{subfigure}{\textwidth}
            \captionsetup{justification=raggedright, singlelinecheck=false}
            \caption{} \label{F_field_2b}
            \includegraphics[width=\textwidth]{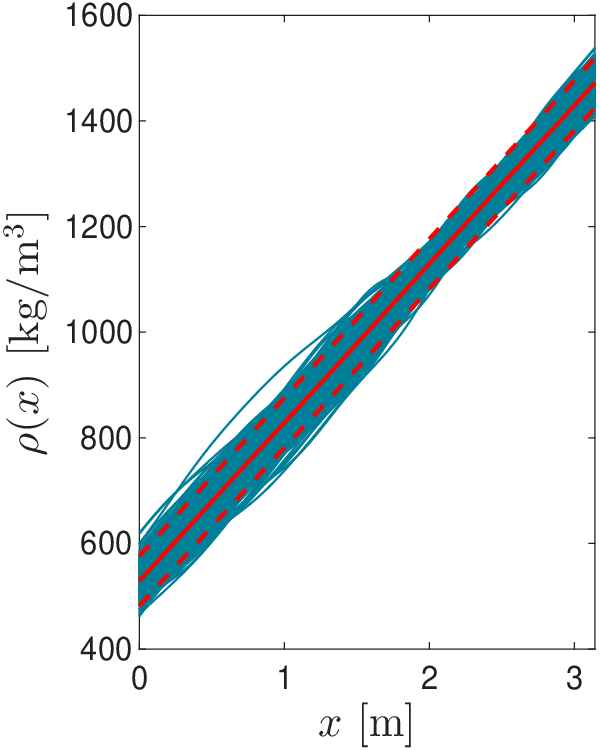}
        \end{subfigure}

        \vspace{0.3cm}

        \begin{subfigure}{\textwidth}
            \captionsetup{justification=raggedright, singlelinecheck=false}
            \caption{} \label{F_field_2c}
            \includegraphics[width=\textwidth]{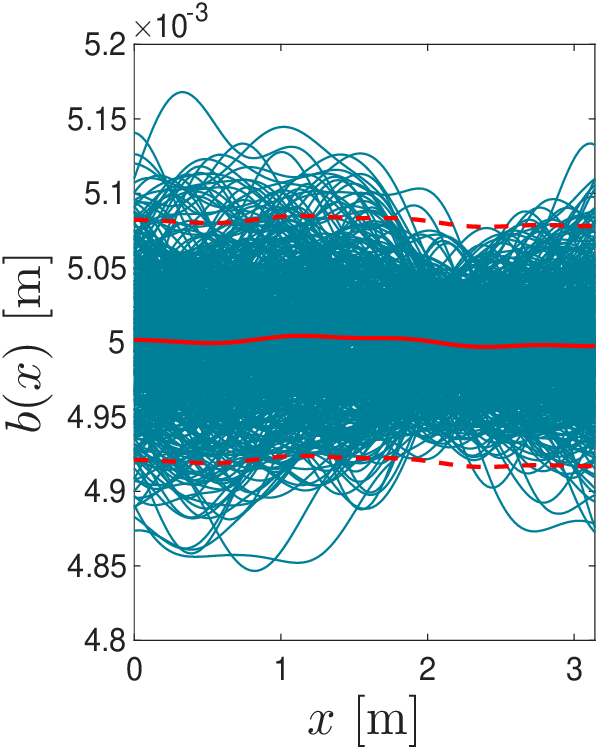}
        \end{subfigure}
    \end{minipage}\hfill 
    \begin{minipage}{.69\textwidth}
        \begin{subfigure}{\textwidth}
            \captionsetup{justification=raggedright, singlelinecheck=false}
            \caption{} \label{F_shaft_4}
            \includegraphics[width=\textwidth]{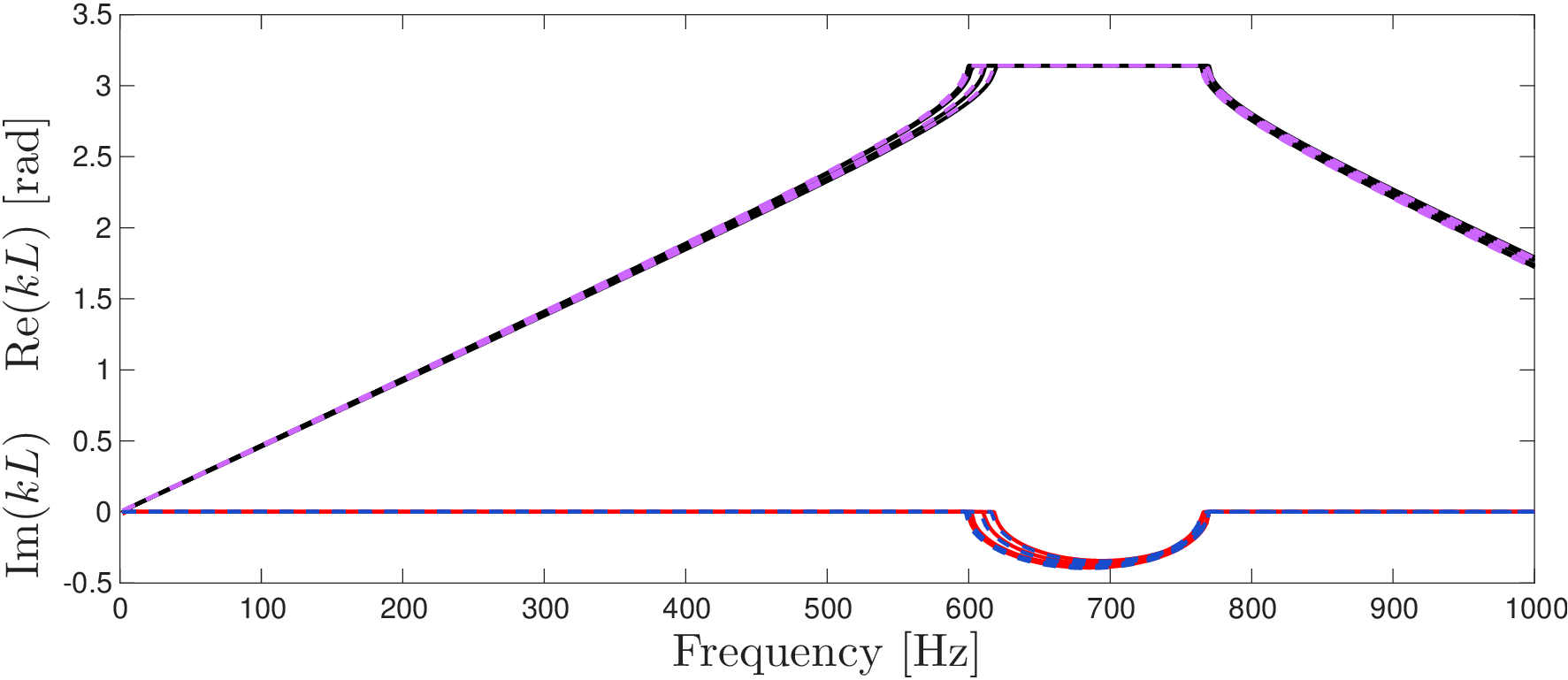}
        \end{subfigure}
        
        \vspace{0.3cm} 

        \begin{subfigure}{\textwidth}
            \captionsetup{justification=raggedright, singlelinecheck=false}
            \caption{} \label{F_shaft_5}
            \includegraphics[width=\textwidth]{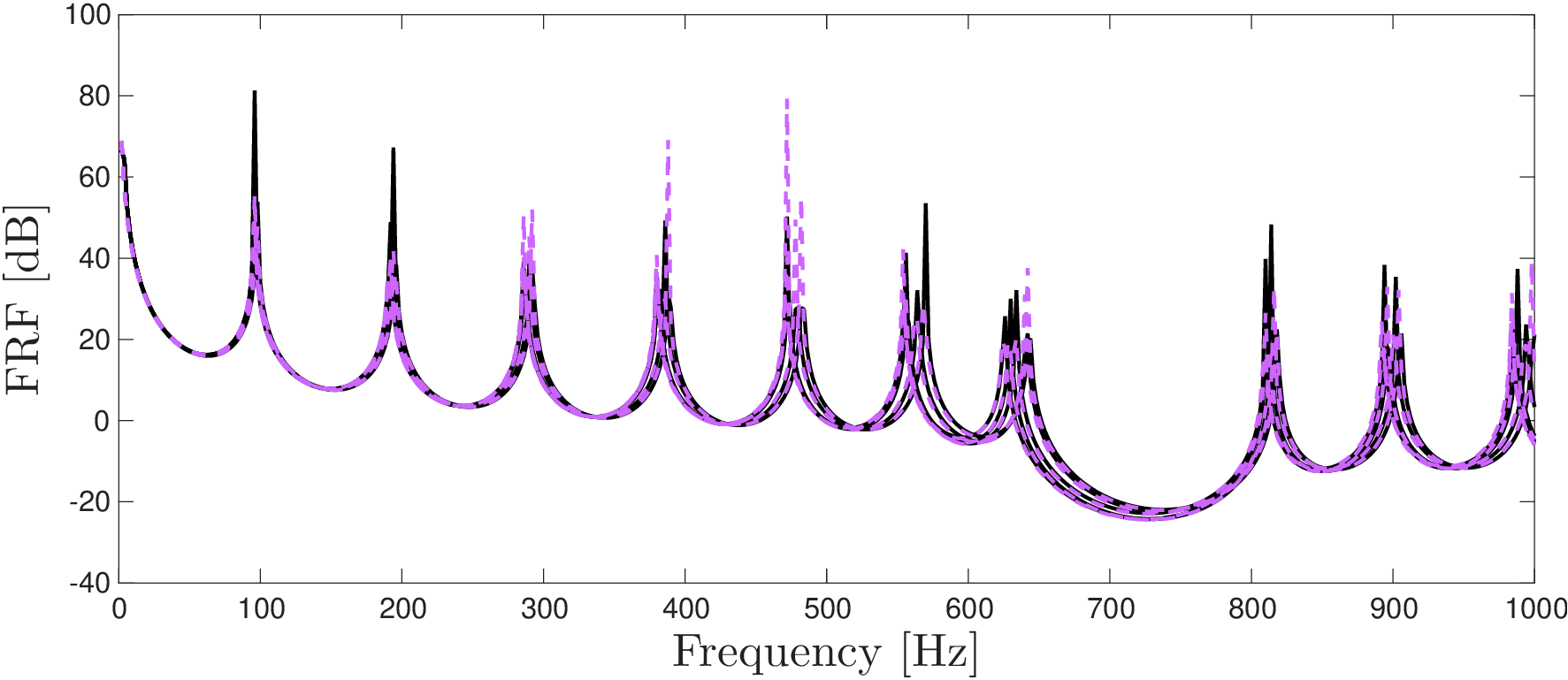}
        \end{subfigure}

        \vspace{0.3cm}

        \begin{subfigure}{\textwidth}
            \captionsetup{justification=raggedright, singlelinecheck=false}
            \caption{} \label{F_shaft_6}
            \includegraphics[width=\textwidth]{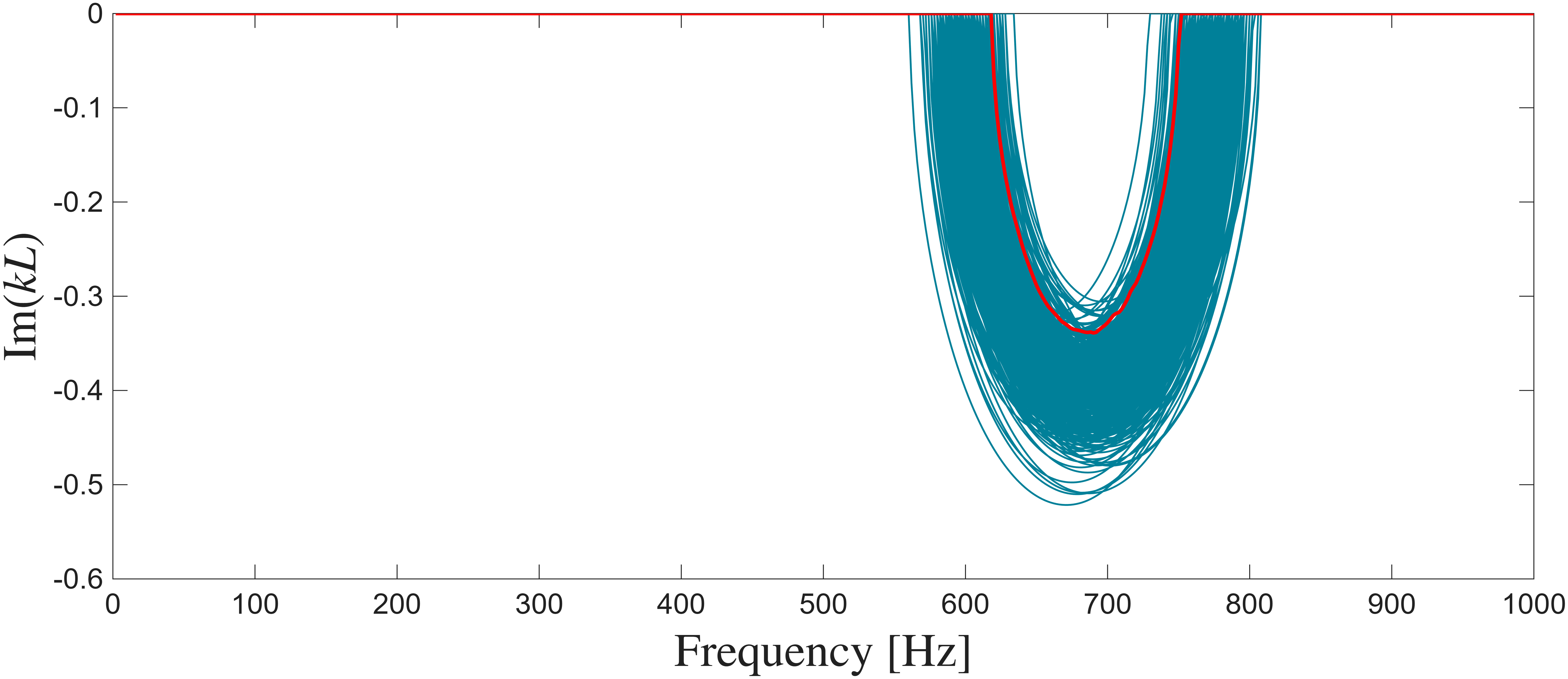}
        \end{subfigure}
    \end{minipage}%
    
    \caption{Simulated stochastic fields (samples in purple, mean as a blue continuous line and 95\% confidence interval as blue dashed lines) for the spatially varying (a) shear modulus, (b) mass density, and (c) length of the side for the assumed square Saint-Venant shaft.
    (d) Illustration of five samples of the stochastic dispersion diagram computed via the SEM (black and red continuous lines) compared with the one computed from the SLTV-based method (purple and blue dashed lines) for the Saint-Venant shaft unit cell and (e) five samples of the stochastic forced response computed via the SEM (black continuous lines) compared with the one computed from the SLTV-based method (purple dashed lines) for a Saint-Venant shaft finite structure. (f) Illustration of five hundred samples of the stochastic dispersion diagram computed via the SLTV-based method (green lines) and the 95\% confidence interval robust attenuation band (red line) for an elementary Saint-Venant shaft unit cell with spatially varying properties presented in Table \ref{Table_1} and the variability is defined in Table \ref{Table_2}. The blue line represents the inference performed on the stochastic results, indicating the attenuation that will occur in 95\% of the cases.}
    \label{F_disp_2}
\end{figure}

Here, the stochastic fields are modeled using the KLE, and a shaft element presenting spatially varying geometry and mechanical properties following the stochastic fields illustrated in Figs. \ref{F_field_2a}--\ref{F_field_2c} are assumed. Five samples of the stochastic dispersion diagram obtained using the SLTV-based approach are shown in Fig. \ref{F_shaft_4} to demonstrate the coincidence with the results obtained from the SEM and TMM. Furthermore, five samples of the stochastic FRF are shown in Fig. \ref{F_shaft_5} to illustrate that the FRF computed via the SLTV-based approach and the SEM using a conventional assembly method are equivalent. In this case, the single occurring bandgap is robust against the proposed variability in geometry and mechanical properties, as indicated by the inference performed on the stochastic results of the minimum part of the imaginary wavenumbers indicated in Fig. \ref{F_shaft_6}.
 
\subsection{Euler-Bernoulli beam theory}

\begin{figure}[H]
    \centering 

    \begin{minipage}{.28\textwidth}
        \begin{subfigure}{\textwidth}
            \captionsetup{justification=raggedright, singlelinecheck=false}
            \caption{} \label{F_field_3a} 
            \includegraphics[width=\textwidth]{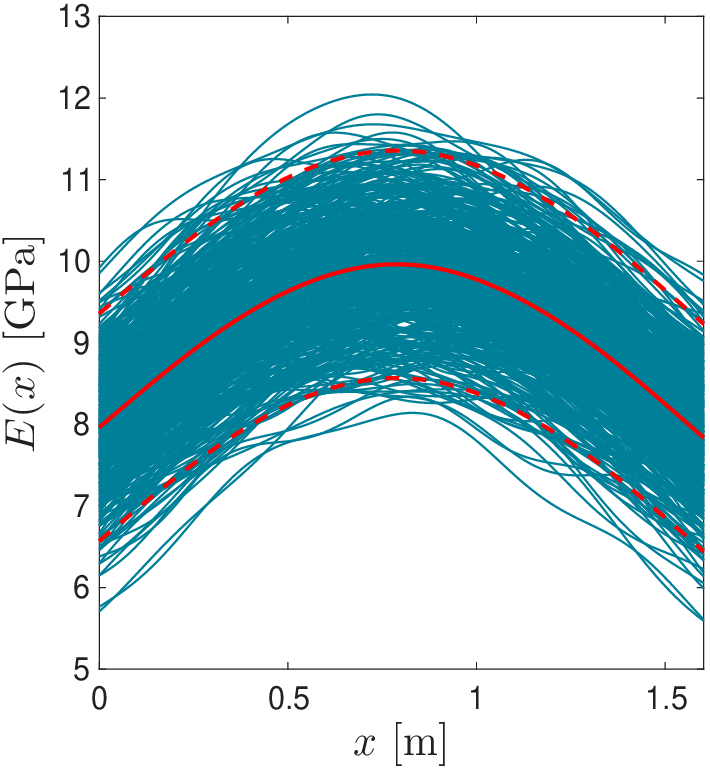}
        \end{subfigure}
        
        \vspace{0.3cm} 

        \begin{subfigure}{\textwidth}
            \captionsetup{justification=raggedright, singlelinecheck=false}
            \caption{} \label{F_field_3b}
            \includegraphics[width=\textwidth]{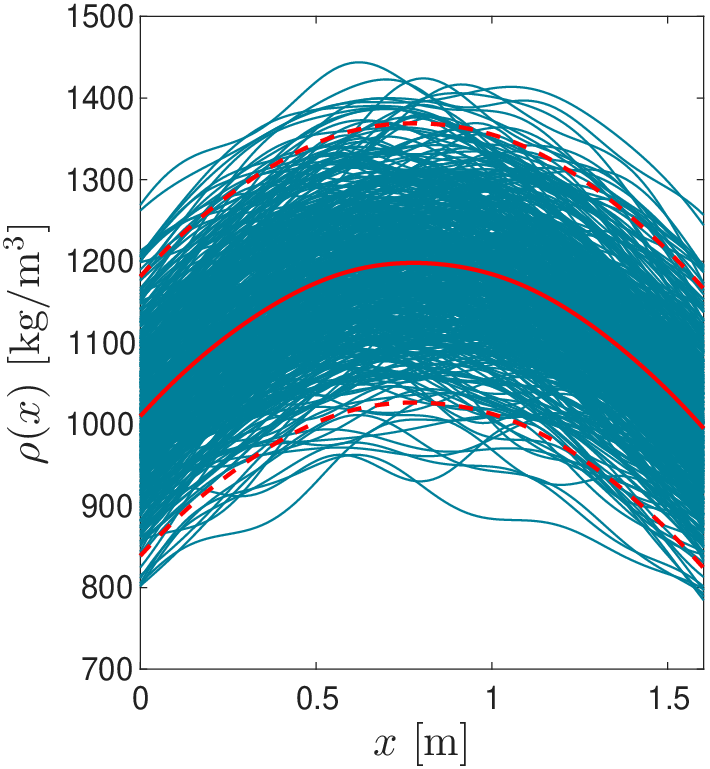}
        \end{subfigure}

        \vspace{0.3cm}

        \begin{subfigure}{\textwidth}
            \captionsetup{justification=raggedright, singlelinecheck=false}
            \caption{} \label{F_field_3c}
            \includegraphics[width=\textwidth]{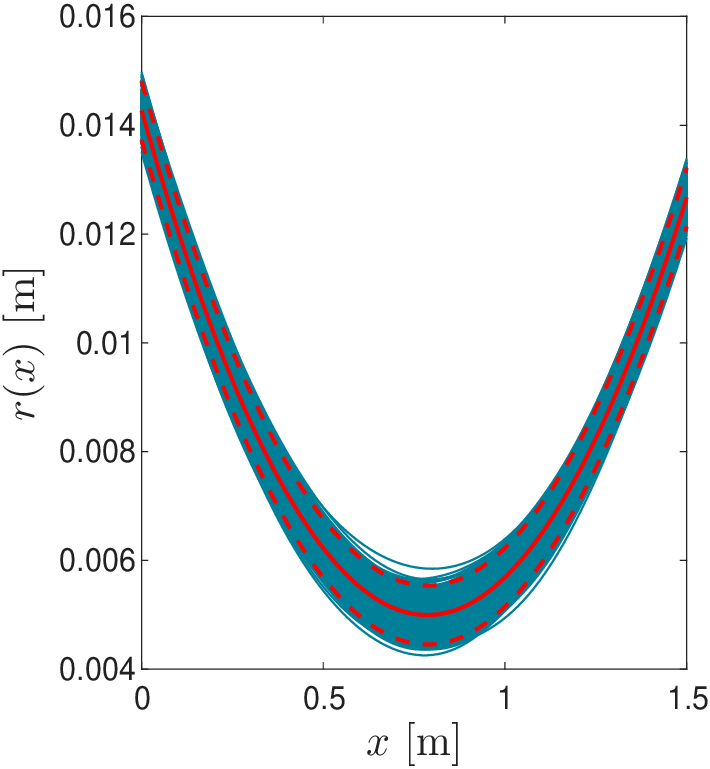}
        \end{subfigure}
    \end{minipage}\hfill 
    \begin{minipage}{.69\textwidth}
        \begin{subfigure}{\textwidth}
            \captionsetup{justification=raggedright, singlelinecheck=false}
            \caption{} \label{F_EB_4}
            \includegraphics[width=\textwidth]{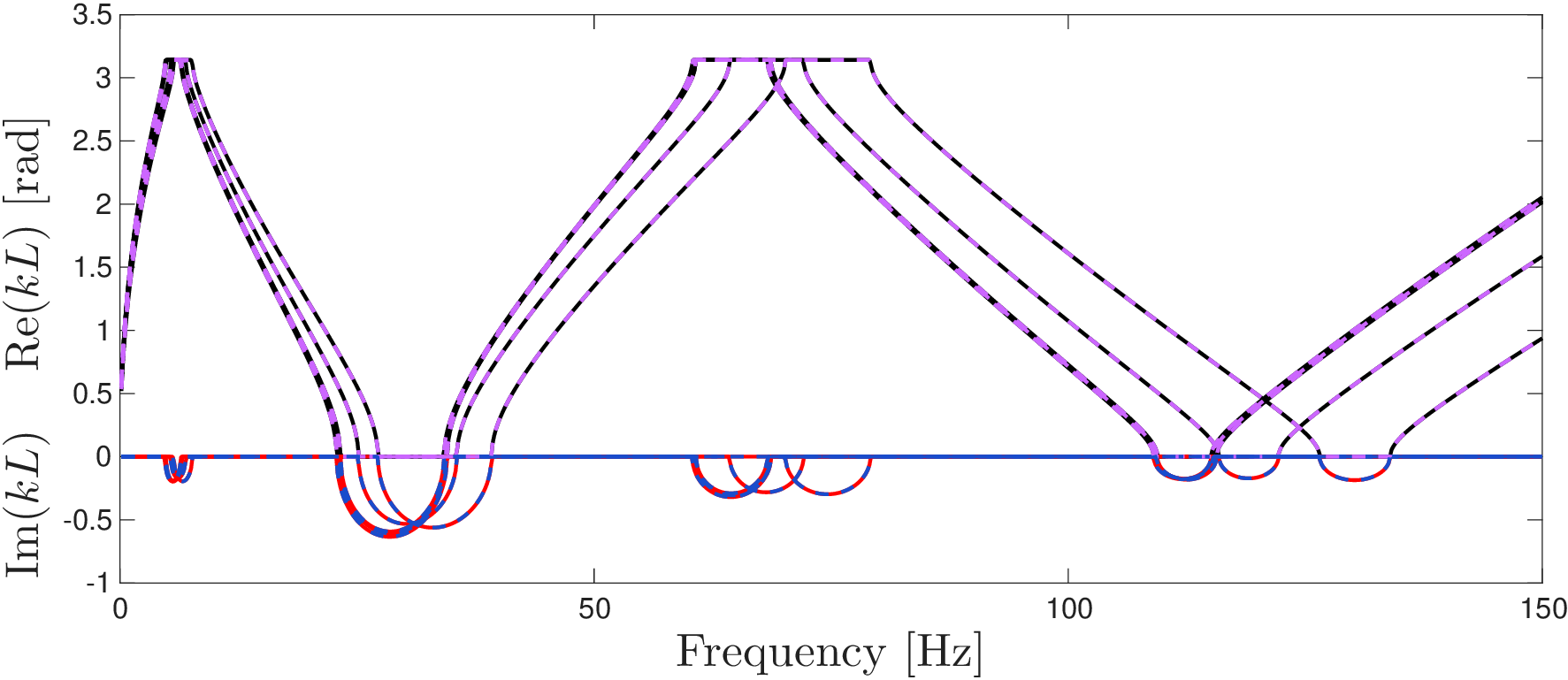}
        \end{subfigure}
        
        \vspace{0.3cm} 

        \begin{subfigure}{\textwidth}
            \captionsetup{justification=raggedright, singlelinecheck=false}
            \caption{} \label{F_EB_5}
            \includegraphics[width=\textwidth]{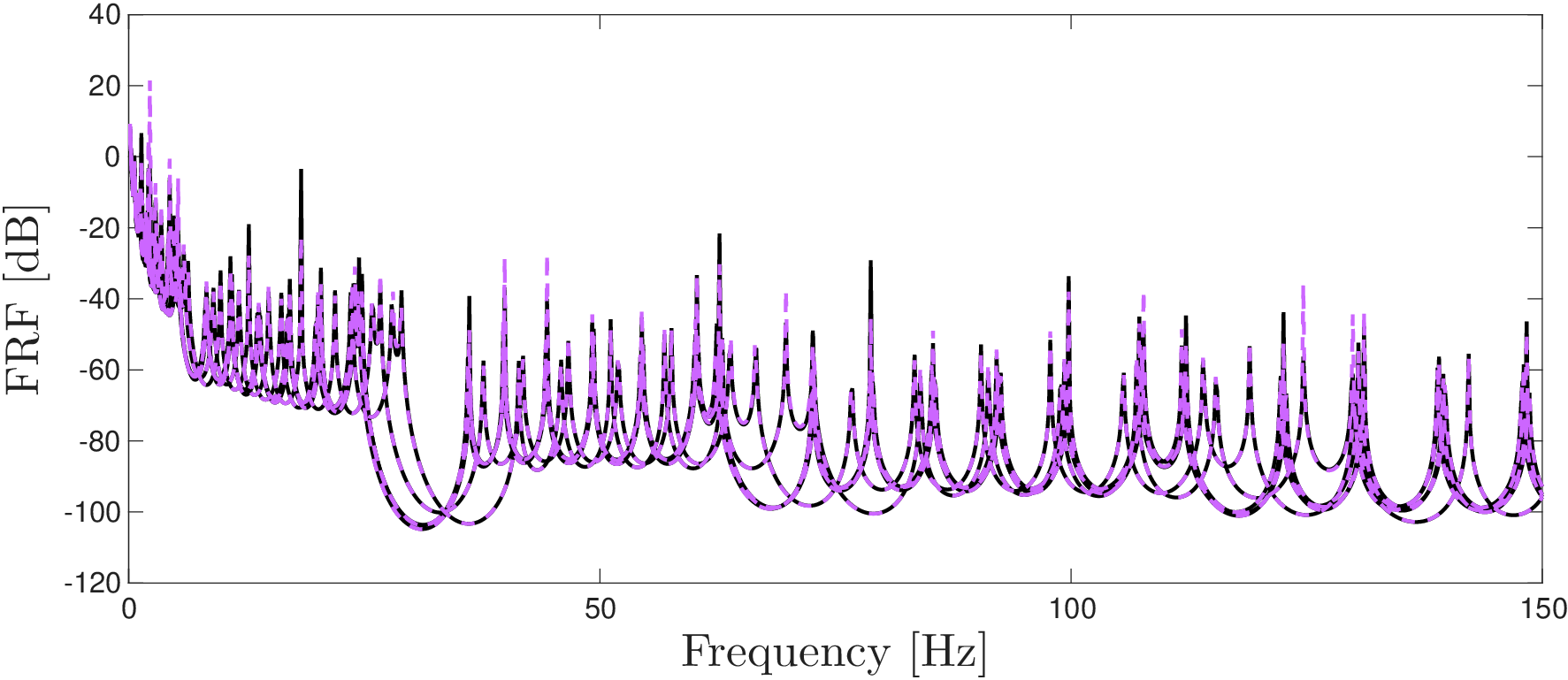}
        \end{subfigure}

        \vspace{0.3cm}

        \begin{subfigure}{\textwidth}
            \captionsetup{justification=raggedright, singlelinecheck=false}
            \caption{} \label{F_EB_6}
            \includegraphics[width=\textwidth]{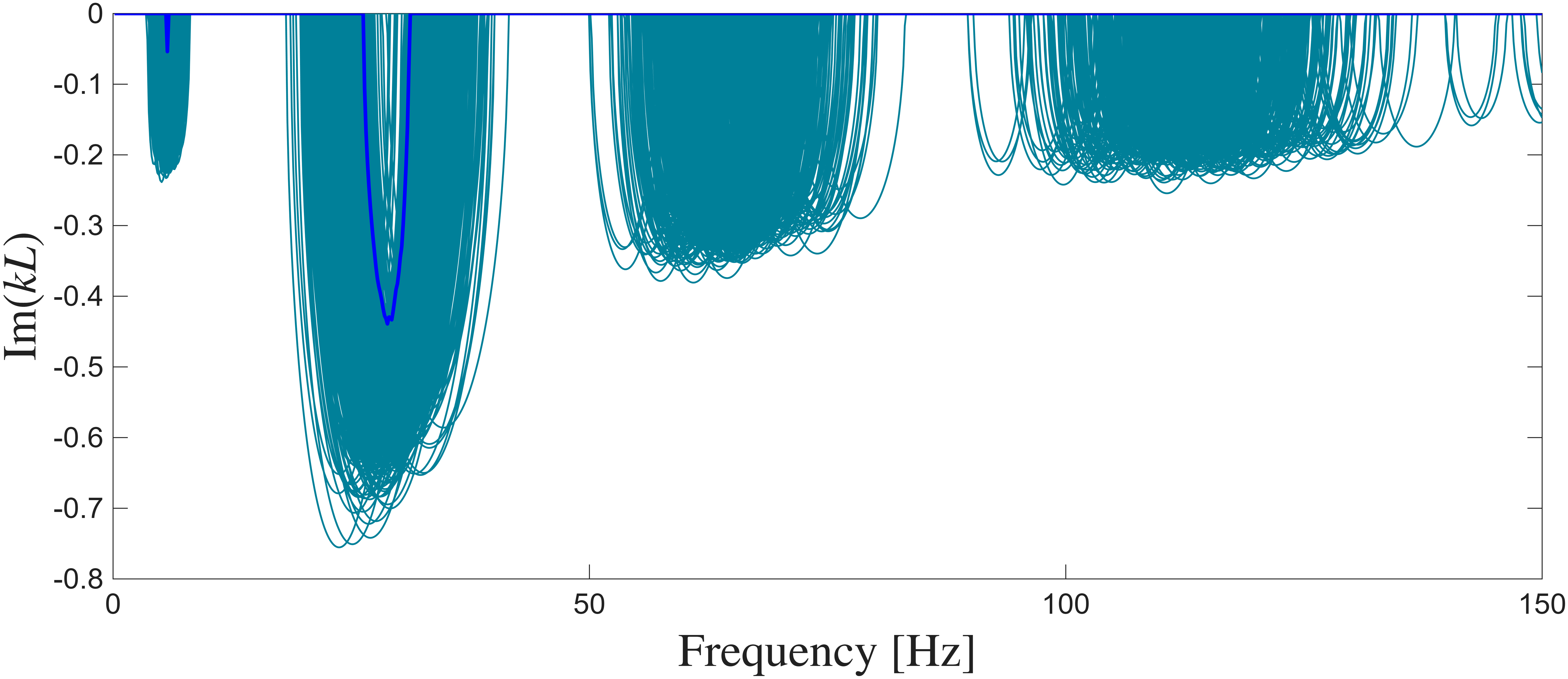}
        \end{subfigure}
    \end{minipage}

    \caption{Simulated stochastic fields (samples in purple, mean as a blue continuous line and 95\% confidence interval as blue dashed lines) for the spatially varying (a) Young's modulus, (b) mass density, (c) and radius for the assumed Euler-Bernoulli beam. (d) Illustration of five samples of the stochastic dispersion diagram computed via the SEM (black and red continuous lines) compared with the one computed from the SLTV-based method (purple and blue dashed lines) for the Euler-Bernoulli beam unit cell and (e) five samples of the stochastic FRF computed via the SEM (black continuous lines) compared with the one computed from the SLTV-based method (purple dashed lines) for an Euler-Bernoulli beam metastructure made of seven unit cells with spatially varying properties presented in Table \ref{Table_1} and the variability is defined in Table \ref{Table_2}. (f) (f) Illustration of five hundred samples of the stochastic dispersion diagram computed via the SLTV-based method (green lines) and the 95\% confidence interval robust attenuation band (red line) for the Euler-Bernoulli beam unit cell with spatially varying properties presented in Table \ref{Table_1}, and the variability is defined in Table \ref{Table_2}. The blue line represents the inference performed on the stochastic results, indicating the attenuation that will occur in 95\% of the cases.}
    \label{F_disp_3}
\end{figure}

In the current section, the stochastic fields were modeled using the SFS and are illustrated in Figs. \ref{F_field_3a}--\ref{F_field_3c}.
Five samples of the stochastic dispersion diagram are used to illustrate that the SLTV-based approach coincides with the SEM concerning the stochastic dispersion diagram obtained via the TMM, and are presented in Fig. \ref{F_EB_4}. Five samples of the stochastic FRF are shown in Fig. \ref{F_EB_5} to illustrate that the FRF computed via the SLTV-based approach combined with the recursive condensation method also coincides with the FRF computed via SEM and the conventional assembly for the stochastic modeling for an Euler-Bernoulli beam.
The stochastic dispersion relation for Euler-Bernoulli beams and the inference performed on this stochastic result to present the robust bandgap is presented in Fig. \ref{F_EB_6} for a beam element with spatially varying geometry and mechanical properties following the stochastic field illustrated in Fig. \ref{F_disp_3}. 
Using a similar analysis as done for the rod, one can notice that the first and second bandgaps of the proposed structure are robust against the proposed variability, as evidenced by comparing Figs.~\ref{F_EB_6} and \ref{F_EB_23}.

\subsection{SSH Topological structures}

\begin{figure}[H]
    \centering 

    \begin{minipage}{.31\textwidth}
        \begin{subfigure}{\textwidth}
            \captionsetup{justification=raggedright, singlelinecheck=false}
            \caption{} \label{Field_rod_zak_1} 
            \includegraphics[width=\textwidth]{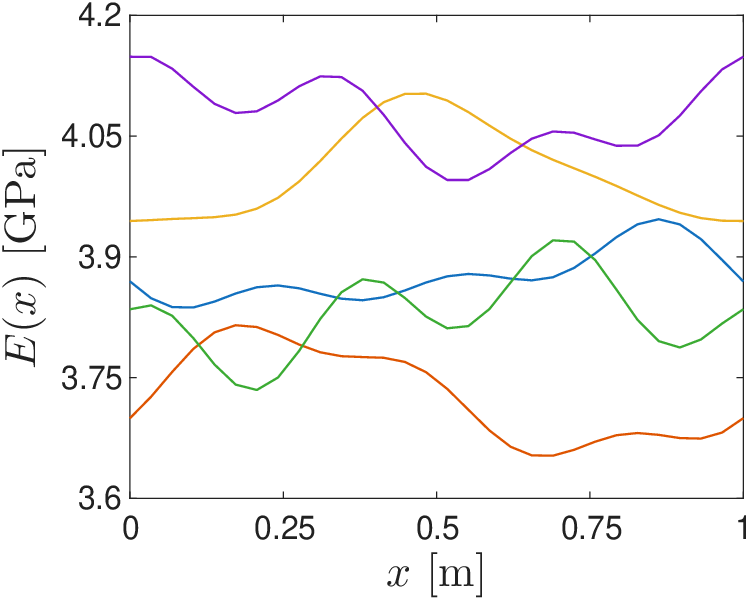}
        \end{subfigure}    
        
        \vspace{0.3cm} 

        \begin{subfigure}{\textwidth}
            \captionsetup{justification=raggedright, singlelinecheck=false}
            \caption{} \label{Field_rod_zak_4}
            \includegraphics[width=\textwidth]{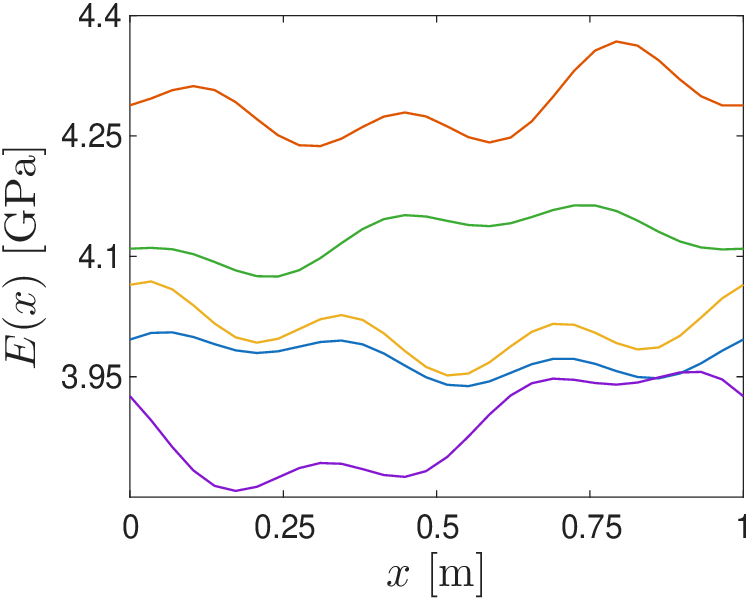}
        \end{subfigure}
    \end{minipage}\hfill 
    \begin{minipage}{.31\textwidth}
        \begin{subfigure}{\textwidth}
            \captionsetup{justification=raggedright, singlelinecheck=false}
            \caption{} \label{Field_rod_zak_2}
            \includegraphics[width=\textwidth]{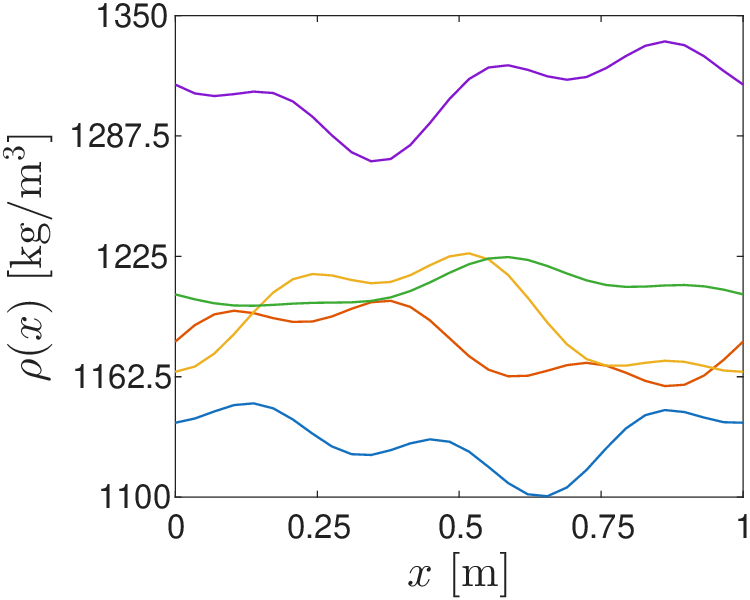}
        \end{subfigure}    
        
        \vspace{0.3cm}

        \begin{subfigure}{\textwidth}
            \captionsetup{justification=raggedright, singlelinecheck=false}
            \caption{} \label{Field_rod_zak_5}
            \includegraphics[width=\textwidth]{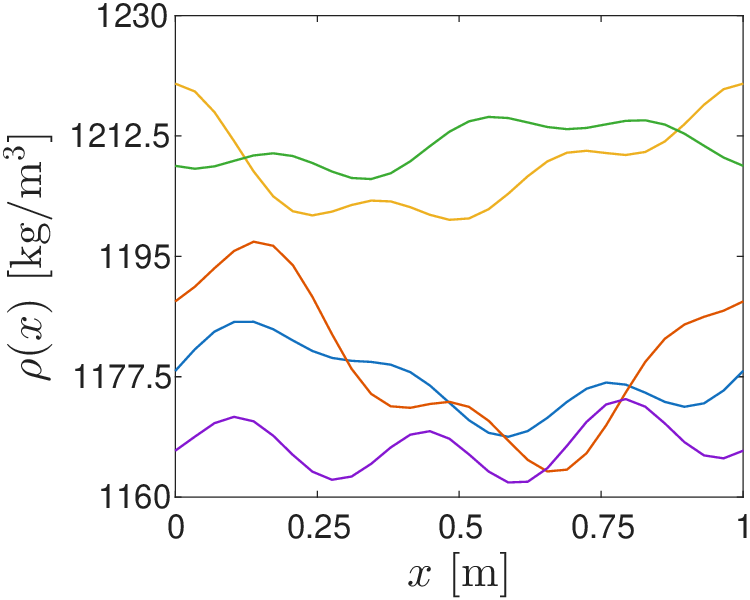}
        \end{subfigure}
    \end{minipage}\hfill
    \begin{minipage}{.31\textwidth}
        \begin{subfigure}{\textwidth}
            \captionsetup{justification=raggedright, singlelinecheck=false}
            \caption{} \label{Field_rod_zak_3}
            \includegraphics[width=\textwidth]{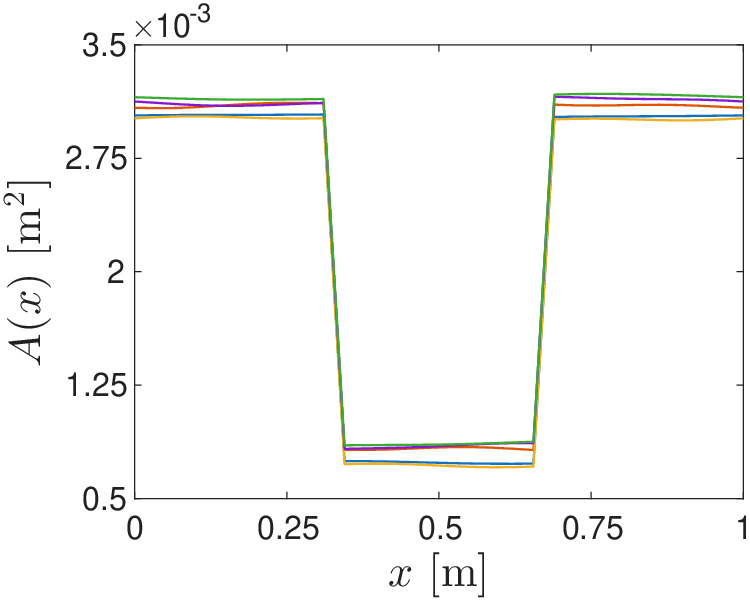}
        \end{subfigure}
        
        \vspace{0.3cm}

        \begin{subfigure}{\textwidth}
            \captionsetup{justification=raggedright, singlelinecheck=false}
            \caption{} \label{Field_rod_zak_6}
            \includegraphics[width=\textwidth]{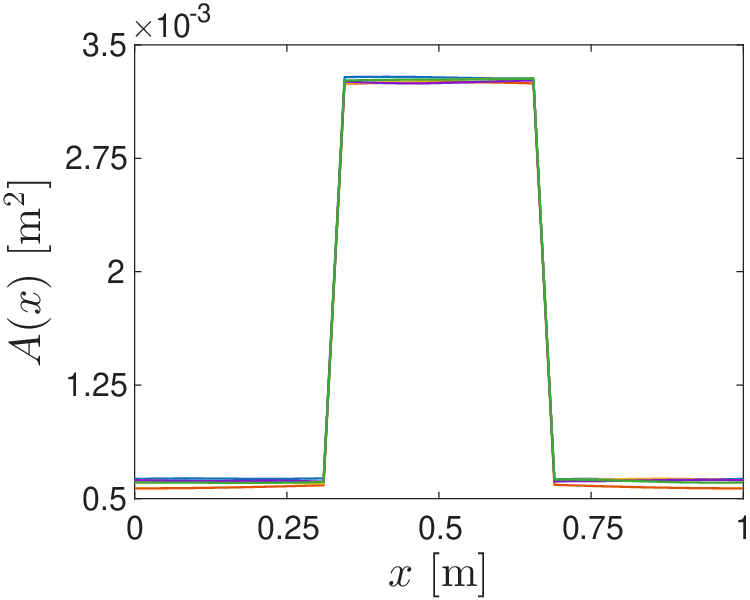}
        \end{subfigure}
    \end{minipage}

    \caption{Robustness of the Zak phase under mechanical and geometric variability for the rod waveguide: (a)-(f) spatial field distributions and mode profiles across five stochastic realization samples for modulation parameters $\Delta_A = -0.0023$ (first vertical line of Fig. \ref{F_Top_rod_2}) and $\Delta_A = 0.0027$ (second vertical line of Fig. \ref{F_Top_rod_2}). The Zak phases computed across the first five passbands rigorously match the deterministic benchmark ($\Theta_n^{\text{Zak}} = \pi$ for bands below transition points at $\Delta_A = -0.0023$, and $\Theta_n^{\text{Zak}} = 0$ for all bands at $\Delta_A = 0.0027$).}
\label{F_Zak_rod}
\end{figure}

To assess the sensitivity of the topological features with respect to spatially correlated variability, random variability in the mechanical and geometric properties were introduced for a rod using the stochastic Fourier series per segment, with $\sigma^2=0.5$ for $E(x)$, $\sigma^2=0.3$ for $\rho(x)$, and $\sigma^2=0.2$ for $A(x)$. Five representative stochastic realization samples were evaluated and compared against the nominal baseline at two distinct modulation parameters arbitrarily chosen: $\Delta_A = -0.0023$ and $\Delta_A = 0.0027$. Figure~\ref{F_Zak_rod} illustrates these sample realizations alongside the Zak phase calculations for the first five passbands of the rod element at two distinct modulation parameters: $\Delta_A = -0.0023$ (first vertical line of Fig. \ref{F_Top_rod_2}) and $\Delta_A = 0.0027$ (second vertical line of Fig. \ref{F_Top_rod_2}). Remarkably, despite the presence of spatial variability, the computed topological invariants remain fully consistent with the deterministic predictions shown in Fig.~\ref{Fig_Top_red}. Specifically, for $\Delta_A = -0.0023$, the Zak phases for the passbands situated below the topological phase transition points are non-trivial ($\Theta_n^{\text{Zak}} = \pi$), whereas the remaining bands yield a trivial Zak phase ($\Theta_n^{\text{Zak}} = 0$). Conversely, for $\Delta_A = 0.0027$, all five evaluated passbands exhibit trivial topological properties ($\Theta_n^{\text{Zak}} = 0$).

\begin{figure}[H]
    \centering 

    \begin{minipage}{.31\textwidth}
        \begin{subfigure}{\textwidth}
            \captionsetup{justification=raggedright, singlelinecheck=false}
            \caption{} \label{Field_shaft_zak_1} 
            \includegraphics[width=\textwidth]{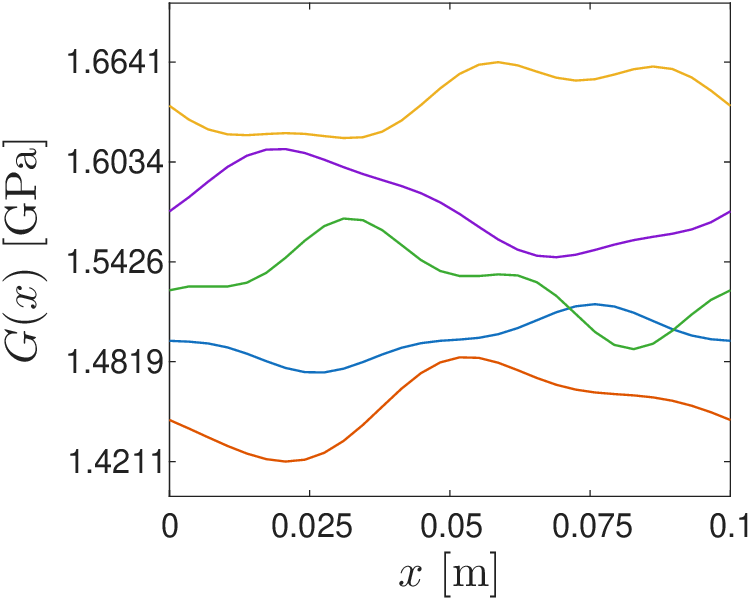}
        \end{subfigure}    
        
        \vspace{0.3cm} 

        \begin{subfigure}{\textwidth}
            \captionsetup{justification=raggedright, singlelinecheck=false}
            \caption{} \label{Field_shaft_zak_4}
            \includegraphics[width=\textwidth]{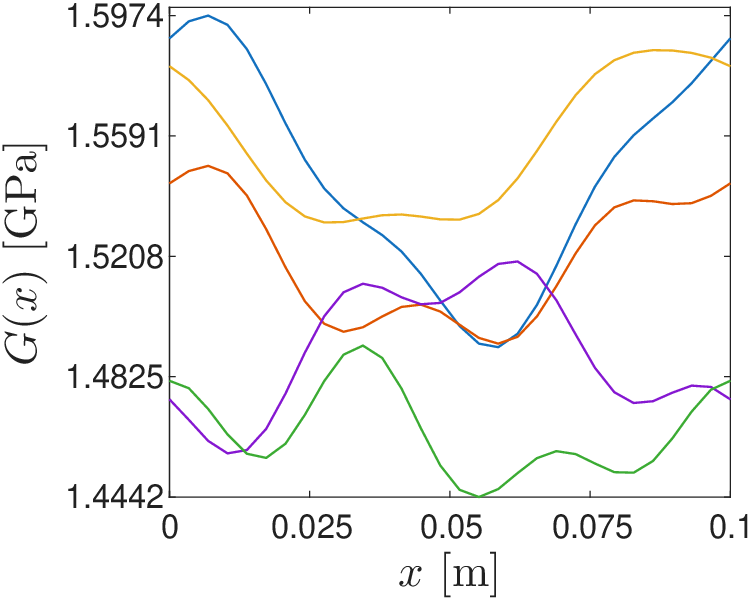}
        \end{subfigure}
    \end{minipage}\hfill 
    \begin{minipage}{.31\textwidth}
        \begin{subfigure}{\textwidth}
            \captionsetup{justification=raggedright, singlelinecheck=false}
            \caption{} \label{Field_shaft_zak_2}
            \includegraphics[width=\textwidth]{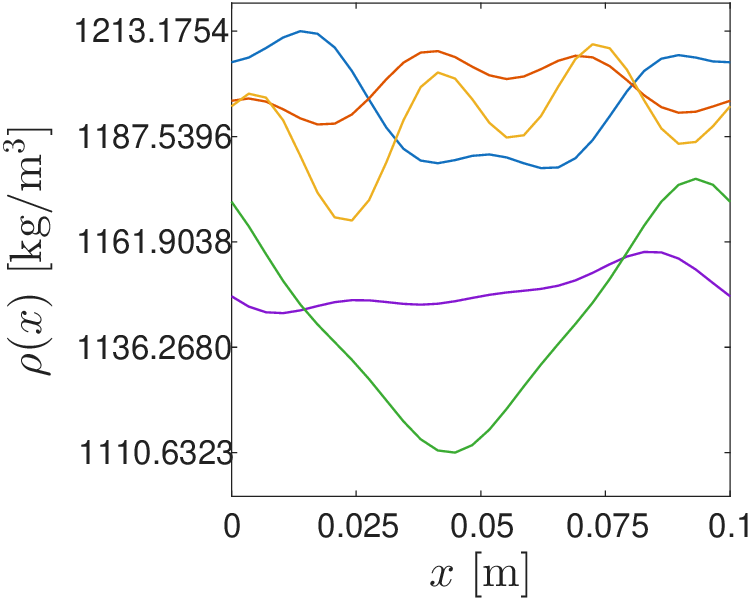}
        \end{subfigure}    
        
        \vspace{0.3cm}

        \begin{subfigure}{\textwidth}
            \captionsetup{justification=raggedright, singlelinecheck=false}
            \caption{} \label{Field_shaft_zak_5}
            \includegraphics[width=\textwidth]{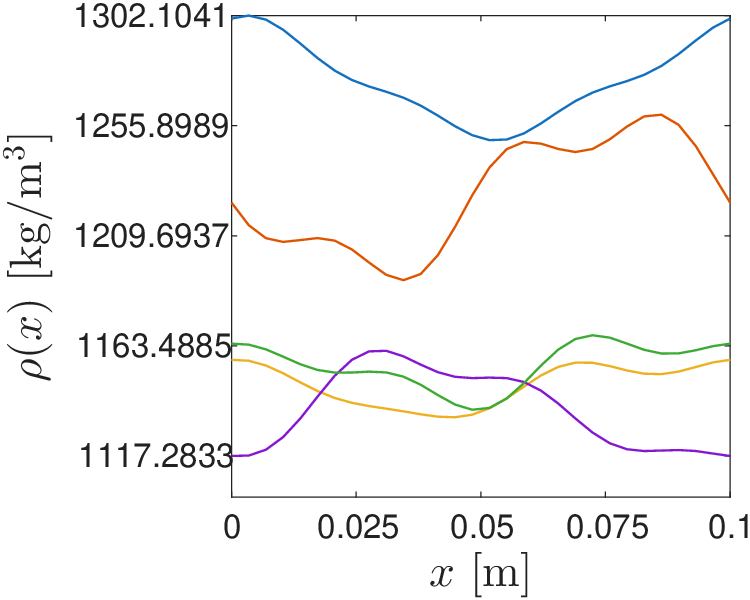}
        \end{subfigure}
    \end{minipage}\hfill
    \begin{minipage}{.31\textwidth}
        \begin{subfigure}{\textwidth}
            \captionsetup{justification=raggedright, singlelinecheck=false}
            \caption{} \label{Field_shaft_zak_3}
            \includegraphics[width=\textwidth]{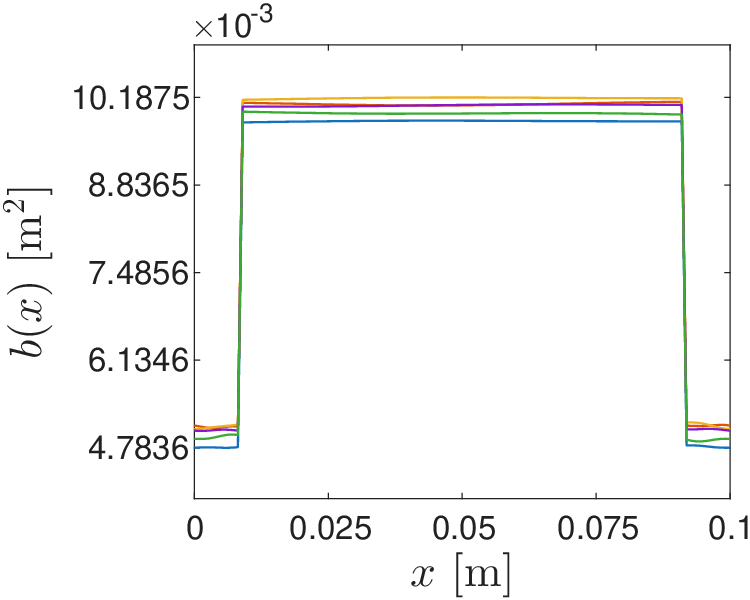}
        \end{subfigure}

        \vspace{0.3cm}

        \begin{subfigure}{\textwidth}
            \captionsetup{justification=raggedright, singlelinecheck=false}
            \caption{} \label{Field_shaft_zak_6}
            \includegraphics[width=\textwidth]{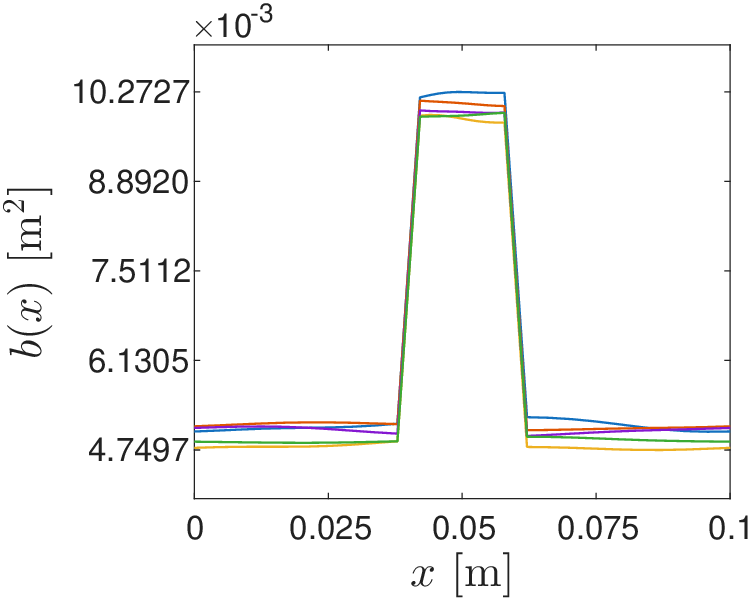}
        \end{subfigure}
    \end{minipage}

    \caption{Robustness of the Zak phase under parameter variability for the shaft waveguide: (a)--(f) spatial field responses and mode structures across five stochastic realization samples evaluated at $\Delta_L = -0.0334$ and $\Delta_L = 0.0328$. The Zak phases computed for the first three passbands fully confirm agreement with the deterministic invariants ($\Theta_n^{\text{Zak}} = \pi$ for the second band at $\Delta_L = -0.0334$, and $\Theta_n^{\text{Zak}} = 0$ across all bands at $\Delta_L = 0.0328$).}
    \label{F_Zak_shaft}
\end{figure}

A similar uncertainty quantification procedure was applied to a square shaft waveguide, using the stochastic Fourier series per segment using $\sigma^2=0.5$ for $G(x)$, $\sigma^2=0.3$ for $\rho(x)$, and $\sigma^2=0.2$ for $b(x)$. Figure~\ref{F_Zak_shaft} presents the five random realizations and the corresponding Zak phase evaluations for the first three passbands at modulation states at two distinct modulation parameters arbitrarily chosen: $\Delta_L = -0.0334$ and $\Delta_L = 0.0328$. Again, the topological invariants demonstrate remarkable resilience against parameter fluctuations, perfectly matching the deterministic results reported in Fig.~\ref{Fig_Top_red}. For $\Delta_L = 0.0328$, all three passbands display trivial topological characteristics ($\Theta_n^{\text{Zak}} = 0$). In contrast, for $\Delta_L = -0.0334$, the second passband—which immediately precedes the topological phase transition point—exhibits a non-trivial Zak phase ($\Theta_n^{\text{Zak}} = \pi$), while the remaining bands remain trivial.

\begin{figure}[H]
    \centering 

    \begin{minipage}{.31\textwidth}
        \begin{subfigure}{\textwidth}
            \captionsetup{justification=raggedright, singlelinecheck=false}
            \caption{} \label{Field_EB_zak_1} 
            \includegraphics[width=\textwidth]{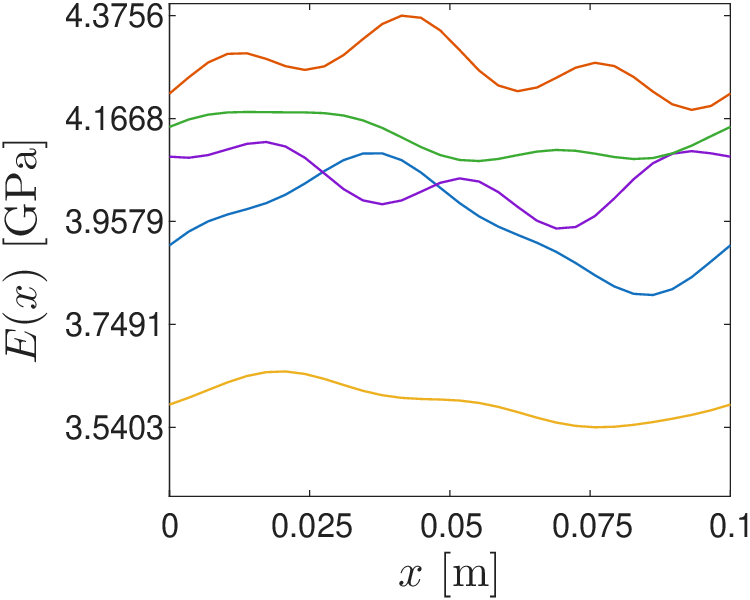}
        \end{subfigure}    
        
        \vspace{0.3cm} 

        \begin{subfigure}{\textwidth}
            \captionsetup{justification=raggedright, singlelinecheck=false}
            \caption{} \label{Field_EB_zak_4}
            \includegraphics[width=\textwidth]{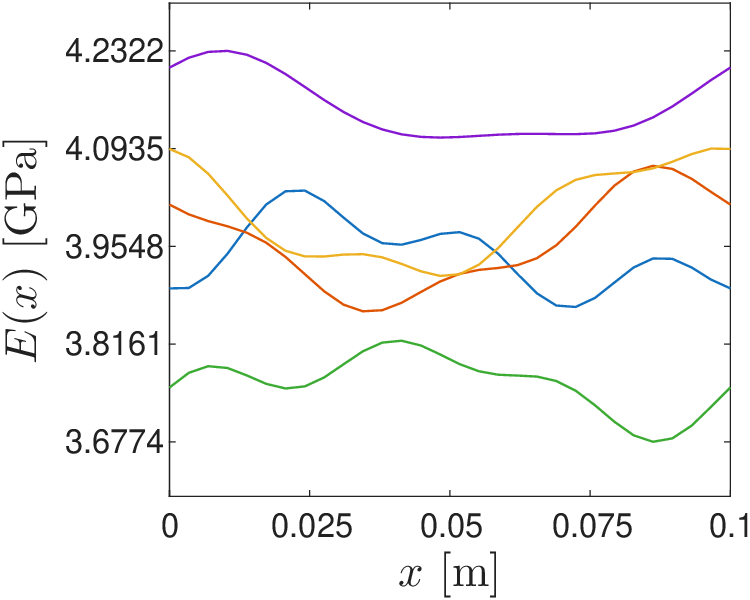}
        \end{subfigure}
    \end{minipage}\hfill 
    \begin{minipage}{.31\textwidth}
        \begin{subfigure}{\textwidth}
            \captionsetup{justification=raggedright, singlelinecheck=false}
            \caption{} \label{Field_EB_zak_2}
            \includegraphics[width=\textwidth]{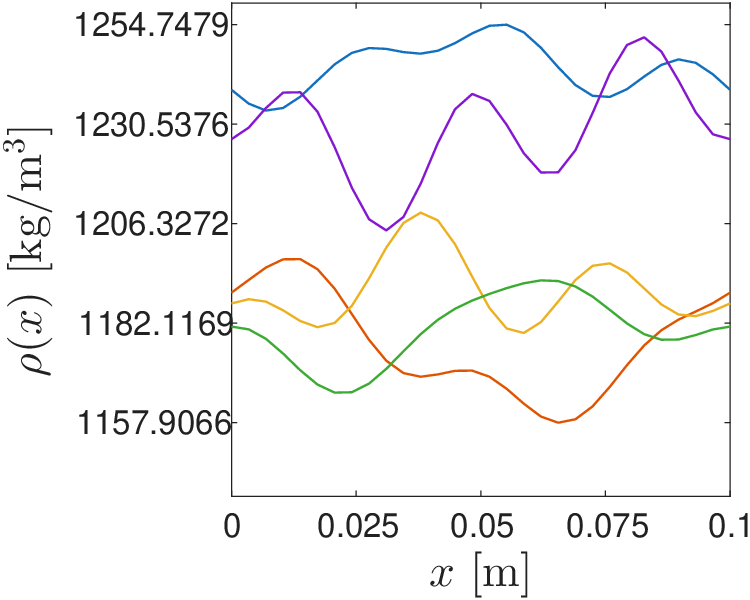}
        \end{subfigure}    
        
        \vspace{0.3cm}

        \begin{subfigure}{\textwidth}
            \captionsetup{justification=raggedright, singlelinecheck=false}
            \caption{} \label{Field_EB_zak_5}
            \includegraphics[width=\textwidth]{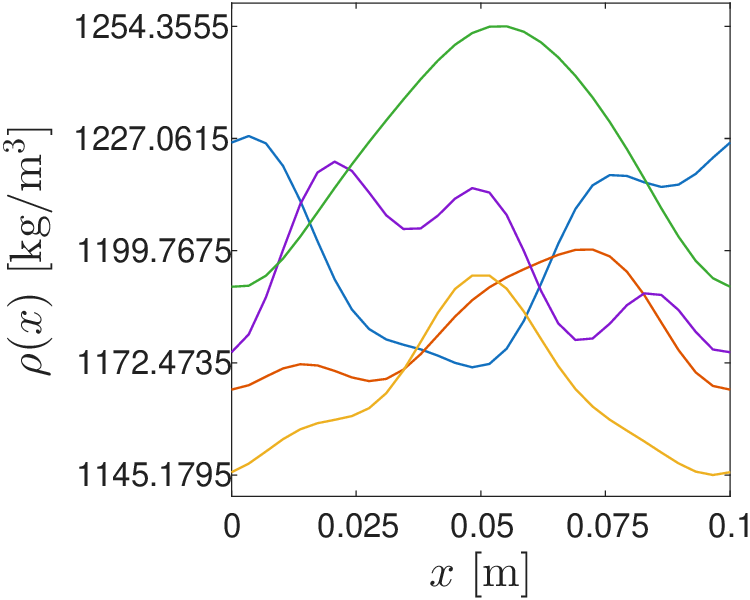}
        \end{subfigure}
    \end{minipage}\hfill
    \begin{minipage}{.31\textwidth}
        \begin{subfigure}{\textwidth}
            \captionsetup{justification=raggedright, singlelinecheck=false}
            \caption{} \label{Field_EB_zak_3}
            \includegraphics[width=\textwidth]{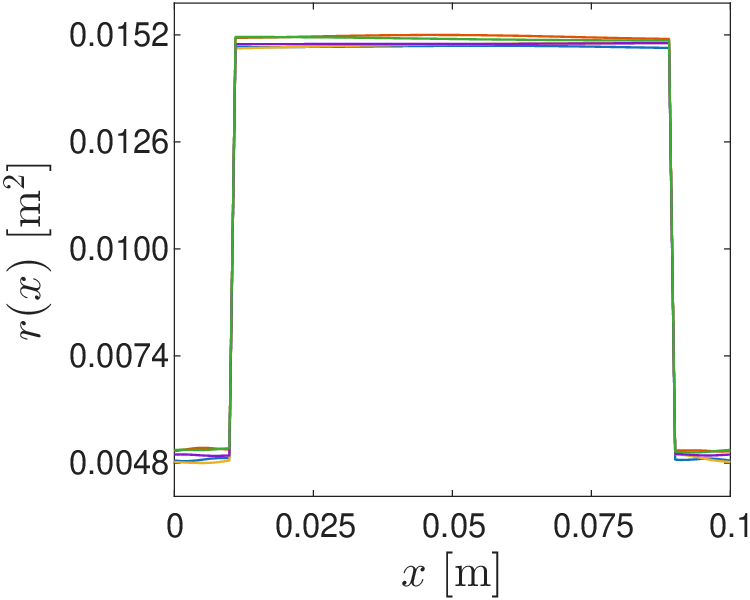}
        \end{subfigure}

        \vspace{0.3cm}

        \begin{subfigure}{\textwidth}
            \captionsetup{justification=raggedright, singlelinecheck=false}
            \caption{} \label{Field_EB_zak_6}
            \includegraphics[width=\textwidth]{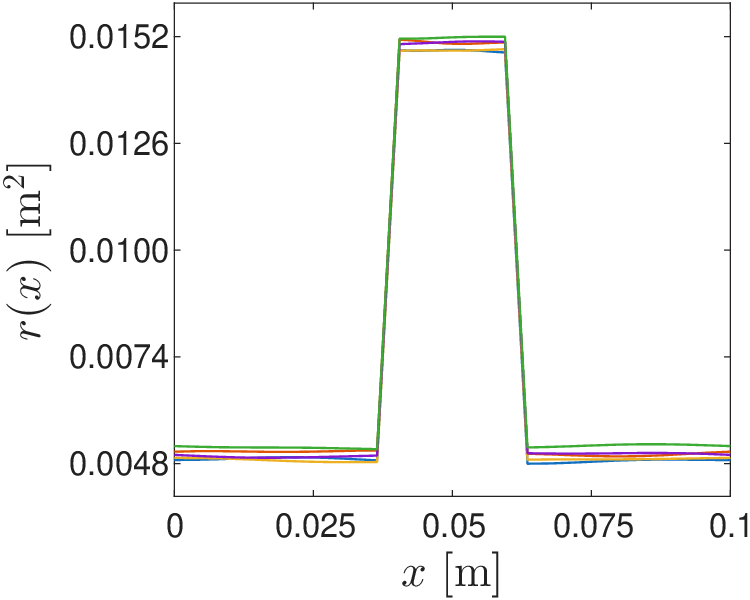}
        \end{subfigure}
    \end{minipage}

    \caption{Robustness of the Zak phase under parameter variability for the Euler-Bernoulli beam: (a)--(f) mode shapes and field distributions for five stochastic realization samples evaluated at $\Delta_L = -0.0147$ and $\Delta_L = 0.0149$. The Zak phase evaluated for the first passband fully matches the deterministic behavior ($\Theta_n^{\text{Zak}} = \pi$ for $\Delta_L = -0.0147$ and $\Theta_n^{\text{Zak}} = 0$ for $\Delta_L = 0.0149$).}
    \label{F_Zak_EB}
\end{figure}

Finally, the impact of spatial variability on the flexural behavior of a circular Euler-Bernoulli beam was investigated using the stochastic Fourier series per segment with $\sigma^2=0.5$ for $E(x)$, $\sigma^2=0.3$ for $\rho(x)$, and $\sigma^2=0.2$ for $r(x)$.  Five stochastic realization samples were analyzed for the first passband at two distinct modulation parameters arbitrarily chosen: $\Delta_L = -0.0147$ (first vertical line of Fig. \ref{F_Top_EB_2}) and $\Delta_L = 0.0149$ (second vertical line of Fig. \ref{F_Top_EB_2}), as depicted in Fig.~\ref{F_Zak_EB}. Despite the structural disorder, the topological characterization fully coincides with the deterministic benchmark in Fig.~\ref{Fig_Top_red}. For $\Delta_L = -0.0147$, the first passband yields a non-trivial Zak phase ($\Theta_n^{\text{Zak}} = \pi$). Conversely, at $\Delta_L = 0.0149$, which is located beyond the topological transition point, the computed Zak phase is strictly trivial ($\Theta_n^{\text{Zak}} = 0$). These findings confirm that the topological invariants governed by the Zak phase are inherently protected against moderate material and geometric perturbations across all three structural models.

These stochastic evaluations show that the topological invariants governed by the Zak phase remain the same as the deterministic ones under substantive small material and geometric perturbations across all three structural models. Even when subjected to continuous small spatial disorder, the topological phase transitions strictly dictate the nature of the adjacent bandgaps, indicating the fundamental wave localization properties. The proposed method can be also use in investigations on the robustness of SSH topological crystals under different circumstances, where the robustness of topological edge states can be lost.

\section{Conclusions} \label{Sec7}

In this work, we have demonstrated the equivalence between dispersion relations computed via a linear time-varying (LTV) method and the spectral element method (SEM) combined with the transfer matrix method (TMM). This equivalence was mathematically established and numerically verified through deterministic analyses of band diagrams and forced responses on finite structures, encompassing elementary rod, Saint-Venant shaft, and Euler-Bernoulli beam elements. The deterministic formulation was subsequently expanded into a stochastic linear time-varying (SLTV) framework. By integrating the SLTV method with closed-form spatial stochastic fields generated via the stochastic Fourier series (SFS) and analytical Karhunen–Loève expansion (KLE), we efficiently computed stochastic dispersion diagrams and forced responses validated by Monte Carlo simulations.  

A primary advantage of this state-space framework is its capacity to isolate individual wavemodes without the computationally expensive mode-tracking step required by conventional eigenproblem formulations. This characteristic proved highly advantageous for evaluating topological invariants, specifically the Zak phase, allowing for a robust assessment of topological phase transitions in Su–Schrieffer–Heeger (SSH) metamaterials subjected to continuous spatial variability.  

Despite these advantages, the current approach is limited to one-dimensional equivalent waveguides. Extending this exact state-space formulation to two- or three-dimensional phononic topologies, or incorporating coupled wavemodes, significantly increases the complexity of the fundamental matrices and the associated computational overhead. Furthermore, the reliance on analytical dynamic equations restricts the framework to spatial variability levels where the underlying structural continuum theories remain valid.

Future work will focus on extending the proposed LTV framework to accommodate two- and three-dimensional periodic structures, as well as higher-order topological insulators. Additionally, the methodology will be further investigated concerning applications in the fabrication of microscale waveguiding devices and experimentally validated using additively manufactured metastructures exhibiting spatially correlated variability.  

\section*{Acknowledgements}
LHMSR and JRFA gratefully acknowledge the financial support of the São Paulo Research Foundation (FAPESP) through processes number 2020/15328-5, 2019/00315-8, and 2018/15894-0 and the Brazilian National Council of Research CNPq (Grant Agreement ID: 305293/2021-4 and 314168/2020-6). LHMSR, AS, and DC gratefully acknowledge the FWO Onderzoeksprojecten PARADIGM: oPtimal Acoustic lenses enabled through a quantum gRAph DesIGn Methodology (G009625N).
	
\bibliographystyle{ieeetr}
\bibliography{References}

\begin{appendix}

    \section{Details on the LTV and SEM formulation for EB beam theory}
    
    \label{AppendixA}
    
    \setcounter{equation}{0}
    \renewcommand{\theequation}{A.\arabic{equation}}
    
    \setcounter{figure}{0} 
    \renewcommand\thefigure{A.\arabic{figure}} 

For the LTV-based method, when the Eq. \eqref{Eq.29} is expanded as
\begin{align}\label{Eq.030}
\begin{split}
\boldsymbol{\psi}_{C_{Eb},n} =& \begin{bmatrix}
	-1/k_{Eb,n}^3 & -i/k_{Eb,n}^3 & i/k_{Eb,n}^3 & 1/k_{Eb,n}^3\\
	1/k_{Eb,n}^2 & -1/k_{Eb,n}^2 & -1/k_{Eb,n}^2 & 1/k_{Eb,n}^2\\
	-1/k_{Eb,n} & i/k_{Eb,n} & -i/k_{Eb,n} & 1/k_{Eb,n}\\
	1 & 1 & 1 & 1
\end{bmatrix}, \hspace{0.5cm}
\boldsymbol{\Lambda}_{C_{Eb},n}= \begin{bmatrix}
	e^{-k_{Eb,n}\Delta_n} & 0 & 0 & 0\\
	0 & e^{-ik_{Eb,n}\Delta_n} & 0 & 0\\
	0 & 0 & e^{ik_{Eb,n}\Delta_n} & 0\\
	0 & 0 & 0 & e^{k_{Eb,n}\Delta_n}
\end{bmatrix}\\
	\boldsymbol{\psi}_{C_{Eb},n}^{-1}  = \frac{1}{4}&
\begin{bmatrix}
	-k_{Eb,n}^3 & k_{Eb,n}^2 & -k_{Eb,n} & 1\\
	ik_{Eb,n}^3 & -k_{Eb,n}^2 & -ik_{Eb,n} & 1\\
	-ik_{Eb,n}^3 & -k_{Eb,n}^2 & ik_{Eb,n} & 1\\
	k_{Eb,n}^3 & k_{Eb,n}^2 & k_{Eb,n} & 1
\end{bmatrix}.
\end{split}
\end{align}
Hence, the transfer matrix computed from the LTV-based method for the $n$-th homogeneous segment of the Euler-Bernoulli beam element is defined as
\begin{equation}\label{Eq.031}
\mathbf{T}_{\text{LTV}_{Eb},n} = \frac{1}{4}
\begin{bmatrix}
	T_{\text{LTV}_{Eb},n,11} & T_{\text{LTV}_{Eb},n,12} & T_{\text{LTV}_{Eb},n,13} & T_{\text{LTV}_{Eb},n,14} \\
	T_{LTV,n,21} & T_{LTV,n,22} & T_{LTV,n,23} & T_{LTV,n,24} \\
	T_{LTV,n,31} & T_{LTV,n,32} & T_{LTV,n,33} & T_{LTV,n,34} \\
	T_{LTV,n,41} & T_{LTV,n,42} & T_{LTV,n,43} & T_{LTV,n,44} 
\end{bmatrix}
\end{equation}
with the matrix entries given by 
\begin{align}\label{Eq.032}
\begin{split}
	T_{\text{LTV}_{Eb},n,11} &= \left( e^{k_{Eb,n}\Delta_n} + e^{-k_{Eb,n}\Delta_n}\right)  + \left( e^{ik_{Eb,n}\Delta_n} + e^{-ik_{Eb,n}\Delta_n}\right), \\
	T_{\text{LTV}_{Eb},n,12} &= \frac{1}{k_{Eb,n}}\left[ \left( e^{k_{Eb,n}\Delta_n} - e^{-k_{Eb,n}\Delta_n}\right)  + \left( -ie^{ik_{Eb,n}\Delta_n} + ie^{-ik_{Eb,n}\Delta_n}\right)\right],  \\
	T_{\text{LTV}_{Eb},n,13} &= \frac{1}{E_nI_nk_{Eb,n}^3} \left[ \left( e^{k_{Eb,n}\Delta_n} - e^{-k_{Eb,n}\Delta_n}\right)  + \left( ie^{ik_{Eb,n}\Delta_n} - ie^{-ik_{Eb,n}\Delta_n}\right) \right], \\
	T_{\text{LTV}_{Eb},n,14} &= \frac{1}{E_nI_nk_{Eb,n}^2}\left[ -\left( e^{k_{Eb,n}\Delta_n} + e^{-k_{Eb,n}\Delta_n}\right)  + \left( e^{ik_{Eb,n}\Delta_n} + e^{-ik_{Eb,n}\Delta_n}\right) \right], \\
	T_{\text{LTV}_{Eb},n,21} &= k_{Eb,n} \left[ \left( e^{k_{Eb,n}\Delta_n} - e^{-k_{Eb,n}\Delta_n}\right)  + \left( ie^{k_{Eb,n}\Delta_n} - ie^{-k_{Eb,n}\Delta_n}\right) \right], \\
	T_{\text{LTV}_{Eb},n,22} &= \left( e^{k_{Eb,n}\Delta_n} + e^{-k_{Eb,n}\Delta_n}\right)  + \left( e^{ik_{Eb,n}\Delta_n} + e^{-ik_{Eb,n}\Delta_n}\right), \\
	T_{\text{LTV}_{Eb},n,23} &= \frac{1}{E_nI_nk_{Eb,n}^2}\left[  \left( e^{k_{Eb,n}\Delta_n} + e^{-k_{Eb,n}\Delta_n}\right) -\left( e^{ik_{Eb,n}\Delta_n} + e^{-ik_{Eb,n}\Delta_n}\right) \right],  \\
	T_{\text{LTV}_{Eb},n,24} &=\frac{1}{E_nI_nk_{Eb,n}} \left[ -\left( e^{k_{Eb,n}\Delta_n} - e^{-k_{Eb,n}\Delta_n} \right)  + \left( ie^{ik_{Eb,n}\Delta_n} - ie^{-ik_{Eb,n}\Delta_n}\right) \right], \\
	T_{\text{LTV}_{Eb},n,31} &= E_nI_nk_{Eb,n}^3\left[  \left( e^{k_{Eb,n}\Delta_n} - e^{-k_{Eb,n}\Delta_n}\right)  - \left( ie^{ik_{Eb,n}\Delta_n} - ie^{-ik_{Eb,n}\Delta_n} \right) \right], \\
	T_{\text{LTV}_{Eb},n,32} &= E_nI_nk_{Eb,n}^2\left[  \left( e^{k_{Eb,n}\Delta_n} + e^{-k_{Eb,n}\Delta_n}\right)  - \left( e^{ik_{Eb,n}\Delta_n} +  e^{-ik_{Eb,n}\Delta_n} \right), \right]  \\
	T_{\text{LTV}_{Eb},n,33}  &= \left( e^{k_{Eb,n}\Delta_n} + e^{-k_{Eb,n}\Delta_n}\right)  + \left( e^{ik_{Eb,n}\Delta_n} + e^{-ik_{Eb,n}\Delta_n}\right), \\
	T_{\text{LTV}_{Eb},n,34} &= k_{Eb,n}\left[  -\left( e^{k_{Eb,n}\Delta_n} - e^{-k_{Eb,n}\Delta_n}\right)  - \left( e^{ik_{Eb,n}\Delta_n} - e^{-ik_{Eb,n}\Delta_n} \right) \right],  \\
	T_{\text{LTV}_{Eb},n,41} &= E_nI_nk_{Eb,n}^2\left[  - \left( e^{k_{Eb,n}\Delta_n} + e^{-k_{Eb,n}\Delta_n}\right)  - \left( e^{ik_{Eb,n}\Delta_n} +  e^{-ik_{Eb,n}\Delta_n} \right) \right],  \\
	T_{\text{LTV}_{Eb},n,42} &= E_nI_nk_{Eb,n}\left[  -\left( e^{k_{Eb,n}\Delta_n} - e^{-k_{Eb,n}\Delta_n}\right)  - \left( e^{ik_{Eb,n}\Delta_n} -  e^{-ik_{Eb,n}\Delta_n} \right) \right],  \\
	T_{\text{LTV}_{Eb},n,43} &= \frac{1}{k_{Eb,n}}\left[ -\left( e^{k_{Eb,n}\Delta_n} - e^{-k_{Eb,n}\Delta_n} \right)  + \left( ie^{ik_{Eb,n}\Delta_n} - ie^{-ik_{Eb,n}\Delta_n}\right) \right], \\
	T_{\text{LTV}_{Eb},n,44} &= \left( e^{k_{Eb,n}\Delta_n} + e^{-k_{Eb,n}\Delta_n}\right)  + \left( e^{ik_{Eb,n}\Delta_n} + e^{-ik_{Eb,n}\Delta_n}\right). \\	\end{split}
\end{align}

For the SEM solution, the dynamic stiffness matrix can be obtained by combining Eq. \eqref{Eq.37} and Eq. \eqref{Eq.39} as the product of the matrices $\mathbf{D}_{eb,n} = \mathbf{F} \mathbf{A}^{-1}$.
 Then, the transfer matrix can be obtained using the $2\times2$ partitioned dynamic stiffness matrix for the $n$-th segment for the Euler-Bernoulli beam as \cite{hussein2014dynamics}
\begin{equation}\label{Eq.41}
	\mathbf{T}_{\text{SEM}_{Eb},n}
	= \frac{1}{4} \begin{bmatrix}
		T_{\text{SEM}_{Eb},n,11} & T_{\text{SEM}_{Eb},n,12} & T_{\text{SEM}_{Eb},n,13} & T_{\text{SEM}_{Eb},n,14} \\
		T_{\text{SEM}_{Eb},n,21} & T_{\text{SEM}_{Eb},n,22} & T_{\text{SEM}_{Eb},n,23} & T_{\text{SEM}_{Eb},n,24} \\
		T_{\text{SEM}_{Eb},n,31} & T_{\text{SEM}_{Eb},n,32} & T_{\text{SEM}_{Eb},n,33} & T_{\text{SEM}_{Eb},n,34} \\
		T_{\text{SEM}_{Eb},n,41} & T_{\text{SEM}_{Eb},n,42} & T_{\text{SEM}_{Eb},n,43} & T_{\text{SEM}_{Eb},n,44} 
	\end{bmatrix},
\end{equation}
where
\begin{align}\label{Eq.42}
	\begin{split}
	T_{\text{SEM}_{Eb},n,11} &= \left( e^{k_{Eb,n}\Delta_n}+e^{-k_{Eb,n}\Delta_n}\right)  +\left( e^{ik_{Eb,n}\Delta_n}+e^{-ik_{Eb,n}\Delta_n}\right), \\
	T_{\text{SEM}_{Eb},n,12} &= \frac{1}{k_{Eb,n}}\left[ \left( e^{k_{Eb,n}\Delta_n} - e^{-k_{Eb,n}\Delta_n}\right)  - \left( ie^{ik_{Eb,n}\Delta_n} - ie^{-ik_{Eb,n}\Delta_n} \right)  \right], \\
	T_{\text{SEM}_{Eb},n,13} &= \frac{1}{E_nI_nk_{Eb,n}^3} \left[  -\left( e^{k_{Eb,n}\Delta_n}- e^{-k_{Eb,n}\Delta_n} \right) + \left( ie^{ik_{Eb,n}\Delta_n} - ie^{-ik_{Eb,n}\Delta_n}\right)\right],  \\
	T_{\text{SEM}_{Eb},n,14} &= \frac{1}{E_nI_nk_{Eb,n}^2} \left[  -\left( e^{k_{Eb,n}\Delta_n}+ e^{-k_{Eb,n}\Delta_n} \right) + \left( e^{ik_{Eb,n}\Delta_n} + e^{-ik_{Eb,n}\Delta_n}\right)\right],  \\
	T_{\text{SEM}_{Eb},n,21} &= k_{Eb,n} \left[  \left( e^{k_{Eb,n}\Delta_n}- e^{-k_{Eb,n}\Delta_n} \right) + \left( ie^{ik_{Eb,n}\Delta_n} - ie^{-ik_{Eb,n}\Delta_n}\right)\right], \\
	T_{\text{SEM}_{Eb},n,22} &= \left( e^{k_{Eb,n}\Delta_n}+e^{-k_{Eb,n}\Delta_n}\right)  +\left( e^{ik_{Eb,n}\Delta_n}+e^{-ik_{Eb,n}\Delta_n}\right), \\
	T_{\text{SEM}_{Eb},n,23} &= \frac{1}{E_nI_nk_{Eb,n}^2} \left[  \left( e^{k_{Eb,n}\Delta_n}+ e^{-k_{Eb,n}\Delta_n} \right) - \left( e^{ik_{Eb,n}\Delta_n} + e^{-ik_{Eb,n}\Delta_n}\right)\right], \\
	T_{\text{SEM}_{Eb},n,24} &= \frac{1}{E_nI_nk_{Eb,n}} \left[  -\left(e^{k_{Eb,n}\Delta_n}- e^{-k_{Eb,n}\Delta_n} \right) + \left( ie^{ik_{Eb,n}\Delta_n} - ie^{-ik_{Eb,n}\Delta_n}\right)\right], \\
	T_{\text{SEM}_{Eb},n,31} &= E_nI_nk_{Eb,n}^3 \left[  \left(e^{k_{Eb,n}\Delta_n}- e^{-k_{Eb,n}\Delta_n} \right) - \left( ie^{ik_{Eb,n}\Delta_n} - ie^{-ik_{Eb,n}\Delta_n}\right)\right], \\
	T_{\text{SEM}_{Eb},n,32} &= E_nI_nk_{Eb,n}^2 \left[  \left(e^{k_{Eb,n}\Delta_n}+ e^{-k_{Eb,n}\Delta_n} \right) - \left( e^{ik_{Eb,n}\Delta_n} + e^{-ik_{Eb,n}\Delta_n}\right)\right] \\
	T_{\text{SEM}_{Eb},n,33}  &= \left( e^{k_{Eb,n}\Delta_n}+e^{-k_{Eb,n}\Delta_n}\right)  +\left( e^{ik_{Eb,n}\Delta_n}+e^{-ik_{Eb,n}\Delta_n}\right), \\
	T_{\text{SEM}_{Eb},n,34} &= k_{Eb,n} \left[  \left(e^{k_{Eb,n}\Delta_n}- e^{-k_{Eb,n}\Delta_n} \right) - \left( ie^{ik_{Eb,n}\Delta_n} - ie^{-ik_{Eb,n}\Delta_n}\right)\right], \\
	T_{\text{SEM}_{Eb},n,41} &= E_nI_nk_{Eb,n}^2 \left[  -\left(e^{k_{Eb,n}\Delta_n}+ e^{-k_{Eb,n}\Delta_n} \right) + \left( ie^{ik_{Eb,n}\Delta_n} + ie^{-ik_{Eb,n}\Delta_n}\right)\right], \\
	T_{\text{SEM}_{Eb},n,42} &= E_nI_nk_{Eb,n} \left[  -\left(e^{k_{Eb,n}\Delta_n}- e^{-k_{Eb,n}\Delta_n} \right) - \left( ie^{ik_{Eb,n}\Delta_n} - ie^{-ik_{Eb,n}\Delta_n}\right)\right], \\
	T_{\text{SEM}_{Eb},n,43} &= \frac{1}{k_{Eb,n}} \left[  -\left(e^{k_{Eb,n}\Delta_n}- e^{-k_{Eb,n}\Delta_n} \right) - \left( ie^{ik_{Eb,n}\Delta_n} + ie^{-ik_{Eb,n}\Delta_n}\right)\right], \\
	T_{\text{SEM}_{Eb},n,44} &= \left( e^{k_{Eb,n}\Delta_n}+e^{-k_{Eb,n}\Delta_n}\right) + \left( e^{ik_{Eb,n}\Delta_n}+e^{-ik_{Eb,n}\Delta_n}\right).
	\end{split}
	\end{align}

    \section{Wavenumber and wavemode computation using SEM}
    
    \label{AppendixB}
    
    \setcounter{equation}{0}
    \renewcommand{\theequation}{B.\arabic{equation}}
    
    \setcounter{figure}{0} 
    \renewcommand\thefigure{B.\arabic{figure}}

The dynamic condensation \cite{leung1978accurate} can be applied to $\mathbf{D}_{G}$ of the unit cell to obtain the condensed dynamic stiffness matrix presented in Eq. \ref{Eq_cond_1}, assuming the edge degrees-of-freedom (DOF) are active ($A$), and the internal DOF are slaves ($S$) (for details, please see \cite{ribeiro2022investigating}).
 
	\begin{equation}\label{Eq_cond_1}
		\mathbf{D}_{C}=\mathbf{D}_{G,AA}-\mathbf{D}_{G,AS}\mathbf{D}_{G,SS}^{-1} \mathbf{D}_{G,SA} =         \begin{bmatrix}
			\mathbf{D}_{LL}&  \mathbf{D}_{LR}\\ 
			\mathbf{D}_{RL} & \mathbf{D}_{RR}
		\end{bmatrix}.
	\end{equation}
	 The condensed matrix relates $F_1$ and $F_n$ to $u_1$ and $u_n$:
	\begin{equation}\label{Eq_Dc_1}
		\begin{Bmatrix}
			F_{1}\\
			F_{N}
		\end{Bmatrix}
		= \mathbf{D}_{C}
		\begin{Bmatrix}
			u_{1}\\
			u_{N}
		\end{Bmatrix}.
	\end{equation}
	
 From the above equation, it is possible to obtain the transfer matrix in terms of dynamic stiffness sub-matrices. The transfer matrix relates displacements and internal forces at the two ends of the one-dimensional unit cell \cite{mace2005finite, zhong1995direct}.
	\begin{equation}\label{Eq_Dg_1}
		\begin{Bmatrix}
			u_{1}\\
			F_{1}
		\end{Bmatrix}
		= \mathbf{T}
		\begin{Bmatrix}
			u_{N}\\
			F_{N}
		\end{Bmatrix}, 
		\hspace{0.5cm} \text{where}
		\hspace{0.5cm}
		\mathbf{T}=\begin{bmatrix}
			-\mathbf{D}_{LR}^{-1} \mathbf{D}_{LL}& \mathbf{D}_{LR}^{-1}\\ 
			-\mathbf{D}_{RL}+\mathbf{D}_{RR}\mathbf{D}_{LR}^{-1}\mathbf{D}_{LL}&-\mathbf{D}_{RR}\mathbf{D}_{LR}^{-1}
		\end{bmatrix}.
		\hspace{0.5cm}
	\end{equation}

	The exponential term associated with the Bloch solution $e^{-ik_xL_x}$, where $k_x$ is the complex wavenumber and $L_x$ is the unit cell length, also relates the quantities at the ends of the unit cell of the periodic structure. By combining these two relationships, one obtains an eigenproblem where the eigenvalues contain the wavenumbers for a given frequency $\omega$ \cite{cicirello2020sensitivity}:
	\begin{equation}\label{Eq.w19_2}
		\mathbf{T} \begin{Bmatrix}
                 u_{N}\\ 
                F_{N}
		\end{Bmatrix}  = e^{ik_xL_x} 
            \begin{Bmatrix}
                 u_{N}\\ 
                F_{N}
		\end{Bmatrix}.
	\end{equation}
	with eigenvectors containing the wavemode shapes condensed at the ends, active DOF. Because of the system's characteristics, the eigenvalues are given in complex conjugate pairs, and the associated eigenvector represents positive-going and negative-going wavemodes \cite{mace2005finite, mencik2015wave}. The internal, slave DOF of the wavemodes are obtained from the expansion:
	\begin{equation}
		\begin{Bmatrix}
			u_2\\
			\vdots\\
			u_{N-1}
		\end{Bmatrix} =
		-\mathbf{D}_{G,SS}^{-1}\mathbf{D}_{G,SA}
		\begin{Bmatrix}
			u_1\\
			u_{N}
		\end{Bmatrix}.
	\end{equation}

    \end{appendix}

\end{document}